\documentclass{fundam}
\usepackage{amsmath}
\usepackage{amsfonts}
\usepackage{cleveref}
\usepackage{thm-restate}
\usepackage{graphicx}
\usepackage[style=base]{caption}
\usepackage{subcaption}
\usepackage{tikz}
\usetikzlibrary{positioning,calc,shapes.geometric,arrows.meta}
\definecolor{SAfigureRandom}{RGB}{0,114,178}
\definecolor{SAfigureSwitch}{RGB}{0,158,115}
\definecolor{SAfigureSwitchLight}{RGB}{86,180,153}
\definecolor{SAfigurePlayerOne}{RGB}{213,94,0}
\definecolor{SAfigurePlayerTwo}{RGB}{204,121,167}
\definecolor{SAfigureWarm}{RGB}{213,94,0}
\definecolor{SAedgeReturn}{RGB}{120,90,20}
\definecolor{SAedgeConsequence}{RGB}{0,120,130}
\definecolor{SAedgeFailure}{RGB}{170,60,90}
\tikzset{
  every picture/.style={line width=0.9pt},
  sa random/.style={draw=SAfigureRandom},
  sa switch/.style={draw=SAfigureSwitch},
  sa player one/.style={draw=SAfigurePlayerOne},
  sa player two/.style={draw=SAfigurePlayerTwo},
  sa node label/.style={anchor=center,inner sep=0pt,font=\large,align=center},
  sa node label large/.style={anchor=center,inner sep=0pt,font=\Large,align=center},
  sa node label compact/.style={anchor=center,inner sep=0pt,font=\normalsize,align=center},
  sa triangle label/.style={sa node label,yshift=-1pt},
  sa edge label/.style={anchor=center,inner sep=1pt,font=\normalsize,align=center,fill=white,fill opacity=.85,text opacity=1},
  sa gadget label/.style={anchor=center,inner sep=0pt,font=\normalsize,align=center},
  sa external label/.style={anchor=north west,inner sep=0.75pt,font=\normalsize,align=left},
  sa route return/.style={draw=SAedgeReturn},
  sa route consequence/.style={draw=SAedgeConsequence},
  sa route failure/.style={draw=SAedgeFailure}
}
\usepackage{makecell} 
\usepackage{complexity} 

\usepackage{preamble}

\begin{document}
\emergencystretch=2em 
\title{The Stochastic Arrival Problem}
%
%
\author{Thomas Webster}
\address{tw423@cantab.ac.uk}
\setcounter{page}{1}
\publyear{2021}
\papernumber{0001}
\volume{178}
\issue{1}
\runninghead{T. Webster}{The Stochastic Arrival Problem}

\maketitle              
\begin{abstract}
{\tt Arrival} is a deterministic reachability problem whose local transition rules cycle through prescribed switching orders. Its decision complexity remains unresolved: no polynomial-time algorithm is known, and neither is a \P{}-hardness result. We study stochastic Arrival, which augments switching vertices with random and player-controlled vertices and thereby connects Arrival with reachability switching games and finite stochastic games.

For the basic model with switching and random vertices, positive and almost-sure reachability are respectively \NP{}-complete and \coNP{}-complete. End-component compression shows that the almost-sure problem is polynomial-time equivalent to the corresponding Player~2 reachability switching game. For quantitative reachability, deciding whether the value exceeds one half is \PP{}-hard, while exact comparison with one half is in \PSPACE{}. Adding player-controlled vertices makes the quantitative problems \PSPACE{}-hard; almost-sure reachability is also \PSPACE{}-hard whenever Player~1 is present, and every qualitative variant is in \EXPTIME{}. After polynomial-time preprocessing that collapses target-dead behaviour to a losing sink, the switching-and-random model has an exponential expected absorption-time bound.
\end{abstract}
\begin{list}{}{\small\rm\leftmargin=1cm\rightmargin=1cm\labelwidth=0pt\itemindent=0pt}
\item[]\textbf{ACM Subject Classification:} Theory of computation---Problems, reductions and completeness; Random walks and Markov chains; Algorithmic game theory.
\end{list}

\section{Introduction}
\label{sec:introduction}

{\tt Arrival} is a simple-to-state decision problem introduced by Dohrau, G\"{a}rtner, Kohler, Matou\v{s}ek and Welzl \cite{DGKMW16}. Informally, it asks whether a token moving through the vertices of a finite directed graph, starting at a designated origin, eventually reaches a given target. Each vertex is equipped with a fixed cyclic ordering of its outgoing edges. On successive visits to a vertex the token takes successive edges in this ordering, returning to the beginning once the ordering is exhausted. This process is known as ``switching''. It may be viewed as a deterministic analogue of a random walk, or as a transition system equipped with local deterministic cyclic schedulers.

Dohrau et al.\ showed that {\tt Arrival} lies in $\NP\cap\coNP$, but it remains open whether the problem is in \P{}. Subsequent work has placed the associated search problem in several subclasses of \TFNP{}, including \PLS{} \cite{Kar17}, \CLS{} \cite{GHH+18}, and \UniqueEOPL{} \cite{FGMS19}, and has shown that {\tt Arrival} lies in $\UP\cap\coUP$ \cite{GHH+18}. Known lower bounds include \PL{}-hardness and \CC{}-hardness \cite{Man21}. On the algorithmic side, G\"{a}rtner, Haslebacher and Hoang gave the first subexponential algorithm, running in time $2^{\bigO(\sqrt{n}\log n)}$, together with a polynomial-time algorithm for almost-acyclic instances \cite{GHH21}. Auger et al.\ subsequently gave a polynomial-time algorithm on tree-like multigraphs \cite{ACD22}. More recently, Haslebacher gave an algorithm running in $2^{\bigO(k\log^2 n)}$ time on graphs of treewidth $k$, derived from a recursive framework for the more general {\tt G-Arrival} problem, and related {\tt G-Arrival} to approximate fixed points of $\ell_1$-contractions \cite{Has25}. For the related multi-token problem, Ghorbani, Hoff and Mnich gave a quasi-polynomial-time algorithm for {\tt Multi-Arrival} on tree-like multigraphs, together with polynomial-time algorithms for further restricted cases \cite{GHM25}. G\"{a}rtner, Haslebacher and Hoang have also shown that {\tt Arrival} is polynomial-time solvable on ladder graphs \cite{GHH26}.

The complexity of {\tt Arrival} is particularly interesting in the context of games on graphs. Simple stochastic games, mean-payoff games, and parity games are prominent examples for which the two-player decision problems have strong complexity upper bounds, while their corresponding one-player variants are polynomial-time solvable \cite{Con92,ZP96,Jur98}. {\tt Arrival}, by contrast, is a zero-player problem for which no polynomial-time algorithm is known. Fearnley, Gairing, Mnich and Savani introduced Reachability Switching Games by adding player-controlled vertices to switching graphs \cite{FGMS21}. They showed, in particular, that the one-player reachability problem is \NP{}-complete and that the two-player problem is \PSPACE{}-hard.

Recent work has also made the relationship between {\tt Arrival} and simple stochastic games more concrete. Haslebacher's treewidth algorithm and contraction formulation exhibit parallel phenomena for {\tt Arrival} and simple stochastic games \cite{Has25}, while G\"{a}rtner, Haslebacher and Hoang obtain a common tractability result for {\tt Arrival} and stopping binary simple stochastic games on ladders \cite{GHH26}. These similarities provide further motivation for studying models in which switching and probabilistic behaviour occur together.

We build upon Reachability Switching Games by introducing probabilistic vertices, at which the successor is chosen according to a specified probability distribution. Our generalised {\tt Arrival} graphs may therefore contain four kinds of vertices: probabilistic vertices, switching vertices, vertices controlled by a maximising player, and vertices controlled by a minimising player. Different choices of which vertex types are present recover several familiar models. Switching vertices alone give the original {\tt Arrival} problem; switching together with controlled vertices gives Reachability Switching Games; and in the absence of switching vertices we recover Markov chains, Markov decision processes, and simple stochastic games.

The conference version of this paper introduced the stochastic Arrival framework and proved the principal quantitative hardness results \cite{W22}. The present journal version develops and streamlines those reductions, adds the qualitative-$1$ \PSPACE{}-hardness result and general upper bounds, proves an exponential expected-absorption bound after value-preserving preprocessing, and gives a complete \PSPACE{} upper bound for {\tt $\{R,S\}$-Arrival-Quant}. Results obtained directly by expanding the switching positions into an exponential finite stochastic game are separated from the new results proved specifically for stochastic Arrival.

Our main results concern mixtures of switching and randomisation. For the basic $\{R,S\}$ model, positive reachability is \NP{}-complete and almost-sure reachability is \coNP{}-complete. End-component compression further shows that the latter is polynomial-time equivalent to {\tt $\{S,2\}$-Arrival} and to {\tt $\{R,S,2\}$-Arrival-Qual-1}, making all three \coNP{}-complete. The quantitative threshold problem is \PP{}-hard and belongs to \PSPACE{}. Since $\NP\subseteq\PP$, the \PP{}-hardness also implies \NP{}-hardness and thus rules out a polynomial-time algorithm under $\P\ne\NP$. After polynomial preprocessing that redirects $d$-dead behaviour to an absorbing losing sink, we obtain an exponential expected-absorption bound for the resulting value-preserving instance. For variants with player-controlled nodes, the quantitative problems are \PSPACE{}-hard. Qualitative-$1$ is also \PSPACE{}-hard whenever Player~1 is available, while every qualitative variant is in \EXPTIME{}. Table~\ref{tab:complexity} gives an overview, including the classical switching-only variants.

The remainder of the paper is organised as follows. Section~2 introduces generalised {\tt Arrival} graphs and the qualitative and quantitative decision problems considered in the paper, together with preliminary reductions and upper bounds. Section~3 proves \PSPACE{}-hardness of the quantitative problems containing random, switching, and player-controlled vertices. Section~4 considers qualitative problems for random and switching vertices, including the expected-absorption-time bound and the \NP{}- and \coNP{}-completeness results. Section~5 establishes \PP{}-hardness of quantitative {\tt $\{R,S\}$-Arrival}, and Section~6 gives the \PSPACE{} upper bound currently known for this problem. We conclude with the remaining complexity gaps and directions for further work.

\section{Preliminaries}\label{sec:preliminaries}

Reachability Switching Games (RSGs) extend {\tt Arrival} with player-controlled nodes \cite{FGMS21}. We add random nodes and use one notation for all resulting combinations. An instance consists of a finite directed graph $G$ with start vertex $o$, target vertex $d$, and a partition of the vertices by node type. The four allowed node types are as follows:
\begin{itemize}
    \item\textbf{Random nodes} - We denote the set of random nodes by $V_R \in \mathcal{V}$ and require a probability distribution $\Prob$ on their outgoing edges.
    \item\textbf{Switching nodes} - We call the set of switching nodes $V_S\in\mathcal{V}$, and require an ordering, $\Ord$, to be given on their outgoing edges.
    \item\textbf{Player~1 nodes} - We denote by $V_1\in\mathcal{V}$ the nodes controlled by Player~1, whose objective is to reach the target $d$.
    \item\textbf{Player~2 nodes} - We denote by $V_2\in\mathcal{V}$ the nodes controlled by Player~2, whose objective is to prevent the play from reaching the target $d$.
\end{itemize}

We use a set $\mathcal{B}\subseteq\{R,S,1,2\}$ to denote the node types that are \emph{allowed} in an instance; a set $V_\sigma$ with $\sigma\in\mathcal B$ is permitted to be empty, while types outside $\mathcal B$ are absent. This convention makes the natural inclusions between problem classes literal: for example, every $\{S,1\}$-instance is also an $\{R,S,1\}$-instance by taking $V_R=\emptyset$. The original {\tt Arrival} model studied in \cite{DGKMW16} corresponds to $\mathcal{B} = \{ S \}$. Reachability Switching Games defined in \cite{FGMS21} correspond to $\mathcal{B} = \{S , 1\}$ and $\mathcal{B} = \{S , 1, 2\}$. Taking $\mathcal{B}\subseteq\{R,1,2\}$ gives the usual Markov-chain, Markov-decision-process, and simple-stochastic-game special cases. Generalised Arrival graphs have the following structure.

\begin{definition}
\label{sa:def:basicarrivalstruct}
A generalised Arrival graph is a tuple
\[G:=(V,E,\mathcal{V},\Prob,\Ord),\]
where:
\begin{itemize}
    \item $(V,E)$ is a finite directed graph.
    \item For all $v \in V$, we require $\outdeg(v) \geq 1$, and we allow self-loop edges of the form $(v,v)$.
    \item $\mathcal{V}\subseteq\mathcal{P}(V)$ is a partition of the vertices of $V$ into different node types. Often we will take $\mathcal{V}=\{V_R,V_S,V_1,V_2\}$, omitting empty sets, with each of these sets as described above.
    \item A function $\Prob:V_R\cross V\to[0,1]\cap\mathbb Q$ with the properties that for any $v\in V_R$ we have $\sum_{w\in V}\Prob(v,w)=1$ and where $\Prob(v,w)>0$ if and only if $(v,w)\in E$. Thus, for fixed $v\in V_R$, the function $\Prob(v,\cdot)$ is a probability distribution on the successors of $v$. Every probability is part of the finite input and is encoded as a binary rational (a pair of binary integers giving numerator and positive denominator).
    \item A function $\Ord: V_S\to V^+$ from switching nodes to a finite sequence of vertices, represented explicitly as part of the input. We require that, for $v\in V_S$,  $(v,w) \in E$ if and only if there exists an $i$ such that $w=\Ord(v)_i$. So every outgoing edge from $v$ is used in $\Ord(v)$, possibly more than once.
\end{itemize}
\end{definition}

For $v\in V$ write $\mathrm{Succ}(v):=\{w:(v,w)\in E\}$ for its successors; for $C\subseteq V$, the notation $G[C]$ denotes the subgraph induced by $C$. In particular, the switching-order condition for a vertex of $C$ is tested against its original order in $G$.

To define our problems we also require vertices $o,d\in V$: $o$ is called the {\em start}; and, $d$ the {\em target} node. \footnote{We use $o$ for ``origin'' and $d$ for ``destination''. Other sources use $s$ for ``source'' and $t$ for ``target''.} Since our objective is reachability, replacing all outgoing behaviour of $d$ by a self-loop does not alter any value. We therefore assume throughout, without loss of generality, that $d$ is absorbing. Whenever we introduce $\overline d$, it denotes an absorbing \emph{losing sink}. We reserve \emph{$d$-dead} for vertices from which the target cannot be reached under the reachability-game definition below; a $d$-dead vertex need not itself be a sink.

 Given such a model, we wish to define a play of the game. To do so we first need to define the current state. Due to how switching nodes work we will also include the current positions of those nodes in our game state.
 
\begin{definition}
\label{sa:def:swithcingposition}
Given a set of switching nodes $V_S$ the {\em current switching node position} is a function $q: V_S\to\nat$, i.e., a function from vertices to natural numbers, where we require that $\forall v\in V_S$, $q(v) < \abs{\Ord(v)}$.
We call the set of all such position functions $Q$. If there are no switching vertices then $Q$ is a singleton containing only the empty function.
\end{definition}
\begin{definition}
\label{sa:def:arrivalstate}
A {\em state of the game} consists of an ordered pair $(v,q)\in V\cross Q$ with $v\in V$ denoting the current vertex, and $q\in Q$, denoting the current position of the switching nodes. Thus we call the set $V \cross Q$ our state space.
\end{definition}

Now that we have a state space we can define valid transitions between states. 

\begin{definition}
\label{sa:def:validtransitions}
For a generalised Arrival graph $G$ we let $\Valid_{G}:V\cross Q\to\mathcal{P}(V\cross Q)$ be the function defined as follows:
\begin{itemize}
    \item For $v\in V_S$ and any $q\in Q$, where by definition $q:V_S\to\nat$, we define $\Valid_{G}(v,q)$ as the singleton $\{(u,q^\prime)\}$, where $u$ and $q^\prime$ are defined as follows:
    \begin{itemize}
        \item Suppose $\Ord(v) =(u_0,\ldots,u_{k-1})$. We let $u:=u_{q(v)}$.  Note that this is well defined, i.e., $0\leq q(v)<\abs{\Ord(v)}=k$, because $(v,q)$ is a state.
        \item For $x\in V_S$ with $x\neq v$ we let $q^\prime(x):=q(x)$.
        \item Furthermore, we let $q^\prime(v):= (q(v)+1 \mod k)$.
    \end{itemize}
    \item For $v\in V_1\cup V_2$ and any $q\in Q$, we let $\Valid(v,q):=\{(u,q) : (v,u)\in E\}$.
    \item For $v\in V_R$ and any $q\in Q$ we let $\Valid(v,q):=\{(u,q): \Prob(v,u)>0\}$
\end{itemize}
We call a transition from a state $(v,q)$ to a state $(u,q') \in \Valid(v,q)$  {\em valid}, and otherwise we call it {\em invalid}.
\end{definition}
 It follows directly from the definitions that for any state $(v,q)$, $\Valid(v,q)\neq\emptyset$.

We call an infinite sequence $\pi = (v_0,q_0) (v_1,q_1) (v_2,q_2)\ldots \in(V\cross Q)^\omega$ over the state space $V\cross Q$ a {\em play} if for every $i \in \nat$ we have $(v_{i+1},q_{i+1})\in\Valid(v_i,q_i)$. We use $\Omega$ to denote the set of all (infinite) plays. A {\em partial play} of the game is a finite initial prefix $w \in(V\cross Q)^*$ of a play. For a partial play $w$, we define its {\em basic cylinder},  $\mathsf{C}(w)\subseteq w(V\cross Q)^\omega$, as the set of all plays with $w$ as an initial segment.   We use $\Pi \subseteq (V \times Q)^*$ to denote the set of all finite partial plays. We say a play $\pi$ is {\em winning} for Player~1 if there exist an index $i$ and a switching position $q\in Q$ with $\pi_i=(d,q)$. Otherwise, it is a losing play (winning for Player~2).

We use standard word notation throughout: for an alphabet $\Sigma$, $\Sigma^*$ denotes the finite words, $\Sigma^\omega$ the infinite words, and $\Sigma^+=\Sigma^*\setminus\{\epsilon\}$. Reachability objectives such as $\Reach{d}$ are $\omega$-regular; no result below requires more general $\omega$-regular objectives.

By positional determinacy for finite stochastic reachability games \cite{Con92}, for all our generalised Arrival games it suffices to consider deterministic ``essentially memoryless'' strategies for a player $i$ given by $\Strat_i:(V_i\cross Q) \rightarrow V$, which ignore the history in a partial play $\pi$, and only considers the current state $(v,q)$ in order to choose (deterministically) a move to the next vertex, $v'$, such that $(v',q) \in \Valid(v,q)$. (Note that switching positions only change during transitions out of switching nodes.) Indeed, we can view our instances of generalised Arrival as defining exponentially larger simple stochastic games over the state space $V\cross Q$, because of the deterministic way the switching position $q$ updates with each transition.

Fixing a start state $o$, and strategies $\sigma_1$ and $\tau_2$ for the two players, naturally determines a probability space $(\Omega,\mathcal{F},\mathbb{P}_{o,\sigma_1,\tau_2})$ on the set $\Omega$ of (infinite) plays starting from state $(o,q^0)$.
Here  $\mathcal{F}$ denotes the Borel $\sigma$-algebra of events generated by the set of basic cylinders $\{ \mathsf{C}(w) \mid w \in \Pi\}$, and $\mathbb{P}_{o,\sigma_1,\tau_2}$ denotes the probability measure defined on events in $\mathcal{F}$ uniquely determined by probabilities of basic cylinders, which are defined inductively in the standard way, starting with the base case given by $\mathbb{P}(\mathsf{C}((o,q^0))):=1$, where by definition  $q^0(v):= 0$ for all $v \in V_S$.
In other words, all plays begin, with probability 1, with state $(o,q^0)$ as the initial state.

\begin{definition}
\label{sa:def:value}
Given a generalised Arrival graph $G = (V,E,\{V_R,V_S, V_1,V_2\},\Prob,\Ord)$, a start $o\in V$ and target $d\in V$ we define the {\em value of the instance} as follows. Let $\Reach{d}\in\mathcal{F}$ be the event $\Reach{d}:=\{\pi = (v_0,q_0) (v_1,q_1) (v_2,q_2)\ldots \in \Omega : \exists i \in \nat , v_i =d\}$ and let $\sigma_1$ and $\tau_2$ range over strategies for each player:
\begin{equation*}
    \val(G,o,d) := \max_{\sigma_1}\min_{\tau_2}\mathbb{P}_{o,\sigma_1,\tau_2}(\Reach{d})
\end{equation*}
We use \emph{value} exclusively for $\val(G,o,d)$ below.
\end{definition}

It follows from known results for simple stochastic games that these games are determined, meaning that $\val(G,o,d) = \min_{\tau_2}\max_{\sigma_1}\mathbb{P}_{o,\sigma_1,\tau_2}(\Reach{d})$ and that these maxima and minima are obtained.

We begin by generalising the notion of a ``hopeful edges'' given by Dohrau et al. \cite{DGKMW16}:

\begin{definition}
\label{sa:def:hopeful}
Given a $\mathcal{B}$-Arrival graph $G=(V,E,\mathcal{V},\Prob,\Ord)$ and target $d$, form the ordinary two-player reachability game on $(V,E)$ in which Player~1 controls $V_R\cup V_S\cup V_1$, Player~2 controls $V_2$, and Player~1 wins by reaching $d$. A vertex $v$ is \emph{$d$-hopeful} if Player~1 has a winning strategy from $v$ in this game. An edge $(v,w)$ is \emph{$d$-hopeful} if $w$ is $d$-hopeful. Vertices and edges that are not $d$-hopeful are \emph{$d$-dead}. When preprocessing introduces a unique absorbing $d$-dead vertex, we denote that losing sink by $\overline d$.
\end{definition}

Here a \emph{reachability game} means the ordinary finite two-player game on a directed graph in which Player~1 chooses at vertices in $V_1^\prime$, Player~2 chooses at vertices in $V_2$, and Player~1 wins exactly by reaching $d$. Thus the definition captures that, even against an adversarial Player~2, there is some choice of random outcomes and switching successors under which the target can still be reached.

The $d$-hopeful vertices can be computed in \NL{} when $V_2=\emptyset$, and in \P{} in general, by solving this ordinary reachability game. We now define different versions of the computational problems we wish to study, using a common notation. As above, a subset $\mathcal{B} \subseteq \{R,S,1,2\}$ denotes the node types allowed in the instances for the problem in question; allowed type sets may be empty. With randomisation there are three key decision problems to study, the first pair are our qualitative decision problems:

\problemStatement{$\mathcal{B}$-Arrival-Qual-0}{\label{prob:B-arrival-qual-0}
Instance={A $\mathcal{B}$-Arrival Graph $G:=(V,E,\{V_\sigma : \sigma\in \mathcal{B}\},\Prob,\Ord)$ and vertices $o,d\in V$.},
Problem={Decide whether or not $\val(G,o,d) > 0$.}
}

\problemStatement{$\mathcal{B}$-Arrival-Qual-1}{\label{prob:B-arrival-qual-1}
Instance={A $\mathcal{B}$-Arrival Graph $G:=(V,E,\{V_\sigma : \sigma\in \mathcal{B}\},\Prob,\Ord)$ and vertices $o,d\in V$.},
Problem={Decide whether or not $\val(G,o,d) = 1$.}
}

Our third decision problem is our quantitative problem, which takes an additional input probability:

\problemStatement{$\mathcal{B}$-Arrival-Quant}{\label{prob:B-arrival-quant}
Instance={A $\mathcal{B}$-Arrival Graph $G:=(V,E,\{V_\sigma : \sigma\in \mathcal{B}\},\Prob,\Ord)$, vertices $o,d\in V$ and a (rational) probability $p\in(0,1)$.},
Problem={Decide whether or not $\val(G,o,d) > p$.}
}

We also use the corresponding non-strict quantitative problem:
\problemStatement{$\mathcal{B}$-Arrival-Quant-Eq}{\label{prob:B-arrival-quant-eq}
Instance={A $\mathcal{B}$-Arrival Graph $G:=(V,E,\{V_\sigma : \sigma\in \mathcal{B}\},\Prob,\Ord)$, vertices $o,d\in V$ and a (rational) probability $p\in(0,1)$.},
Problem={Decide whether or not $\val(G,o,d) \ge p$.}
}

The original {\tt Arrival} problem studied in \cite{DGKMW16} corresponds to the above definition with $\mathcal{B} = \{ S \}$. 
Reachability Switching Games defined in \cite{FGMS21} correspond to $\mathcal{B} = \{S , 1\}$ and $\mathcal{B} = \{S , 1, 2\}$.
Taking $\mathcal{B}\subseteq\{R,1,2\}$ corresponds to Markov Chains, Markov Decision Processes, and Simple Stochastic Games.

When $R\notin \mathcal{B}$ these problems all coincide, since in that case $\val(G,o,d)\in\{0,1\}$ and such instances constitute an (exponentially large) deterministic problem.  In such a case we use {\tt $\mathcal{B}$-Arrival} to refer to the problem of deciding if $\val(G,o,d)=1$. Several of these deterministic problems have previously known complexity. Throughout this work, we aim to show complexity results for the cases when $R \in \mathcal{B}$. 

\begin{proposition}\label[proposition]{sa:res:rec-quant-only-p-half}
For a subset $\mathcal{B}\subseteq\{R,S,1,2\}$ with $R\in \mathcal{B}$, given a $\mathcal{B}$-arrival graph
$G=(V,E,\{V_\sigma:\sigma\in\mathcal B\},\Prob,\Ord)$, vertices $o,d\in V$, and a rational
threshold $p\in(0,1)$, the problem of deciding whether $\val(G,o,d)>p$ is polynomial-time equivalent
to the same problem with threshold $1/2$.
\end{proposition}
\begin{proof}
All constructions add a single random start vertex and use only rational probabilities whose encodings have
size polynomial in the encoding of $p$. We spell out both directions.

To reduce threshold $p$ to threshold $1/2$, if $p>1/2$ let the new start move to $o$ with probability
$1/(2p)$ and to an absorbing losing state with the remaining probability. Then
\[
    \val(G',o',d)=\frac{1}{2p}\val(G,o,d),
\]
so $\val(G,o,d)>p$ iff $\val(G',o',d)>1/2$. If $p<1/2$, let the new start move to $o$ with probability
$1/(2(1-p))$ and directly to $d$ with probability $(1-2p)/(2(1-p))$. Then
\[
    \val(G',o',d)=\frac{1}{2(1-p)}\val(G,o,d)+\frac{1-2p}{2(1-p)},
\]
and again $\val(G,o,d)>p$ iff $\val(G',o',d)>1/2$. When $p=1/2$, no transformation is needed.

Conversely, to reduce threshold $1/2$ to threshold $p$, if $p>1/2$ let the new start move to $o$ with
probability $2(1-p)$ and directly to $d$ with probability $2p-1$. This gives
\[
    \val(G',o',d)=2(1-p)\val(G,o,d)+(2p-1),
\]
which is greater than $p$ iff $\val(G,o,d)>1/2$. If $p<1/2$, let the new start move to $o$ with
probability $2p$ and to an absorbing losing state otherwise. Then
$\val(G',o',d)=2p\,\val(G,o,d)>p$ iff $\val(G,o,d)>1/2$.

The affine transformations above are strictly increasing; the same constructions therefore preserve the corresponding non-strict comparisons with the threshold.

\begin{figure}[htbp]
    \centering
\resizebox{0.7\textwidth}{!}{\begin{tikzpicture}[x=0.75pt,y=0.75pt,yscale=-1,xscale=1]

\draw    (47.97,63.03) -- (152.92,37.28) ;
\fill [shift={(154.87,36.8)}, rotate = 166.21] [color={rgb, 255:red, 0; green, 0; blue, 0 }  ] (0,0) -- (8.5,-3.5) -- (8.5,3.5) -- cycle ;
\draw [sa switch]   (75.87,119) .. controls (75.87,110.72) and (82.58,104) .. (90.87,104) .. controls (99.15,104) and (105.87,110.72) .. (105.87,119) .. controls (105.87,127.28) and (99.15,134) .. (90.87,134) .. controls (82.58,134) and (75.87,127.28) .. (75.87,119) -- cycle ;
\draw [sa switch]   (154.87,36.8) .. controls (154.87,28.52) and (161.58,21.8) .. (169.87,21.8) .. controls (178.15,21.8) and (184.87,28.52) .. (184.87,36.8) .. controls (184.87,45.08) and (178.15,51.8) .. (169.87,51.8) .. controls (161.58,51.8) and (154.87,45.08) .. (154.87,36.8) -- cycle ;
\draw  [dash pattern={on 4.5pt off 4.5pt}] (122.2,18.7) -- (273.8,18.7) -- (273.8,108.2) -- (122.2,108.2) -- cycle ;
\draw [sa random]   (27.72,42.66) -- (47.97,63.03) -- (27.72,83.41) -- (7.47,63.03) -- cycle ;
\draw    (47.97,63.03) -- (74.97,117.21) ;
\fill [shift={(75.87,119)}, rotate = 243.5] [color={rgb, 255:red, 0; green, 0; blue, 0 }  ] (0,0) -- (8.5,-3.5) -- (8.5,3.5) -- cycle ;
\draw    (105.87,119) .. controls (121.93,103.83) and (120.73,130.87) .. (107.51,124.57) ;
\fill [shift={(105.8,123.6)}, rotate = 32.97] [color={rgb, 255:red, 0; green, 0; blue, 0 }  ] (0,0) -- (8.5,-3.5) -- (8.5,3.5) -- cycle ;

\draw (90.87,119.00) node [sa node label large] {$\displaystyle \overline{d}$};
\draw (27.72,63.03) node [sa node label large] {$\displaystyle o^{\prime }$};
\draw (169.87,36.80) node [sa node label large] {$\displaystyle o$};
\draw (48.6,9.8) node [anchor=north west][inner sep=0.75pt]   [align=left] {$\displaystyle \frac{1}{2p}$};
\draw (124.8,56.6) node [anchor=north west][inner sep=0.75pt]  [font=\LARGE] [align=left] {G};
\draw (14.2,83.4) node [anchor=north west][inner sep=0.75pt]   [align=left] {$\displaystyle 1-\frac{1}{2p}$};

\end{tikzpicture}}
    \caption[{\tt $\mathcal{B}$-Arrival-Quant} is polynomial-time equivalent to {\tt $\mathcal{B}$-Arrival-Quant} where $p=\frac{1}{2}$.]{One of the affine threshold transformations used above: the case $p>\frac12$ reducing threshold $p$ to threshold $\frac12$.}
    \label{sa:fig:b-arrival-quant-half}
\end{figure}
\end{proof}

Hence we will use {\tt $\mathcal{B}$-Arrival-Quant} to refer to the quantitative arrival problem when $p=\frac{1}{2}$, and it suffices to only consider this quantitative decision problem. With these instances, all of {\tt $\mathcal{B}$-Arrival-Quant}, {\tt $\mathcal{B}$-Arrival-Qual-0} and {\tt $\mathcal{B}$-Arrival-Qual-1} take the form $(G,o,d)$ for a $\mathcal{B}$-Arrival Graph $G:=(V,E,\{V_\sigma : \sigma\in \mathcal{B}\},\Prob,\Ord)$ and vertices $o,d\in V$. We use the expression ``instance of a generalised $\mathcal{B}$-arrival problem'' to refer to any instance $(G,o,d)$ which could be given as input to any of these problems. Thus the problems listed in \Cref{tab:complexity} represent all the possible cases we could define. 
\begin{table}[t]
\centering
\caption{Complexity of {\tt Arrival} variants with different node types.}
\footnotesize
\begin{tabular}{|p{0.30\textwidth}|p{0.33\textwidth}|p{0.23\textwidth}|}
\hline
 {\bf Problem Name} & {\bf Known Complexity} & {\bf Reference} \\
\hline \hline {\tt $\{S\}$-Arrival} & \makecell[l]{\PL-hard, \CC-hard (explicit input)\\\P-hard (succinct input)\\in \UEOPL,\\in \CLS,\\in \PLS,\\in $\UP\cap\coUP$} & \makecell[l]{\cite{Man21}\\\cite{FGMS21}\\\cite{FGMS19}\\\cite{GHH+18}\\\cite{Kar17}\\\cite{GHH+18}} \\
\hline 
 {\tt $\{S,1\}$-Arrival} & \NP-complete & \cite{FGMS21} \\
\hline 
 {\tt $\{S,2\}$-Arrival} & \coNP-complete & \Cref{sa:res:S2-is-coNP-com} \\
\hline 
 {\tt $\{S,1,2\}$-Arrival} & \makecell[l]{\PSPACE-hard\\in \EXPTIME} & \makecell[l]{\cite{FGMS21}\\ \cite{FGMS21}\\} \\
\hline 
\hline
 {\tt $\{R,S\}$-Arrival-Qual-0} & \NP{}-complete & \Cref{sa:res:rs(1)-Qual-0-NPc}  \\
\hline 
 {\tt $\{R,S\}$-Arrival-Qual-1} & \coNP{}-complete & \Cref{sa:res:player2-qual1-equivalence}  \\
\hline 
 {\tt $\{R,S\}$-Arrival-Quant} & \makecell[l]{\PP{}-hard,\\ in \PSPACE{}} & \makecell[l]{\Cref{sa:res:rs-quant-is-PP-hard},\\ \Cref{sa:res:rs-quant-in-pspace}}  \\
\hline 
\hline
 {\tt $\{R,S,1\}$-Arrival-Qual-0} & \NP{}-complete & \Cref{sa:res:rs(1)-Qual-0-NPc} \\
\hline 
 {\tt $\{R,S,1\}$-Arrival-Qual-1} & \makecell[l]{\PSPACE{}-hard,\\ in \EXPTIME{}} & \makecell[l]{\Cref{sa:res:rs-qual-1-is-pspace-hard}\\ \Cref{sa:res:all-qual-in-exptime}} \\
\hline 
 {\tt $\{R,S,1\}$-Arrival-Quant} & \makecell[l]{\PSPACE{}-hard,\\ in \EXPTIME{}} & \makecell[l]{\Cref{sa:res:rs1-quant-is-pspace-hard}\\ \Cref{sa:res:expansion-upper-bounds}} \\
\hline
\hline 
 {\tt $\{R,S,2\}$-Arrival-Qual-0} & equiv {\tt $\{S,1,2\}$-Arrival} & \Cref{sa:res:rs(1)2-qual-0-equiv-to-s12} \\
\hline 
 {\tt $\{R,S,2\}$-Arrival-Qual-1} & \coNP{}-complete & \Cref{sa:res:player2-qual1-equivalence} \\
\hline 
 {\tt $\{R,S,2\}$-Arrival-Quant} & \makecell[l]{\PSPACE{}-hard,\\ in \EXPTIME{}}  & \makecell[l]{\Cref{sa:res:rs2-quant-is-pspace-hard}\\ \Cref{sa:res:expansion-upper-bounds}} \\
\hline 
\hline 
 {\tt $\{R,S,1,2\}$-Arrival-Qual-0} &  equiv {\tt $\{S,1,2\}$-Arrival} &  \Cref{sa:res:rs(1)2-qual-0-equiv-to-s12}  \\
 \hline 
 {\tt $\{R,S,1,2\}$-Arrival-Qual-1} & \makecell[l]{\PSPACE{}-hard,\\in \EXPTIME{}} &  \makecell[l]{\Cref{sa:res:rs-qual-1-is-pspace-hard}\\\Cref{sa:res:all-qual-in-exptime}}  \\
\hline 
 {\tt $\{R,S,1,2\}$-Arrival-Quant} &  \makecell[l]{\PSPACE{}-hard,\\  in $\NEXPTIME\cap\coNEXPTIME$} &  \makecell[l]{\Cref{sa:res:rs1-quant-is-pspace-hard},\\ \Cref{sa:res:expansion-upper-bounds}}  \\
 \hline
\end{tabular}\label{tab:complexity}
\end{table}

When drawing generalised Arrival graphs we follow the prior conventions for drawing instances of {\tt Arrival}, with the following new additions. At probabilistic nodes we assume there is a uniform distribution over outgoing edges, otherwise, we label each edge with the probability assigned to it. We also introduce new shapes, as shown in \Cref{back:fig:drawing-node-types}, for the new node types: diamonds for random nodes in $V_R$; circles for switching nodes in $V_S$; squares for player one nodes in $V_1$; and, triangles for Player~2 nodes in $V_2$. 
We use a consistent colour accent for the four node types throughout the figures; colour is redundant with shape, so no argument depends on colour reproduction.

\begin{figure}[htbp]
    \centering
    \resizebox{0.54\textwidth}{!}{\begin{tikzpicture}[x=1cm,y=1cm]
  \node[sa player one,minimum width=8mm,minimum height=8mm] at (0,0) {};
  \node[sa random,diamond,minimum width=9mm,minimum height=9mm,inner sep=0pt] at (3,0) {};
  \node[sa switch,circle,minimum size=9mm] at (6,0) {};
  \node[sa player two,regular polygon,regular polygon sides=3,shape border rotate=0,minimum size=11mm,inner sep=0pt] at (9,0) {};

  \node[anchor=north,font=\large,align=center] at (0,-0.65) {$V_1$ -- Player 1};
  \node[anchor=north,font=\large,align=center] at (3,-0.65) {$V_R$ -- Random};
  \node[anchor=north,font=\large,align=center] at (6,-0.65) {$V_S$ -- Switching};
  \node[anchor=north,font=\large,align=center] at (9,-0.65) {$V_2$ -- Player 2};
\end{tikzpicture}}
    \caption[Shape convention for drawing $\mathcal{B}$-Arrival graphs.]{Shape convention for drawing nodes of different types in $\mathcal{B}$-Arrival graphs, i.e. $V_R,V_S,V_1$ or $V_2$.}
    \label{back:fig:drawing-node-types}
\end{figure}

\subsection{Preliminary Results}

We will repeatedly use the following normal form. A switching vertex with a single successor is regarded as a deterministic wire and has a switching order of length one.

\begin{lemma}[Normal form]\label[lemma]{sa:lem:nice-form}
Given a generalised $\mathcal B$-arrival instance $(G,o,d)$, one can construct in polynomial time an instance $(G^{\mathrm{nf}},o^{\mathrm{nf}},d^{\mathrm{nf}})$ using only node types allowed by $\mathcal B$, with
\[
   \val(G,o,d)=\val(G^{\mathrm{nf}},o^{\mathrm{nf}},d^{\mathrm{nf}}),
\]
and whose encoding size is polynomial in that of $G$, such that:
\begin{enumerate}
\item the target $d^{\mathrm{nf}}$ is absorbing, as is any losing sink $\overline d$ introduced by a later construction;
\item every non-absorbing vertex has out-degree at most two;
\item every random vertex either has one successor with probability $1$, or two successors chosen with probability $1/2$ each;
\item every switching vertex either has a switching order of length one, or has an order of length two with two distinct successors; and
\item every player vertex has at most two successors.
\end{enumerate}
We say that such an instance is in \emph{normal form}.
\end{lemma}
\begin{proof}
Making the target absorbing is value preserving for a reachability objective. A player vertex of out-degree $r>2$ can be replaced by a binary decision tree controlled by the same player, of size $O(r)$, whose leaves are the original successors.

For a random vertex, first reduce an arbitrary rational binary choice $a/b$ to fair coins. Let $k=\lceil\log_2 b\rceil$. Repeatedly generate a uniformly random $k$-bit integer $U$. If $U<a$ take the first outcome, if $a\le U<b$ take the second, and if $U\ge b$ restart the gadget. The comparison with the two fixed binary integers $a$ and $b$ can be performed while the bits of $U$ are generated, keeping only the bit position and whether the prefix read so far is below, equal to, or above the corresponding prefix of $a$ and $b$. Each trial is accepted with probability $b/2^k>0$, so retries terminate almost surely (though a probability-zero infinite run is possible). Thus the gadget uses $O(k)$ fair-coin and deterministic states and implements the probability $a/b$ exactly. For a random vertex with successors of probabilities $p_1,\ldots,p_r$, choose the first successor with probability $p_1$; conditional on not doing so, choose the second with probability $p_2/(1-p_1)$; and continue. The required conditional probabilities are rational numbers whose binary encodings have polynomial length in the original input. This gives a polynomial-size fair-coin implementation of the original distribution.

It remains to normalise a switching order $\Ord(v)=(u_0,\ldots,u_{m-1})$. Orders of length one are already deterministic. For $m\ge2$, take $k=\lceil\log_2 m\rceil$ and replace $v$ by a complete depth-$k$ binary tree of switching vertices, each initially pointing to its first child. Repeated entries into the root visit the $2^k$ leaves in a fixed cyclic order and, after all $2^k$ leaf visits, every switch in the tree returns to its initial position. Label the first $m$ leaves in this visitation order by $u_0,\ldots,u_{m-1}$; redirect every remaining leaf back to the root. Hence one external entry into the gadget follows any intervening rejected leaves until it reaches the next labelled leaf. The first $m$ external entries output $u_0,\ldots,u_{m-1}$. On the next external entry, the gadget traverses the remaining unlabelled leaves and redirects each one to the root. After the last unlabelled leaf, the complete $2^k$-leaf cycle has restored every internal switch to its initial position; a further traversal from the root, still within that same external entry, then exits through the leaf labelled $u_0$. Thus the externally visible output sequence is exactly $(u_0,\ldots,u_{m-1})^\omega$. Since $2^k<2m$, the gadget has $O(m)$ vertices. Any degenerate binary switch whose two successors coincide can be contracted to a deterministic length-one switch. Thus the replacement is polynomial because the original switching order is explicitly represented.

All replacements preserve the sequence, choices, and probabilities of transitions between original vertices when each gadget is viewed as a macro-transition. Applying them to every vertex therefore preserves the reachability value and increases the encoding size only polynomially.
\end{proof}

We may also view a generalised Arrival instance, $G$, as a concise way of specifying an expanded (exponentially larger) game, $Exp(G)$, without switching. These results are analogous to Fearnley et al. who reduce a 2-player reachability switching game to an exponentially large reachability game \cite[Lemma~4.6]{FGMS21}.
From this expanded-game viewpoint, we can derive analogues of many of the results for simple stochastic games obtained by Condon \cite{Con92}. including lower bounds on how close the value can be to zero without being zero. In \emph{normal form}, if $n=|V|$, then a positive value is at least $4^{-n2^n}$, while there are instances of linear size whose positive value is $2^{-(2^n-1)}$.

\begin{corollary}\label[corollary]{sa:res:value-is-rational-bounded}
Let $(G,o,d)$ be in \emph{normal form} with $n=|V|$ vertices, and let
$N\leq n2^n$ be the number of states in $\mathrm{Exp}(G)$. Then
$\val(G,o,d)=p/q$ for integers $0\leq p\leq q\leq 4^N\leq4^{n2^n}$. In particular,
$\val(G,o,d)>0$ implies $\val(G,o,d)\geq4^{-n2^n}$.
\end{corollary}

\begin{proof}
Optimal positional strategies exist in $\mathrm{Exp}(G)$; fix such strategies. The value is then a hitting
probability in the resulting Markov chain, whose transition probabilities lie in $\{0,1/2,1\}$.
Let $S'$ be the non-target states from which the target is reachable, let $A$ be the transition matrix
restricted to $S'$, and let $b$ be the vector of one-step probabilities into the target. Then $A^k\to0$,
$I-A$ is invertible, and the hitting probabilities $x$ satisfy $(I-A)x=b$. Put $B=2(I-A)$. This is an
integer matrix whose rows have at most three non-zero entries, each of absolute value at most $2$.
Hadamard's inequality gives $1\leq|\det B|\leq4^N$. By Cramer's rule,
$x_s=\det B_s/\det B$, where $B_s$ has column $s$ replaced by $2b$; its rows have at most four non-zero
entries of absolute value at most $2$, so $|\det B_s|\leq4^N$. Thus the claimed bound follows.
\end{proof}

However, we can show that we can actually obtain a value of this small magnitude, even in the case where we only have $\mathcal{B}=\{R,S\}$. We do so through modification of the example given by Dohrau et. al. (\cite[Figure~1]{DGKMW16}) showing that a (purely switching) arrival instance can require exponentially many steps to reach the target.

\begin{proposition}\label[proposition]{sa:res:instance-with-double-exp-value}
For every positive integer $n$, we can construct an instance $(G,o,d)$ of the generalised
$\mathcal{B}$-arrival problem, for any $\mathcal{B}$ with $R,S\in\mathcal{B}$, such that $G$ has
encoding size $O(n)$ and
\[
    \val(G,o,d)=2^{-(2^n-1)}>0.
\]
\end{proposition}
\begin{proof}
Take switching nodes $v_1,\ldots,v_n$ with $\Ord(v_i)=(x,v_{i+1})$ for $i<n$ and
$\Ord(v_n)=(x,d)$, a single edge $o\to v_1$, and a random node $x$ with probability $1/2$ to $o$
(Heads) and probability $1/2$ to $\overline d$ (Tails). The only edge into $v_1$ is from $o$.
If the target is reached, let $N_i$ be the number of visits to $v_i$. Visits alternate between $x$
and $v_{i+1}$, starting with $x$, and the run ends at the second visit to $v_n$. Thus $N_n=2$ and
$N_i=2N_{i+1}$ for $i<n$, so $N_1=2^n$. Visits to $x$ come from the odd-numbered visits to each
$v_i$, giving $\sum_i N_i/2=2^n-1$ visits to $x$. Every one of these visits must produce Heads,
since Tails leads immediately to $\overline d$. Conversely, if all $2^n-1$ visits to $x$ produce Heads, the switching counters advance exactly as above and the second visit to $v_n$ takes the play to $d$. Hence
\[
    \val(G,o,d)=\left(\frac12\right)^{2^n-1}=2^{-(2^n-1)}.
\]
\end{proof}

\begin{figure}[htbp]
    \centering
    \resizebox{0.75\textwidth}{!}{\begin{tikzpicture}[x=0.75pt,y=0.75pt,yscale=-1,xscale=1]

\draw [sa random]   (208.35,88.3) -- (228.6,108.68) -- (208.35,129.05) -- (188.1,108.68) -- cycle ;
\draw    (34.9,20.5) -- (59.67,20.96) ;
\fill [shift={(61.67,21)}, rotate = 181.07] [color={rgb, 255:red, 0; green, 0; blue, 0 }  ] (0,0) -- (8.5,-3.5) -- (8.5,3.5) -- cycle ;
\draw    (321.17,24.23) -- (350,24.02) ;
\fill [shift={(352,24)}, rotate = 179.57] [color={rgb, 255:red, 0; green, 0; blue, 0 }  ] (0,0) -- (8.5,-3.5) -- (8.5,3.5) -- cycle ;
\draw    (131.33,37) -- (195.05,97.37) ;
\fill [shift={(196.5,98.75)}, rotate = 223.46] [color={rgb, 255:red, 0; green, 0; blue, 0 }  ] (0,0) -- (8.5,-3.5) -- (8.5,3.5) -- cycle ;
\draw [sa switch]   (62.67,21) .. controls (62.67,12.72) and (69.38,6) .. (77.67,6) .. controls (85.95,6) and (92.67,12.72) .. (92.67,21) .. controls (92.67,29.28) and (85.95,36) .. (77.67,36) .. controls (69.38,36) and (62.67,29.28) .. (62.67,21) -- cycle ;
\draw [sa switch]   (116.33,22) .. controls (116.33,13.72) and (123.05,7) .. (131.33,7) .. controls (139.62,7) and (146.33,13.72) .. (146.33,22) .. controls (146.33,30.28) and (139.62,37) .. (131.33,37) .. controls (123.05,37) and (116.33,30.28) .. (116.33,22) -- cycle ;
\draw [sa switch]   (179.67,23.33) .. controls (179.67,15.05) and (186.38,8.33) .. (194.67,8.33) .. controls (202.95,8.33) and (209.67,15.05) .. (209.67,23.33) .. controls (209.67,31.62) and (202.95,38.33) .. (194.67,38.33) .. controls (186.38,38.33) and (179.67,31.62) .. (179.67,23.33) -- cycle ;
\draw [sa switch]   (291.17,24.23) .. controls (291.17,15.95) and (297.88,9.23) .. (306.17,9.23) .. controls (314.45,9.23) and (321.17,15.95) .. (321.17,24.23) .. controls (321.17,32.52) and (314.45,39.23) .. (306.17,39.23) .. controls (297.88,39.23) and (291.17,32.52) .. (291.17,24.23) -- cycle ;
\draw    (92.67,21) -- (115.33,21.92) ;
\fill [shift={(117.33,22)}, rotate = 182.32] [color={rgb, 255:red, 0; green, 0; blue, 0 }  ] (0,0) -- (8.5,-3.5) -- (8.5,3.5) -- cycle ;
\draw    (147.33,22) -- (178.67,23.25) ;
\fill [shift={(180.67,23.33)}, rotate = 182.29] [color={rgb, 255:red, 0; green, 0; blue, 0 }  ] (0,0) -- (8.5,-3.5) -- (8.5,3.5) -- cycle ;
\draw    (209.67,23.33) -- (236,22.09) ;
\fill [shift={(238,22)}, rotate = 177.31] [color={rgb, 255:red, 0; green, 0; blue, 0 }  ] (0,0) -- (8.5,-3.5) -- (8.5,3.5) -- cycle ;
\draw    (91.67,21) -- (190.47,103.96) ;
\fill [shift={(192,105.25)}, rotate = 220.02] [color={rgb, 255:red, 0; green, 0; blue, 0 }  ] (0,0) -- (8.5,-3.5) -- (8.5,3.5) -- cycle ;
\draw    (194.67,38.33) -- (202.69,89.77) ;
\fill [shift={(203,91.75)}, rotate = 261.13] [color={rgb, 255:red, 0; green, 0; blue, 0 }  ] (0,0) -- (8.5,-3.5) -- (8.5,3.5) -- cycle ;
\draw    (306.17,39.23) -- (228.04,103.97) ;
\fill [shift={(226.5,105.25)}, rotate = 320.35] [color={rgb, 255:red, 0; green, 0; blue, 0 }  ] (0,0) -- (8.5,-3.5) -- (8.5,3.5) -- cycle ;
\draw    (263,23.75) -- (289.17,24.2) ;
\fill [shift={(291.17,24.23)}, rotate = 180.98] [color={rgb, 255:red, 0; green, 0; blue, 0 }  ] (0,0) -- (8.5,-3.5) -- (8.5,3.5) -- cycle ;
\draw    (255.67,37.73) -- (220.04,96.04) ;
\fill [shift={(219,97.75)}, rotate = 301.42] [color={rgb, 255:red, 0; green, 0; blue, 0 }  ] (0,0) -- (8.5,-3.5) -- (8.5,3.5) -- cycle ;
\draw [sa switch]   (350.67,24.4) .. controls (350.67,16.12) and (357.38,9.4) .. (365.67,9.4) .. controls (373.95,9.4) and (380.67,16.12) .. (380.67,24.4) .. controls (380.67,32.68) and (373.95,39.4) .. (365.67,39.4) .. controls (357.38,39.4) and (350.67,32.68) .. (350.67,24.4) -- cycle ;
\draw [sa switch]   (3.17,20.9) .. controls (3.17,12.62) and (9.88,5.9) .. (18.17,5.9) .. controls (26.45,5.9) and (33.17,12.62) .. (33.17,20.9) .. controls (33.17,29.18) and (26.45,35.9) .. (18.17,35.9) .. controls (9.88,35.9) and (3.17,29.18) .. (3.17,20.9) -- cycle ;
\draw [sa switch]   (348.67,88.4) .. controls (348.67,80.12) and (355.38,73.4) .. (363.67,73.4) .. controls (371.95,73.4) and (378.67,80.12) .. (378.67,88.4) .. controls (378.67,96.68) and (371.95,103.4) .. (363.67,103.4) .. controls (355.38,103.4) and (348.67,96.68) .. (348.67,88.4) -- cycle ;
\draw    (198.5,119) .. controls (128.7,174.69) and (53.52,90.47) .. (19.19,37.5) ;
\fill [shift={(18.17,35.9)}, rotate = 57.37] [color={rgb, 255:red, 0; green, 0; blue, 0 }  ] (0,0) -- (8.5,-3.5) -- (8.5,3.5) -- cycle ;
\draw    (219.5,118) .. controls (254.33,152.58) and (294.6,140.87) .. (347.86,89.18) ;
\fill [shift={(348.67,88.4)}, rotate = 135.71] [color={rgb, 255:red, 0; green, 0; blue, 0 }  ] (0,0) -- (8.5,-3.5) -- (8.5,3.5) -- cycle ;
\draw[->,line width=.7pt] (376.2,16.7) .. controls (394.0,1.4) and (400.0,43.7) .. (376.2,31.7);
\draw[->,line width=.7pt] (374.2,80.7) .. controls (392.0,65.4) and (398.0,107.7) .. (374.2,95.7);

\draw (208.35,108.68) node [sa node label large] {$\displaystyle x$};
\draw (149.6,109.9) node [anchor=north west][inner sep=0.75pt]   [align=left] {Head};
\draw (230.6,111.68) node [anchor=north west][inner sep=0.75pt]   [align=left] {Tail};
\draw (244.7,17) node [anchor=north west][inner sep=0.75pt]   [align=left] {...};
\draw (92.67,27.5) node [anchor=north west][inner sep=0.75pt]   [align=left] {$\displaystyle 0$};
\draw (91,3.27) node [anchor=north west][inner sep=0.75pt]   [align=left] {$\displaystyle 1$};
\draw (77.67,21.00) node [sa node label large] {$\displaystyle v_{1}$};
\draw (131.33,22.00) node [sa node label large] {$\displaystyle v_{2}$};
\draw (194.67,23.33) node [sa node label large] {$\displaystyle v_{3}$};
\draw (306.17,24.23) node [sa node label large] {$\displaystyle v_{n}$};
\draw (365.67,24.40) node [sa node label large] {$\displaystyle d$};
\draw (18.17,20.90) node [sa node label large] {$\displaystyle o$};
\draw (363.67,88.40) node [sa node label large] {$\displaystyle \overline{d}$};
\draw (141.17,30.5) node [anchor=north west][inner sep=0.75pt]   [align=left] {$\displaystyle 0$};
\draw (185.67,36) node [anchor=north west][inner sep=0.75pt]   [align=left] {$\displaystyle 0$};
\draw (238.17,33.5) node [anchor=north west][inner sep=0.75pt]   [align=left] {$\displaystyle 0$};
\draw (283.67,35) node [anchor=north west][inner sep=0.75pt]   [align=left] {$\displaystyle 0$};
\draw (147,5) node [anchor=north west][inner sep=0.75pt]   [align=left] {$\displaystyle 1$};
\draw (210,6.5) node [anchor=north west][inner sep=0.75pt]   [align=left] {$\displaystyle 1$};
\draw (261.5,7) node [anchor=north west][inner sep=0.75pt]   [align=left] {$\displaystyle 1$};
\draw (319,7) node [anchor=north west][inner sep=0.75pt]   [align=left] {$\displaystyle 1$};
\draw (401.0,23.0) node [sa edge label,anchor=west] {$0,1$};
\draw (399.0,87.0) node [sa edge label,anchor=west] {$0,1$};

\end{tikzpicture}}
    \caption[An exponentially small positive-value instance.]{The $\{R,S\}$-Arrival instance used in \Cref{sa:res:instance-with-double-exp-value}. Reaching $d$ requires $2^n-1$ consecutive Heads outcomes at the random vertex $x$, and hence the value is $2^{-(2^n-1)}$.}
    \label{sa:fig:double-exp-value}
\end{figure}

As in simple stochastic games, one could force these games to terminate, i.e., reach either the target $d$ or the losing sink $\overline{d}$, by modifying them by applying a small discount, ending the game with a small probability after each step (e.g. as in \cite[{Lemma~8}]{Con92}). However, unlike the situation with simple stochastic games, even applying a very small discount of the form $\frac{1}{2^{\mathit{poly}(n)}}$ can change the value of the game drastically (taking a value close to 1 down to a value close to zero). While we can construct smaller doubly-exponential probabilities as in \Cref{sa:res:instance-with-double-exp-value}, where we use this construction for discounting a vertex, allowing the play to continue if it hit the target and stopping early at the losing sink we have to contend with the switch position. After each visit, the nodes $o,v_1,\ldots,v_n$ are set in some switch position, which increases the probability of reaching $\overline{d}$ after the next visit to $o$. We can, however, use \Cref{sa:res:instance-with-double-exp-value} to show that the non-strict problem {\tt $\mathcal B$-Arrival-Quant-Eq} is polynomial-time equivalent to the strict problem {\tt $\mathcal B$-Arrival-Quant}:

\begin{proposition}\label[proposition]{sa:res:rec-quant-equal-prob}
Given a generalised $\mathcal{B}$-arrival graph $G = (V,E,\{V_\sigma : \sigma\in \mathcal{B}\},\Prob,\Ord)$ with $R,S\in \mathcal{B}$, vertices $o,d\in V$, and any rational $p\in(0,1)$, deciding whether $\val(G,o,d)\geq p$ is polynomial-time equivalent to {\tt $\mathcal{B}$-Arrival-Quant} . 
\end{proposition}
\begin{proof}
As observed after \Cref{sa:res:rec-quant-only-p-half}, the strictly increasing affine threshold transformations also preserve non-strict comparisons. Hence it is enough to treat the case of deciding whether $\val(G,o,d)\geq\frac12$. 

Given some instance $(G,o,d)$ we reduce the case of deciding $\val(G,o,d)\geq\frac{1}{2}$ to deciding $\val(G,o,d)>\frac{1}{2}$. Apply \Cref{sa:lem:nice-form} first and let $n$ be the number of vertices of the resulting instance. By \Cref{sa:res:value-is-rational-bounded} we have $N\leq n2^n$ and $\val(G,o,d)=a/b$ with $b\leq4^N$. If $\val=a/b<\frac12$, then
$\frac12-\val=\frac{b-2a}{2b}\geq\frac1{2b}\geq2^{-(2N+1)}$. We use the doubly-exponentially small construction above with $l=2n+2$, so $\epsilon=2^{-(2^l-1)}$ and $2N+2\leq n2^{n+1}+2\leq2^{2n+2}=2^l$. Hence $\epsilon\leq2^{-(2N+1)}$.
We construct a new $\mathcal{B}$-arrival graph $G^\prime$ as shown in \Cref{sa:fig:equality-construction}. We will show that $\val(G^\prime,o^\prime,d)>\frac{1}{2}$ if and only if $\val(G,o,d) \geq \frac{1}{2}$. In $G^\prime$ we have a new start vertex $o^\prime$ and in it we run a single copy of the construction in \Cref{sa:res:instance-with-double-exp-value} as a two-outcome gadget used only once: its rare outcome (of probability $\epsilon$) is redirected to the target $d$, and its other outcome to the original start $o$. Neither outcome re-enters the gadget. By \Cref{sa:res:instance-with-double-exp-value} we know the value of this instance, with $l$ nodes, is $\epsilon=2^{-(2^{l}-1)}$, where we take $l=2n+2$. Thus we have that 
\begin{equation*}
    \val(G^\prime,o^\prime,d)=(1-\epsilon)\val(G,o,d)+\epsilon=\val(G,o,d)+\epsilon\cdot(1-\val(G,o,d))
\end{equation*}

\begin{figure}[tbp]
    \centering
    \resizebox{0.7\textwidth}{!}{\begin{tikzpicture}[x=0.75pt,y=0.75pt,yscale=-1,xscale=1]

\draw [sa random]   (208.35,88.3) -- (228.6,108.68) -- (208.35,129.05) -- (188.1,108.68) -- cycle ;
\draw    (34.9,20.5) -- (59.67,20.96) ;
\fill [shift={(61.67,21)}, rotate = 181.07] [color={rgb, 255:red, 0; green, 0; blue, 0 }  ] (0,0) -- (8.5,-3.5) -- (8.5,3.5) -- cycle ;
\draw    (321.17,24.23) -- (528.68,51.14) ;
\fill [shift={(530.67,51.4)}, rotate = 187.39] [color={rgb, 255:red, 0; green, 0; blue, 0 }  ] (0,0) -- (8.5,-3.5) -- (8.5,3.5) -- cycle ;
\draw    (131.33,36) -- (195.05,97.37) ;
\fill [shift={(196.5,98.75)}, rotate = 223.46] [color={rgb, 255:red, 0; green, 0; blue, 0 }  ] (0,0) -- (8.5,-3.5) -- (8.5,3.5) -- cycle ;
\draw [sa switch]   (61.67,21) .. controls (61.67,12.72) and (68.38,6) .. (76.67,6) .. controls (84.95,6) and (91.67,12.72) .. (91.67,21) .. controls (91.67,29.28) and (84.95,36) .. (76.67,36) .. controls (68.38,36) and (61.67,29.28) .. (61.67,21) -- cycle ;
\draw [sa switch]   (116.33,22) .. controls (116.33,13.72) and (123.05,7) .. (131.33,7) .. controls (139.62,7) and (146.33,13.72) .. (146.33,22) .. controls (146.33,30.28) and (139.62,37) .. (131.33,37) .. controls (123.05,37) and (116.33,30.28) .. (116.33,22) -- cycle ;
\draw [sa switch]   (179.67,23.33) .. controls (179.67,15.05) and (186.38,8.33) .. (194.67,8.33) .. controls (202.95,8.33) and (209.67,15.05) .. (209.67,23.33) .. controls (209.67,31.62) and (202.95,38.33) .. (194.67,38.33) .. controls (186.38,38.33) and (179.67,31.62) .. (179.67,23.33) -- cycle ;
\draw [sa switch]   (291.17,24.23) .. controls (291.17,15.95) and (297.88,9.23) .. (306.17,9.23) .. controls (314.45,9.23) and (321.17,15.95) .. (321.17,24.23) .. controls (321.17,32.52) and (314.45,39.23) .. (306.17,39.23) .. controls (297.88,39.23) and (291.17,32.52) .. (291.17,24.23) -- cycle ;
\draw    (91.67,21) -- (114.33,21) ;
\fill [shift={(116.33,21)}, rotate = 180] [color={rgb, 255:red, 0; green, 0; blue, 0 }  ] (0,0) -- (8.5,-3.5) -- (8.5,3.5) -- cycle ;
\draw    (146.33,21) -- (177.67,21) ;
\fill [shift={(179.67,21)}, rotate = 180] [color={rgb, 255:red, 0; green, 0; blue, 0 }  ] (0,0) -- (8.5,-3.5) -- (8.5,3.5) -- cycle ;
\draw    (209.67,21) -- (236,21) ;
\fill [shift={(238,21)}, rotate = 180] [color={rgb, 255:red, 0; green, 0; blue, 0 }  ] (0,0) -- (8.5,-3.5) -- (8.5,3.5) -- cycle ;
\draw    (91.67,21) -- (190.47,103.96) ;
\fill [shift={(192,105.25)}, rotate = 220.02] [color={rgb, 255:red, 0; green, 0; blue, 0 }  ] (0,0) -- (8.5,-3.5) -- (8.5,3.5) -- cycle ;
\draw    (194.67,36) -- (202.69,89.77) ;
\fill [shift={(203,91.75)}, rotate = 261.13] [color={rgb, 255:red, 0; green, 0; blue, 0 }  ] (0,0) -- (8.5,-3.5) -- (8.5,3.5) -- cycle ;
\draw    (306.17,36) -- (228.04,103.97) ;
\fill [shift={(226.5,105.25)}, rotate = 320.35] [color={rgb, 255:red, 0; green, 0; blue, 0 }  ] (0,0) -- (8.5,-3.5) -- (8.5,3.5) -- cycle ;
\draw    (263,21) -- (289.17,21) ;
\fill [shift={(291.17,21)}, rotate = 180] [color={rgb, 255:red, 0; green, 0; blue, 0 }  ] (0,0) -- (8.5,-3.5) -- (8.5,3.5) -- cycle ;
\draw    (255.67,37.73) -- (220.04,96.04) ;
\fill [shift={(219,97.75)}, rotate = 301.42] [color={rgb, 255:red, 0; green, 0; blue, 0 }  ] (0,0) -- (8.5,-3.5) -- (8.5,3.5) -- cycle ;
\draw [sa switch]   (530.67,51.4) .. controls (530.67,43.12) and (537.38,36.4) .. (545.67,36.4) .. controls (553.95,36.4) and (560.67,43.12) .. (560.67,51.4) .. controls (560.67,59.68) and (553.95,66.4) .. (545.67,66.4) .. controls (537.38,66.4) and (530.67,59.68) .. (530.67,51.4) -- cycle ;
\draw [sa switch]   (3.17,20.9) .. controls (3.17,12.62) and (9.88,5.9) .. (18.17,5.9) .. controls (26.45,5.9) and (33.17,12.62) .. (33.17,20.9) .. controls (33.17,29.18) and (26.45,35.9) .. (18.17,35.9) .. controls (9.88,35.9) and (3.17,29.18) .. (3.17,20.9) -- cycle ;
\draw [sa switch]   (371.67,88.4) .. controls (371.67,80.12) and (378.38,73.4) .. (386.67,73.4) .. controls (394.95,73.4) and (401.67,80.12) .. (401.67,88.4) .. controls (401.67,96.68) and (394.95,103.4) .. (386.67,103.4) .. controls (378.38,103.4) and (371.67,96.68) .. (371.67,88.4) -- cycle ;
\draw    (198.5,119) .. controls (128.7,174.69) and (53.52,90.47) .. (19.19,37.5) ;
\fill [shift={(18.17,35.9)}, rotate = 57.37] [color={rgb, 255:red, 0; green, 0; blue, 0 }  ] (0,0) -- (8.5,-3.5) -- (8.5,3.5) -- cycle ;
\draw    (219,122.5) .. controls (266.26,153.84) and (311.54,134.69) .. (370.77,89.09) ;
\fill [shift={(371.67,88.4)}, rotate = 142.31] [color={rgb, 255:red, 0; green, 0; blue, 0 }  ] (0,0) -- (8.5,-3.5) -- (8.5,3.5) -- cycle ;
\draw  [dash pattern={on 4.5pt off 4.5pt}] (355,23.5) -- (592.4,23.5) -- (592.4,113) -- (355,113) -- cycle ;

\draw (208.35,108.68) node [sa node label large] {$\displaystyle x$};
\draw (149.6,109.9) node [anchor=north west][inner sep=0.75pt]   [align=left] {Head};
\draw (228.5,108.25) node [anchor=north west][inner sep=0.75pt]   [align=left] {Tail};
\draw (244.7,17) node [anchor=north west][inner sep=0.75pt]   [align=left] {...};
\draw (91.67,27.5) node [anchor=north west][inner sep=0.75pt]   [align=left] {$\displaystyle 0$};
\draw (90,3.27) node [anchor=north west][inner sep=0.75pt]   [align=left] {$\displaystyle 1$};
\draw (76.67,21.00) node [sa node label large] {$\displaystyle v_{1}$};
\draw (131.33,21.00) node [sa node label large] {$\displaystyle v_{2}$};
\draw (194.67,21.00) node [sa node label large] {$\displaystyle v_{3}$};
\draw (306.17,21.00) node [sa node label large] {$\displaystyle v_{l}$};
\draw (545.67,51.40) node [sa node label large] {$\displaystyle d$};
\draw (18.17,20.90) node [sa node label large] {$\displaystyle o^{\prime }$};
\draw (386.67,88.40) node [sa node label large] {$\displaystyle o$};
\draw (140.17,30.5) node [anchor=north west][inner sep=0.75pt]   [align=left] {$\displaystyle 0$};
\draw (184.67,36) node [anchor=north west][inner sep=0.75pt]   [align=left] {$\displaystyle 0$};
\draw (238.17,33.5) node [anchor=north west][inner sep=0.75pt]   [align=left] {$\displaystyle 0$};
\draw (283.67,35) node [anchor=north west][inner sep=0.75pt]   [align=left] {$\displaystyle 0$};
\draw (146,5) node [anchor=north west][inner sep=0.75pt]   [align=left] {$\displaystyle 1$};
\draw (209,6.5) node [anchor=north west][inner sep=0.75pt]   [align=left] {$\displaystyle 1$};
\draw (261.5,7) node [anchor=north west][inner sep=0.75pt]   [align=left] {$\displaystyle 1$};
\draw (319,7) node [anchor=north west][inner sep=0.75pt]   [align=left] {$\displaystyle 1$};
\draw (358.4,46.2) node [anchor=north west][inner sep=0.75pt]  [font=\LARGE] [align=left] {G};

\end{tikzpicture}}
    \caption[Proof of \ref{sa:res:rec-quant-equal-prob}: reducing deciding equality to inequality.]{Proof of \ref{sa:res:rec-quant-equal-prob}: reducing deciding $\geq \frac{1}{2}$ to deciding $> \frac{1}{2}$: construction of graph $G^\prime$ for a given instance $G$.}
    \label{sa:fig:equality-construction}
\end{figure}

Assuming that $\val(G,o,d)=\frac{a}{b}<\frac{1}{2}$, we have $\frac{1}{2}-\val(G,o,d)\geq2^{-(2N+1)}$. By the choice of $l$, $\epsilon\leq2^{-(2N+1)}$. If $\val(G,o,d)>0$, the increase is $\epsilon(1-\val)<\epsilon$, so the new value remains strictly below $1/2$; if $\val(G,o,d)=0$, then $\val(G^\prime,o^\prime,d)=\epsilon<1/2$ directly. Hence in every case $\val(G^\prime,o^\prime,d)<\frac12$. By construction we can also see that $\val(G^\prime,o^\prime,d)\geq\val(G,o,d)$ and this is a strict increase when $\val(G,o,d)\neq 1$, hence if $\val(G,o,d)\geq\frac{1}{2}$ we know we have $\val(G^\prime,o^\prime,d) >\frac{1}{2}$. 

We may also perform a similar reduction from the case of deciding $\val(G,o,d)>\frac{1}{2}$ to deciding $\val(G,o,d)\geq\frac{1}{2}$ by the analogous one-shot construction shown in \Cref{sa:fig:equality-construction-pt2}, this time redirecting the rare outcome to the losing sink $\overline d$ and the other outcome to $o$. In that construction the new value is $(1-\epsilon)\val(G,o,d)$, so the decrease is $\epsilon\val(G,o,d)$ rather than literally $\epsilon$. If $\val(G,o,d)>\frac12$, the same rational gap bound gives $\val(G,o,d)-\frac12\geq2^{-(2N+1)}\geq\epsilon$, and hence $(1-\epsilon)\val(G,o,d)\geq\frac12$; if $\val(G,o,d)\leq\frac12$, the new value is strictly below $\frac12$. This gives the reverse reduction and hence the equivalence. 

\begin{figure}[tbp]
    \centering
    \resizebox{0.7\textwidth}{!}{\begin{tikzpicture}[x=0.75pt,y=0.75pt,yscale=-1,xscale=1]

\draw [sa random]   (208.35,88.3) -- (228.6,108.68) -- (208.35,129.05) -- (188.1,108.68) -- cycle ;
\draw    (34.9,20.5) -- (59.67,20.96) ;
\fill [shift={(61.67,21)}, rotate = 181.07] [color={rgb, 255:red, 0; green, 0; blue, 0 }  ] (0,0) -- (8.5,-3.5) -- (8.5,3.5) -- cycle ;
\draw    (321.17,24.23) -- (505.74,75.86) ;
\fill [shift={(507.67,76.4)}, rotate = 195.63] [color={rgb, 255:red, 0; green, 0; blue, 0 }  ] (0,0) -- (8.5,-3.5) -- (8.5,3.5) -- cycle ;
\draw    (131.33,36) -- (195.05,97.37) ;
\fill [shift={(196.5,98.75)}, rotate = 223.46] [color={rgb, 255:red, 0; green, 0; blue, 0 }  ] (0,0) -- (8.5,-3.5) -- (8.5,3.5) -- cycle ;
\draw [sa switch]   (61.67,21) .. controls (61.67,12.72) and (68.38,6) .. (76.67,6) .. controls (84.95,6) and (91.67,12.72) .. (91.67,21) .. controls (91.67,29.28) and (84.95,36) .. (76.67,36) .. controls (68.38,36) and (61.67,29.28) .. (61.67,21) -- cycle ;
\draw [sa switch]   (116.33,22) .. controls (116.33,13.72) and (123.05,7) .. (131.33,7) .. controls (139.62,7) and (146.33,13.72) .. (146.33,22) .. controls (146.33,30.28) and (139.62,37) .. (131.33,37) .. controls (123.05,37) and (116.33,30.28) .. (116.33,22) -- cycle ;
\draw [sa switch]   (179.67,23.33) .. controls (179.67,15.05) and (186.38,8.33) .. (194.67,8.33) .. controls (202.95,8.33) and (209.67,15.05) .. (209.67,23.33) .. controls (209.67,31.62) and (202.95,38.33) .. (194.67,38.33) .. controls (186.38,38.33) and (179.67,31.62) .. (179.67,23.33) -- cycle ;
\draw [sa switch]   (291.17,24.23) .. controls (291.17,15.95) and (297.88,9.23) .. (306.17,9.23) .. controls (314.45,9.23) and (321.17,15.95) .. (321.17,24.23) .. controls (321.17,32.52) and (314.45,39.23) .. (306.17,39.23) .. controls (297.88,39.23) and (291.17,32.52) .. (291.17,24.23) -- cycle ;
\draw    (91.67,21) -- (114.33,21) ;
\fill [shift={(116.33,21)}, rotate = 180] [color={rgb, 255:red, 0; green, 0; blue, 0 }  ] (0,0) -- (8.5,-3.5) -- (8.5,3.5) -- cycle ;
\draw    (146.33,21) -- (177.67,21) ;
\fill [shift={(179.67,21)}, rotate = 180] [color={rgb, 255:red, 0; green, 0; blue, 0 }  ] (0,0) -- (8.5,-3.5) -- (8.5,3.5) -- cycle ;
\draw    (209.67,21) -- (236,21) ;
\fill [shift={(238,21)}, rotate = 180] [color={rgb, 255:red, 0; green, 0; blue, 0 }  ] (0,0) -- (8.5,-3.5) -- (8.5,3.5) -- cycle ;
\draw    (91.67,21) -- (190.47,103.96) ;
\fill [shift={(192,105.25)}, rotate = 220.02] [color={rgb, 255:red, 0; green, 0; blue, 0 }  ] (0,0) -- (8.5,-3.5) -- (8.5,3.5) -- cycle ;
\draw    (194.67,36) -- (202.69,89.77) ;
\fill [shift={(203,91.75)}, rotate = 261.13] [color={rgb, 255:red, 0; green, 0; blue, 0 }  ] (0,0) -- (8.5,-3.5) -- (8.5,3.5) -- cycle ;
\draw    (306.17,36) -- (228.04,103.97) ;
\fill [shift={(226.5,105.25)}, rotate = 320.35] [color={rgb, 255:red, 0; green, 0; blue, 0 }  ] (0,0) -- (8.5,-3.5) -- (8.5,3.5) -- cycle ;
\draw    (263,21) -- (289.17,21) ;
\fill [shift={(291.17,21)}, rotate = 180] [color={rgb, 255:red, 0; green, 0; blue, 0 }  ] (0,0) -- (8.5,-3.5) -- (8.5,3.5) -- cycle ;
\draw    (255.67,37.73) -- (220.04,96.04) ;
\fill [shift={(219,97.75)}, rotate = 301.42] [color={rgb, 255:red, 0; green, 0; blue, 0 }  ] (0,0) -- (8.5,-3.5) -- (8.5,3.5) -- cycle ;
\draw [sa switch]   (492.67,91.4) .. controls (492.67,83.12) and (499.38,76.4) .. (507.67,76.4) .. controls (515.95,76.4) and (522.67,83.12) .. (522.67,91.4) .. controls (522.67,99.68) and (515.95,106.4) .. (507.67,106.4) .. controls (499.38,106.4) and (492.67,99.68) .. (492.67,91.4) -- cycle ;
\draw [sa switch]   (3.17,20.9) .. controls (3.17,12.62) and (9.88,5.9) .. (18.17,5.9) .. controls (26.45,5.9) and (33.17,12.62) .. (33.17,20.9) .. controls (33.17,29.18) and (26.45,35.9) .. (18.17,35.9) .. controls (9.88,35.9) and (3.17,29.18) .. (3.17,20.9) -- cycle ;
\draw [sa switch]   (371.67,88.4) .. controls (371.67,80.12) and (378.38,73.4) .. (386.67,73.4) .. controls (394.95,73.4) and (401.67,80.12) .. (401.67,88.4) .. controls (401.67,96.68) and (394.95,103.4) .. (386.67,103.4) .. controls (378.38,103.4) and (371.67,96.68) .. (371.67,88.4) -- cycle ;
\draw    (198.5,119) .. controls (128.7,174.69) and (53.52,90.47) .. (19.19,37.5) ;
\fill [shift={(18.17,35.9)}, rotate = 57.37] [color={rgb, 255:red, 0; green, 0; blue, 0 }  ] (0,0) -- (8.5,-3.5) -- (8.5,3.5) -- cycle ;
\draw    (219,122.5) .. controls (266.26,153.84) and (311.54,134.69) .. (370.77,89.09) ;
\fill [shift={(371.67,88.4)}, rotate = 142.31] [color={rgb, 255:red, 0; green, 0; blue, 0 }  ] (0,0) -- (8.5,-3.5) -- (8.5,3.5) -- cycle ;
\draw  [dash pattern={on 4.5pt off 4.5pt}] (355,23.5) -- (592.4,23.5) -- (592.4,113) -- (355,113) -- cycle ;

\draw (208.35,108.68) node [sa node label large] {$\displaystyle x$};
\draw (149.6,109.9) node [anchor=north west][inner sep=0.75pt]   [align=left] {Head};
\draw (228.5,108.25) node [anchor=north west][inner sep=0.75pt]   [align=left] {Tail};
\draw (244.7,17) node [anchor=north west][inner sep=0.75pt]   [align=left] {...};
\draw (91.67,27.5) node [anchor=north west][inner sep=0.75pt]   [align=left] {$\displaystyle 0$};
\draw (90,3.27) node [anchor=north west][inner sep=0.75pt]   [align=left] {$\displaystyle 1$};
\draw (76.67,21.00) node [sa node label large] {$\displaystyle v_{1}$};
\draw (131.33,21.00) node [sa node label large] {$\displaystyle v_{2}$};
\draw (194.67,21.00) node [sa node label large] {$\displaystyle v_{3}$};
\draw (306.17,21.00) node [sa node label large] {$\displaystyle v_{l}$};
\draw (507.67,91.40) node [sa node label large] {$\displaystyle \overline{d}$};
\draw (18.17,20.90) node [sa node label large] {$\displaystyle o^{\prime }$};
\draw (386.67,88.40) node [sa node label large] {$\displaystyle o$};
\draw (140.17,30.5) node [anchor=north west][inner sep=0.75pt]   [align=left] {$\displaystyle 0$};
\draw (184.67,36) node [anchor=north west][inner sep=0.75pt]   [align=left] {$\displaystyle 0$};
\draw (238.17,33.5) node [anchor=north west][inner sep=0.75pt]   [align=left] {$\displaystyle 0$};
\draw (283.67,35) node [anchor=north west][inner sep=0.75pt]   [align=left] {$\displaystyle 0$};
\draw (146,5) node [anchor=north west][inner sep=0.75pt]   [align=left] {$\displaystyle 1$};
\draw (209,6.5) node [anchor=north west][inner sep=0.75pt]   [align=left] {$\displaystyle 1$};
\draw (261.5,7) node [anchor=north west][inner sep=0.75pt]   [align=left] {$\displaystyle 1$};
\draw (319,7) node [anchor=north west][inner sep=0.75pt]   [align=left] {$\displaystyle 1$};
\draw (359.4,42.2) node [anchor=north west][inner sep=0.75pt]  [font=\LARGE] [align=left] {G};

\end{tikzpicture}}
    \caption[Proof of \ref{sa:res:rec-quant-equal-prob}: reducing deciding inequality to equality.]{Proof of \ref{sa:res:rec-quant-equal-prob}: reducing deciding $> \frac{1}{2}$ to deciding $\geq \frac{1}{2}$: construction of graph $G^\prime$ for a given instance $G$.}
    \label{sa:fig:equality-construction-pt2}
\end{figure}

\end{proof}

We can also see that, from interpreting these models as succinct representations of exponentially large MCs, MDPs, and SSGs, respectively, we have the following simple upper bounds on these problems. 

\begin{proposition}\label[proposition]{sa:res:expansion-upper-bounds}
The {\tt $\{R,S,1\}$-Arrival-Quant} and {\tt $\{R,S,2\}$-Arrival-Quant} problems are contained in \EXPTIME{} and the {\tt $\{R,S,1,2\}$-Arrival-Quant} is contained in $\NEXPTIME{}\cap\coNEXPTIME{}$.
\end{proposition}
\begin{proof}
By \Cref{sa:lem:nice-form}, we may first normalise the input with only polynomial blow-up. The expanded state space $V\times Q$ then has size at most $n2^n$, so it can be generated explicitly in exponential time. If only one player is present, the expansion is a finite Markov decision process; quantitative reachability values for an explicitly represented finite MDP can be computed in polynomial time, for example by linear programming \cite[Chapter~10]{BaierKatoen2008}, giving the claimed \EXPTIME{} bounds.

With both players present, the expansion is an explicitly represented simple stochastic game of exponential size. Finite stochastic reachability games have optimal positional strategies \cite{Con92}. For the strict threshold problem, a nondeterministic exponential-time machine guesses a positional strategy for Player~1 and then computes the minimum reachability value in the resulting MDP, accepting exactly when this value is greater than the threshold. This gives membership in \NEXPTIME{}. Dually, for a no-instance one can guess an optimal positional strategy for Player~2 and compute the maximum reachability value in the resulting MDP, giving membership in \coNEXPTIME{}.
\end{proof}

\subsection{The Complexity of {\tt $\{S,2\}$-Arrival}}

While Fearnley et al. do not explicitly consider the {\tt $\{S,2\}$-Arrival} problem in \cite{FGMS21} we are able to deduce \coNP{}-completeness using their results and our generalised notion of $d$-hopefulness.

\problemStatement{3DNF-Tautology \cite{Pap94}}{
Instance={A 3DNF formula $\varphi=T_1\vee\cdots\vee T_m$ on variables $x_1,\ldots,x_n$, where each term is a conjunction of at most three distinct literals.},
Problem={Decide whether $\varphi$ evaluates to true under every Boolean assignment.}
}

\begin{proposition}\label[proposition]{sa:res:S2-is-coNP-com}
The {\tt $\{S,2\}$-Arrival} problem is $\coNP{}$-complete.
\end{proposition}
\begin{proof}
We reduce from \textsc{3DNF-Tautology}, which is \coNP{}-complete: negating a 3DNF formula yields a 3CNF formula, so the claim follows from the NP-completeness of 3SAT \cite{Pap94}. Write $w_l\in\{1,2,3\}$ for the number of literals in term $T_l$. Without loss of generality no term contains
a variable together with its negation (delete such terms; if none remain, output a fixed `no'-instance). Let
$a_\iota$ ($b_\iota$) be the number of terms containing $x_\iota$ ($\neg x_\iota$), and $D=\max_\iota\{a_\iota,b_\iota\}$.

\emph{Construction} $G=(V,E,\{V_S,V_2\},\mathrm{Ord})$, target $d$, other sink $\overline d$.
\begin{itemize}
\item $\mathrm{Ord}(o)=(a,\dots,a,\overline d)$ with $(D+1)n$ copies of $a$; $\mathrm{Ord}(a)=(x_1,\dots,x_n)$.
\item For each $\iota$: $x_\iota\in V_2$ with successors $t_\iota,f_\iota$; $c_\iota\in V_S$ with $\mathrm{Ord}(c_\iota)=(o,d)$;
 $\mathrm{Ord}(t_\iota)=(c_\iota,T_{i_1},\dots,T_{i_{a_\iota}},o,\dots,o)$ listing the terms containing $x_\iota$,
 padded with $o$ to length $D+1$; $\mathrm{Ord}(f_\iota)$ likewise with $c_\iota$ followed by the terms containing $\neg x_\iota$.
\item For each term: $T_l\in V_S$, $\mathrm{Ord}(T_l)=(o,\dots,o,d)$ with $w_l-1$ copies of $o$.
\end{itemize}
Every cycle passes through $o$, and the $((D+1)n+1)$-st visit to $o$ goes to $\overline d$, so all plays
terminate in $d$ or $\overline d$.

\emph{Hardness.} ($\varphi$ not a tautology $\Rightarrow$ Player~2 avoids $d$.) Let $\nu$ falsify $\varphi$; Player~2 moves from $x_\iota$ to $t_\iota$ if
$\nu(x_\iota)=\top$ and to $f_\iota$ otherwise. Each $x_\iota$ is visited $D+1$ times. Only one of $t_\iota,f_\iota$ is
used, so $c_\iota$ is visited once and returns to $o$. The used node steps through its entire list, so each term
containing a true literal of $\iota$ is visited once for it. Hence $T_l$ is visited exactly as many times as it has true literals,
which is $<w_l$ since $\nu$ falsifies $T_l$; all these visits return to $o$. After $(D+1)n$ passes the play reaches $\overline d$.

($\varphi$ a tautology $\Rightarrow$ every play reaches $d$.) Suppose a play reaches $\overline d$. Then $o$ is visited $(D+1)n+1$
times, each $x_\iota$ exactly $D+1$ times, and $d$ is never entered. If both $t_\iota$ and $f_\iota$ were
used, $c_\iota$ would be visited twice (the first visit to each of $t_\iota,f_\iota$ goes to $c_\iota$), sending the play to $d$.
So exactly one is used, $D+1$ times; define $\nu(x_\iota)=\top$ iff $t_\iota$ is used. As above each $T_l$ is visited
once per true literal; since $d$ is not entered, each $T_l$ has fewer than $w_l$ true literals, so $\nu$ falsifies every
term and thus $\varphi$, a contradiction.

\emph{Containment.} Here $\val\in\{0,1\}$ and $\val=0$ iff Player~2 has a strategy avoiding $d$. We give an $\mathsf{NP}$ certificate for $\val=0$.
Since only Player~2 chooses, Player~2 avoids $d$ iff $\mathrm{Exp}(G)$ contains an infinite path from $(o,q^0)$ avoiding $\{d\}\times Q$ (equivalently, a lasso).

\emph{Step 1: preprocessing.} Let $H$ be the set of $d$-hopeful vertices: the attractor of $d$ in the game where $V_S$ is controlled by the
attractor player and $V_2$ by the adversary (computable in polynomial time), with $\mathrm{rank}(v)$ the stage at which $v$ enters. If $o\notin H$, the
complement of $H$ is closed in the following sense: a vertex outside $H$ that is switching has all successors outside $H$, and one in $V_2$ has a successor outside $H$.
So Player~2 can stay outside $H$ forever, whence $\val=0$; in this case a fixed certificate for a losing instance suffices. Otherwise, define $G^\dagger$ on $H\cup\{\overline d\}$ ($d$, $\overline d$ sinks) by
redirecting every entry of $\mathrm{Ord}(v)$ (positions are kept) that leads outside $H$ to $\overline d$. Player~2 nodes in $H$ have all successors in $H$ (by the attractor rule), so only switching nodes are affected.
Player~2 avoids $d$ in $G$ iff in $G^\dagger$: a play of $G$ that first enters $V\setminus H$ can be continued inside $V\setminus H$ (closure above) never reaching $d$, and conversely a play of $G^\dagger$ entering $\overline d$ corresponds to such a continuation.

\emph{Step 2: in $G^\dagger$, every infinite play avoiding $d$ reaches $\overline d$.} Suppose an infinite play avoids both $d$ and $\overline d$. Some state of $\mathrm{Exp}(G^\dagger)$ is visited infinitely often, so
the play contains a closed walk $W$ in $\mathrm{Exp}(G^\dagger)$. Let $U\subseteq H\setminus\{d\}$ be the set of vertices on $W$ and let $u\in U$ have minimal rank (rank $\ge1$).
If $u\in V_2$, all its successors have smaller rank (attractor rule for adversary nodes), and the vertex following $u$ on $W$ lies in $U$: contradiction.
If $u\in V_S$: $W$ returns to the same switch configuration, so the number of visits of $W$ to $u$ is a positive multiple of $|\mathrm{Ord}(u)|$, hence $W$ uses every entry of $\mathrm{Ord}(u)$,
including the entry leading to a vertex of smaller rank (it is not redirected, as its target is in $H$): contradiction.

\emph{Step 3: certificate.} By Steps 1--2, $\val=0$ iff Player~2 can reach $\overline d$ in $G^\dagger$. We claim this holds iff there is $x\in\mathbb Z_{\ge0}^{E}$ with:
\begin{itemize}
\item[(i)] flow conservation: $\mathrm{out}(v)=\mathrm{in}(v)$ for $v\notin\{o,\overline d\}$, $\mathrm{out}(o)=\mathrm{in}(o)+1$, $\mathrm{out}(\overline d)=0$ (so $\mathrm{in}(\overline d)=1$ by summing conservation), and $x_{(\cdot,d)}=0$;
\item[(ii)] switching: for $v\in V_S$ with $\mathrm{Ord}(v)=(u_1,\dots,u_k)$ and $n_v=\mathrm{out}(v)$, the flow on the $j$-th entry equals $\#\{t\in[1,n_v]:t\equiv j\pmod k\}$; flow on an edge is the sum over its entries.
\end{itemize}
($\Rightarrow$) Take a shortest play of $\mathrm{Exp}(G^\dagger)$ from $(o,q^0)$ to $\overline d$; it has at most $N\le|V|\prod_v|\mathrm{Ord}(v)|=2^{O(|G|)}$ steps, and its traversal counts satisfy (i), (ii) (switching nodes follow the cyclic order exactly) and have polynomial bit length.
($\Leftarrow$) Build a play from $x$: at switching nodes follow $\mathrm{Ord}$; at Player~2 nodes choose an edge whose traversal count is still below $x_e$. Invariant: each edge $e$ has been traversed at most $x_e$ times. Suppose the play is at $w\ne\overline d$ after $a$ arrivals (counting the start for $w=o$). Then $a\le\mathrm{in}(w)+[w=o]=\mathrm{out}(w)$, and only $a-1<\mathrm{out}(w)$ exits have been used, so a Player~2 choice with remaining flow exists, and at a switching node the $a$-th cyclic edge has remaining flow by (ii) since $a\le n_w$. The total flow is finite, so the play halts, and it can only halt at $\overline d$; since $x_{(\cdot,d)}=0$ it never enters $d$.
Verifying (i), (ii) takes polynomial time, so the complement of $\{S,2\}$-\textsc{Arrival} is in $\mathsf{NP}$.
\end{proof}

\begin{figure}[htbp]
    \centering
    \resizebox{0.95\textwidth}{!}{\input{figures_tikz_work/fig6.tex}}
    \caption[Proof of \Cref{sa:res:S2-is-coNP-com}: Constructing a graph from a formula $\varphi$.]{Overview of construction of a $\{S,2\}$-Arrival graph corresponding to a given 3DNF formula $\varphi$. Triangles represent Player~2 controlled vertices in $V_2$ and circles switching nodes in $V_S$, with switching order labelled on the edges. Only variable gadgets $\mathrm{X}_1$ and $\mathrm{X}_n$ and clause gadgets $\mathrm{C}_{i_{1}}$ and $\mathrm{C}_{j_{1}}$ are shown in full. Coloured edges visually group the principal return, consequence, and failure routes; the construction and edge labels remain unambiguous without colour.}
    \label{sa:fig:s2-conp:control}
\end{figure}
\begin{figure}[htbp]
    \centering
    \resizebox{0.34\textwidth}{!}{\begin{tikzpicture}[x=0.75pt,y=0.75pt,yscale=-1,xscale=1]

\draw [sa switch]   (102.47,33.13) .. controls (102.47,23.52) and (110.26,15.73) .. (119.87,15.73) .. controls (129.48,15.73) and (137.27,23.52) .. (137.27,33.13) .. controls (137.27,42.74) and (129.48,50.53) .. (119.87,50.53) .. controls (110.26,50.53) and (102.47,42.74) .. (102.47,33.13) -- cycle ;
\draw [sa switch]   (24.47,156.33) .. controls (24.47,146.72) and (32.26,138.93) .. (41.87,138.93) .. controls (51.48,138.93) and (59.27,146.72) .. (59.27,156.33) .. controls (59.27,165.94) and (51.48,173.73) .. (41.87,173.73) .. controls (32.26,173.73) and (24.47,165.94) .. (24.47,156.33) -- cycle ;
\draw  [dash pattern={on 4.5pt off 4.5pt}] (4.6,5.6) -- (159.4,5.6) -- (159.4,126.9) -- (4.6,126.9) -- cycle ;
\draw [sa switch]   (95.5,158.33) .. controls (95.5,148.72) and (103.29,140.93) .. (112.9,140.93) .. controls (122.51,140.93) and (130.3,148.72) .. (130.3,158.33) .. controls (130.3,167.94) and (122.51,175.73) .. (112.9,175.73) .. controls (103.29,175.73) and (95.5,167.94) .. (95.5,158.33) -- cycle ;
\draw [color=SAedgeReturn  ,draw opacity=1 ]   (119.87,50.53) -- (43.19,137.43) ;
\fill [shift={(41.87,138.93)}, rotate = 311.42] [color=SAedgeReturn  ,draw opacity=1 ] (0,0) -- (8.5,-3.5) -- (8.5,3.5) -- cycle ;
\draw [color=SAedgeConsequence  ,draw opacity=1 ]   (119.87,50.53) -- (113.05,138.94) ;
\fill [shift={(112.9,140.93)}, rotate = 274.41] [color=SAedgeConsequence  ,draw opacity=1 ] (0,0) -- (8.5,-3.5) -- (8.5,3.5) -- cycle ;

\draw (6.6,8.6) node [anchor=north west][inner sep=0.75pt]  [font=\Large] [align=left] {C$\displaystyle _{l}$};
\draw (108.8,24.83) node [anchor=north west][inner sep=0.75pt]  [font=\normalsize] [align=left] {$\displaystyle C_{l}$};
\draw (41.87,156.33) node [sa node label] {$o$};
\draw (112.90,158.33) node [sa node label] {$\overline d$};
\draw (139.0,58.0) node [sa edge label] {$w_l$};
\draw (14.8,46.2) node [anchor=north west][inner sep=0.75pt]   [align=left] {$\displaystyle 1,\dotsc ,w_{l} -1$};

\end{tikzpicture}}
    \caption[Proof of \Cref{sa:res:S2-is-coNP-com}: term gadget.]{Switching gadget for a term $T_l$ of width $w_l$. It reaches $d$ on the $w_l$-th visit, so it reaches the target exactly when every literal of the term is true under the selected assignment.}
    \label{sa:fig:s2-conp:term}
\end{figure}\FloatBarrier
\section{\PSPACE{}-hardness of {\tt $\{R,S,1\}$-Arrival-Quant}} \label{sa:sec:pspacehardness}
We show that {\tt $\{R,S,1\}$-Arrival-Quant} and {\tt $\{R,S,2\}$-Arrival-Quant} are \PSPACE{}-hard. By inclusion of node types, so is {\tt $\{R,S,1,2\}$-Arrival-Quant}.

In this section we use the global target $d$ and losing sink $\overline d$ consistently. Clause-local failure vertices are denoted $\overline d_l$; each represents the same losing outcome and may be identified with the global losing sink without changing the reduction. \par
Our proof takes inspiration from Fearnley et al.'s proof of \PSPACE{}-hardness for {\tt $\{S,1,2\}$-Arrival} (\cite[Theorem~4.3]{FGMS21}) and uses a technique from their proof of \NP{}-hardness for {\tt $\{S,1\}$-Arrival} (\cite[Theorem~3.8]{FGMS21}), but requires combining these with some new additional tricks to deal with the randomness. We give a brief overview of their methods and highlight the changes made to gain this result.

In their proof of \PSPACE{}-hardness for {\tt $\{S,1,2\}$-Arrival} (\cite[Theorem~4.3]{FGMS21}) they reduce from the {\tt QBF} problem. A given totally quantified boolean formula is evaluated in the following way: in the first phase (the variable phase) Player~1 picks assignments for existential variables and Player~2 picks assignments for universal variables; in the second phase (the formula phase) players play the standard model checking game for first order logic to determine a literal of the formula; the game concludes based on the truth value of that literal as chosen in the first phase. Our proof however reduces from the closely related {\tt SSAT} problem (\cite{Pap85}). We are also not able to construct the model checking game using random nodes as the Player~2 nodes are essential for the universal choices; instead we have to adapt the techniques of \cite[Theorem~3.8]{FGMS21} to evaluate the boolean formula, including adding a verification phase. Our random nodes also create additional problems with this formula evaluation process. In \cite[Theorem~3.8]{FGMS21} they simulate an existential quantifier by initially asking Player~1 to make an assignment then when evaluating the formula they enforce Player~1 to make the same choice. For random quantification we need to use a mixture of random and player nodes to achieve this.

To show the {\tt $\{R,S,1\}$-Arrival-Quant} is \PSPACE{}-hard we reduce from the {\tt SSAT} problem as defined by Papadimitriou (\cite{Pap85}, \Cref{prob:ssat}). The {\tt SSAT} problem is closely related to the {\tt QBF} problem, however, we replace universal "for all" quantification ($\forall$) with a new "for uniformly random" quantifier ($\RQuant$). This random quantification is simpler to achieve using our random nodes. Formally:

\problemStatement{Stochastic SAT (SSAT) \cite{Pap85}}{\label{prob:ssat}
Instance={A 3CNF Boolean formula $\varphi$ on $n$ variables $x_1,\ldots,x_n$, where $n$ is even, with three distinct literals per clause.},
Problem={Decide whether or not:
\begin{equation}
    \exists x_1\RQuant x_2\exists x_3\ldots\RQuant x_n\left[\mathbb{P}\bigl(\varphi(x_1,\ldots,x_n)=\top\bigr)>\frac{1}{2}\right]
\label{sa:eq:ssat-greater-half}
\end{equation}
}
}

Informally, we are asked whether there is a choice of Boolean value for $x_1$ such that, for a random choice (with the probability of true and false each equal to $\frac{1}{2}$) of truth value for $x_2$, there is a choice for $x_3$, etc., with subsequent choices able to depend on prior random outcomes. By \cite[Theorem~2]{Pap85} this problem is \PSPACE{}-complete. 

Our aim is to take an instance of {\tt SSAT} and construct an instance  $(G^\varphi,o,d)$ of generalised {\tt $\{R,S,1\}$-Arrival} with the following property:
\begin{equation}
    \val(G^\varphi,o,d)=\max_{x_1}[\mathbb{E}_{x_2}[\max_{x_3}[\ldots\mathbb{E}_{x_n}[\chi[\varphi(x_1,\ldots,x_n)=\top]\ldots]
    \label{sa:eq:val-gf-ssat}
\end{equation}
Where $\chi$ represents the indicator function for an event. Hence $\val(G^\varphi,o,d)>\frac{1}{2}$ exactly when (\ref{sa:eq:ssat-greater-half}) holds. We now describe the construction, prove that it has polynomial size, and establish this value correspondence.

Given an instance of {\tt SSAT} with 3CNF $\varphi$, $n$ variables and $m$ clauses where $\varphi = C_1 \wedge C_2 \wedge \ldots \wedge C_m$. We construct the instance $(G^\varphi,o,d)$ of generalised {\tt $\{R,S,1\}$-arrival} shown in \Cref{sa:fig:pspace-rs1-control} where each of the boxes represents the gadgets shown in \Cref{sa:fig:pspace-rs1-quantifiers,sa:fig:pspace-rs1-consequences,sa:fig:pspace-rs1-clause}, respectively and the values $a_i,b_i$ and $D$ are computable from the formula $\varphi$. 

\begin{figure}[htbp]
    \centering
\resizebox{0.78\textwidth}{!}{\input{figures_tikz_work/fig8.tex}}
    \caption[Proof of \Cref{sa:res:rs1-quant-is-pspace-hard}: Overall instance control structure.]{Overall instance ``Control'' structure.}
    \label{sa:fig:pspace-rs1-control}
\end{figure}
\begin{figure}[htbp]
\begin{subfigure}[htbp]{0.95\textwidth }
        \centering
\resizebox{0.8\linewidth}{!}{\begin{tikzpicture}[x=0.75pt,y=0.75pt,yscale=-1,xscale=1]

\draw  [dash pattern={on 4.5pt off 4.5pt}] (22,16) -- (527,16) -- (527,335.39) -- (22,335.39) -- cycle ;
\draw [sa random]   (40.08,26.83) -- (70.08,56.83) -- (40.08,86.83) -- (10.08,56.83) -- cycle ;
\draw [sa player one]   (15,139.5) -- (65,139.5) -- (65,189.5) -- (15,189.5) -- cycle ;
\draw [sa player one]   (16,265.5) -- (66,265.5) -- (66,315.5) -- (16,315.5) -- cycle ;
\draw [sa switch]   (131.83,94.33) .. controls (131.83,80.53) and (143.03,69.33) .. (156.83,69.33) .. controls (170.64,69.33) and (181.83,80.53) .. (181.83,94.33) .. controls (181.83,108.14) and (170.64,119.33) .. (156.83,119.33) .. controls (143.03,119.33) and (131.83,108.14) .. (131.83,94.33) -- cycle ;
\draw [sa switch]   (131.83,224) .. controls (131.83,210.19) and (143.03,199) .. (156.83,199) .. controls (170.64,199) and (181.83,210.19) .. (181.83,224) .. controls (181.83,237.81) and (170.64,249) .. (156.83,249) .. controls (143.03,249) and (131.83,237.81) .. (131.83,224) -- cycle ;
\draw  [dash pattern={on 4.5pt off 4.5pt}] (330,71) -- (400,71) -- (400,111) -- (330,111) -- cycle ;
\draw  [dash pattern={on 4.5pt off 4.5pt}] (331,197) -- (401,197) -- (401,237) -- (331,237) -- cycle ;
\draw    (66,57) -- (130.09,93.35) ;
\fill [shift={(131.83,94.33)}, rotate = 209.56] [color={rgb, 255:red, 0; green, 0; blue, 0 }  ] (0,0) -- (8.5,-3.5) -- (8.5,3.5) -- cycle ;
\draw    (66,57) -- (131.1,222.14) ;
\fill [shift={(131.83,224)}, rotate = 248.49] [color={rgb, 255:red, 0; green, 0; blue, 0 }  ] (0,0) -- (8.5,-3.5) -- (8.5,3.5) -- cycle ;
\draw    (66,164) -- (130.46,95.79) ;
\fill [shift={(131.83,94.33)}, rotate = 133.38] [color={rgb, 255:red, 0; green, 0; blue, 0 }  ] (0,0) -- (8.5,-3.5) -- (8.5,3.5) -- cycle ;
\draw    (66,164) -- (130.36,222.65) ;
\fill [shift={(131.83,224)}, rotate = 222.35] [color={rgb, 255:red, 0; green, 0; blue, 0 }  ] (0,0) -- (8.5,-3.5) -- (8.5,3.5) -- cycle ;
\draw    (65,289) -- (131.18,96.22) ;
\fill [shift={(131.83,94.33)}, rotate = 108.95] [color={rgb, 255:red, 0; green, 0; blue, 0 }  ] (0,0) -- (8.5,-3.5) -- (8.5,3.5) -- cycle ;
\draw    (65,289) -- (130.4,225.39) ;
\fill [shift={(131.83,224)}, rotate = 135.8] [color={rgb, 255:red, 0; green, 0; blue, 0 }  ] (0,0) -- (8.5,-3.5) -- (8.5,3.5) -- cycle ;
\draw    (181.83,94.33) -- (326,92.42) ;
\fill [shift={(328,92.39)}, rotate = 179.24] [color={rgb, 255:red, 0; green, 0; blue, 0 }  ] (0,0) -- (8.5,-3.5) -- (8.5,3.5) -- cycle ;
\draw    (181.83,224) -- (326,221.43) ;
\fill [shift={(328,221.39)}, rotate = 178.98] [color={rgb, 255:red, 0; green, 0; blue, 0 }  ] (0,0) -- (8.5,-3.5) -- (8.5,3.5) -- cycle ;
\draw    (181.83,94.33) .. controls (202,-11.61) and (397.5,42) .. (520.5,94) ;
\fill [shift={(520.5,94)}, rotate = 202.92] [color={rgb, 255:red, 0; green, 0; blue, 0 }  ] (0,0) -- (8.5,-3.5) -- (8.5,3.5) -- cycle ;
\draw    (181.83,224) .. controls (182.99,159.71) and (418.13,160) .. (518.99,94.99) ;
\fill [shift={(520.5,94)}, rotate = 146.58] [color={rgb, 255:red, 0; green, 0; blue, 0 }  ] (0,0) -- (8.5,-3.5) -- (8.5,3.5) -- cycle ;
\draw    (403,91) -- (518.5,93.95) ;
\fill [shift={(520.5,94)}, rotate = 181.46] [color={rgb, 255:red, 0; green, 0; blue, 0 }  ] (0,0) -- (8.5,-3.5) -- (8.5,3.5) -- cycle ;
\draw    (401,218) -- (519.11,95.44) ;
\fill [shift={(520.5,94)}, rotate = 133.94] [color={rgb, 255:red, 0; green, 0; blue, 0 }  ] (0,0) -- (8.5,-3.5) -- (8.5,3.5) -- cycle ;
\draw [sa switch]   (483,217.39) .. controls (483,203.58) and (494.19,192.39) .. (508,192.39) .. controls (521.81,192.39) and (533,203.58) .. (533,217.39) .. controls (533,231.2) and (521.81,242.39) .. (508,242.39) .. controls (494.19,242.39) and (483,231.2) .. (483,217.39) -- cycle ;
\draw  [sa switch,fill=SAfigureSwitch  ,fill opacity=1 ] (328,92.39) .. controls (328,89.35) and (330.46,86.89) .. (333.5,86.89) .. controls (336.54,86.89) and (339,89.35) .. (339,92.39) .. controls (339,95.43) and (336.54,97.89) .. (333.5,97.89) .. controls (330.46,97.89) and (328,95.43) .. (328,92.39) -- cycle ;
\draw  [sa switch,fill=SAfigureSwitch  ,fill opacity=1 ] (397.5,91) .. controls (397.5,87.96) and (399.96,85.5) .. (403,85.5) .. controls (406.04,85.5) and (408.5,87.96) .. (408.5,91) .. controls (408.5,94.04) and (406.04,96.5) .. (403,96.5) .. controls (399.96,96.5) and (397.5,94.04) .. (397.5,91) -- cycle ;
\draw  [sa switch,fill=SAfigureSwitch  ,fill opacity=1 ] (322.5,221.39) .. controls (322.5,218.35) and (324.96,215.89) .. (328,215.89) .. controls (331.04,215.89) and (333.5,218.35) .. (333.5,221.39) .. controls (333.5,224.43) and (331.04,226.89) .. (328,226.89) .. controls (324.96,226.89) and (322.5,224.43) .. (322.5,221.39) -- cycle ;
\draw  [sa switch,fill=SAfigureSwitch  ,fill opacity=1 ] (395.5,218) .. controls (395.5,214.96) and (397.96,212.5) .. (401,212.5) .. controls (404.04,212.5) and (406.5,214.96) .. (406.5,218) .. controls (406.5,221.04) and (404.04,223.5) .. (401,223.5) .. controls (397.96,223.5) and (395.5,221.04) .. (395.5,218) -- cycle ;
\draw    (181.83,224) .. controls (203,262.39) and (337,312.39) .. (483,217.39) ;
\fill [shift={(483,217.39)}, rotate = 146.95] [color={rgb, 255:red, 0; green, 0; blue, 0 }  ] (0,0) -- (8.5,-3.5) -- (8.5,3.5) -- cycle ;
\draw    (181.83,94.33) .. controls (182.99,169.02) and (424.57,161.47) .. (482.15,216.56) ;
\fill [shift={(483,217.39)}, rotate = 225] [color={rgb, 255:red, 0; green, 0; blue, 0 }  ] (0,0) -- (8.5,-3.5) -- (8.5,3.5) -- cycle ;
\draw  [sa switch,fill=SAfigureSwitch  ,fill opacity=1 ] (520.5,94) .. controls (520.5,90.96) and (522.96,88.5) .. (526,88.5) .. controls (529.04,88.5) and (531.5,90.96) .. (531.5,94) .. controls (531.5,97.04) and (529.04,99.5) .. (526,99.5) .. controls (522.96,99.5) and (520.5,97.04) .. (520.5,94) -- cycle ;

\draw (40.08,56.83) node [sa node label large] {$as_i$};
\draw (40.00,164.50) node [sa node label] {$\displaystyle ag_{i}$};
\draw (41.00,290.50) node [sa node label] {$\displaystyle ver_{i}$};
\draw (156.83,94.33) node [sa node label] {$\displaystyle x_{i}^{T}$};
\draw (156.83,224.00) node [sa node label] {$\displaystyle x_{i}^{F}$};
\draw (351,81) node [anchor=north west][inner sep=0.75pt]   [align=left] {$\displaystyle \mathit{Neg}_{i}$};
\draw (352,207) node [anchor=north west][inner sep=0.75pt]   [align=left] {$\displaystyle \mathit{Pos}_{i}$};
\draw (245,72.33) node [sa edge label] {$2,\dotsc,D+1$};
\draw (245,202.00) node [sa edge label] {$2,\dotsc,D+1$};
\draw (501.5,64.5) node [anchor=north west][inner sep=0.75pt]   [align=left] {$\displaystyle ret_{i}$};
\draw (202,50.33) node [sa edge label] {$1$};
\draw (207,182.00) node [sa edge label] {$1$};
\draw (486.5,207.5) node [anchor=north west][inner sep=0.75pt]   [align=left] {$\displaystyle next_{i}$};
\draw (238,120.33) node [sa edge label] {$D+2$};
\draw (238,253.00) node [sa edge label] {$D+2$};

\end{tikzpicture}}
    \caption{Randomly Quantified}
    \label{sa:fig:pspace-rs1-rand-quant}
    \end{subfigure}
\begin{subfigure}[htbp]{0.95\textwidth }
    \centering
   \resizebox{0.8\linewidth}{!}{\begin{tikzpicture}[x=0.75pt,y=0.75pt,yscale=-1,xscale=1]

\draw  [dash pattern={on 4.5pt off 4.5pt}] (22,14.72) -- (527,14.72) -- (527,334.11) -- (22,334.11) -- cycle ;
\draw [sa player one]   (15,138.22) -- (65,138.22) -- (65,188.22) -- (15,188.22) -- cycle ;
\draw [sa player one]   (16,264.22) -- (66,264.22) -- (66,314.22) -- (16,314.22) -- cycle ;
\draw [sa switch]   (131.83,93.05) .. controls (131.83,79.24) and (143.03,68.05) .. (156.83,68.05) .. controls (170.64,68.05) and (181.83,79.24) .. (181.83,93.05) .. controls (181.83,106.86) and (170.64,118.05) .. (156.83,118.05) .. controls (143.03,118.05) and (131.83,106.86) .. (131.83,93.05) -- cycle ;
\draw [sa switch]   (131.83,222.72) .. controls (131.83,208.91) and (143.03,197.72) .. (156.83,197.72) .. controls (170.64,197.72) and (181.83,208.91) .. (181.83,222.72) .. controls (181.83,236.53) and (170.64,247.72) .. (156.83,247.72) .. controls (143.03,247.72) and (131.83,236.53) .. (131.83,222.72) -- cycle ;
\draw  [dash pattern={on 4.5pt off 4.5pt}] (330,69.72) -- (400,69.72) -- (400,109.72) -- (330,109.72) -- cycle ;
\draw  [dash pattern={on 4.5pt off 4.5pt}] (331,195.72) -- (401,195.72) -- (401,235.72) -- (331,235.72) -- cycle ;
\draw    (66,55.72) -- (130.09,92.07) ;
\fill [shift={(131.83,93.05)}, rotate = 209.56] [color={rgb, 255:red, 0; green, 0; blue, 0 }  ] (0,0) -- (8.5,-3.5) -- (8.5,3.5) -- cycle ;
\draw    (66,55.72) -- (131.1,220.86) ;
\fill [shift={(131.83,222.72)}, rotate = 248.49] [color={rgb, 255:red, 0; green, 0; blue, 0 }  ] (0,0) -- (8.5,-3.5) -- (8.5,3.5) -- cycle ;
\draw    (66,162.72) -- (130.46,94.51) ;
\fill [shift={(131.83,93.05)}, rotate = 133.38] [color={rgb, 255:red, 0; green, 0; blue, 0 }  ] (0,0) -- (8.5,-3.5) -- (8.5,3.5) -- cycle ;
\draw    (66,162.72) -- (130.36,221.37) ;
\fill [shift={(131.83,222.72)}, rotate = 222.35] [color={rgb, 255:red, 0; green, 0; blue, 0 }  ] (0,0) -- (8.5,-3.5) -- (8.5,3.5) -- cycle ;
\draw    (65,287.72) -- (131.18,94.94) ;
\fill [shift={(131.83,93.05)}, rotate = 108.95] [color={rgb, 255:red, 0; green, 0; blue, 0 }  ] (0,0) -- (8.5,-3.5) -- (8.5,3.5) -- cycle ;
\draw    (65,287.72) -- (130.4,224.11) ;
\fill [shift={(131.83,222.72)}, rotate = 135.8] [color={rgb, 255:red, 0; green, 0; blue, 0 }  ] (0,0) -- (8.5,-3.5) -- (8.5,3.5) -- cycle ;
\draw    (181.83,93.05) -- (326,91.14) ;
\fill [shift={(328,91.11)}, rotate = 179.24] [color={rgb, 255:red, 0; green, 0; blue, 0 }  ] (0,0) -- (8.5,-3.5) -- (8.5,3.5) -- cycle ;
\draw    (181.83,222.72) -- (326,220.15) ;
\fill [shift={(328,220.11)}, rotate = 178.98] [color={rgb, 255:red, 0; green, 0; blue, 0 }  ] (0,0) -- (8.5,-3.5) -- (8.5,3.5) -- cycle ;
\draw    (181.83,93.05) .. controls (202,-12.89) and (397.5,39.72) .. (520.5,91.72) ;
\fill [shift={(520.5,91.72)}, rotate = 202.92] [color={rgb, 255:red, 0; green, 0; blue, 0 }  ] (0,0) -- (8.5,-3.5) -- (8.5,3.5) -- cycle ;
\draw    (181.83,222.72) .. controls (182.99,158.43) and (418.13,157.72) .. (518.99,92.7) ;
\fill [shift={(520.5,91.72)}, rotate = 146.58] [color={rgb, 255:red, 0; green, 0; blue, 0 }  ] (0,0) -- (8.5,-3.5) -- (8.5,3.5) -- cycle ;
\draw    (403,89.72) -- (518.5,91.68) ;
\fill [shift={(520.5,91.72)}, rotate = 180.98] [color={rgb, 255:red, 0; green, 0; blue, 0 }  ] (0,0) -- (8.5,-3.5) -- (8.5,3.5) -- cycle ;
\draw    (401,216.72) -- (519.12,93.16) ;
\fill [shift={(520.5,91.72)}, rotate = 133.71] [color={rgb, 255:red, 0; green, 0; blue, 0 }  ] (0,0) -- (8.5,-3.5) -- (8.5,3.5) -- cycle ;
\draw [sa switch]   (483,216.11) .. controls (483,202.3) and (494.19,191.11) .. (508,191.11) .. controls (521.81,191.11) and (533,202.3) .. (533,216.11) .. controls (533,229.92) and (521.81,241.11) .. (508,241.11) .. controls (494.19,241.11) and (483,229.92) .. (483,216.11) -- cycle ;
\draw  [sa switch,fill=SAfigureSwitch  ,fill opacity=1 ] (328,91.11) .. controls (328,88.07) and (330.46,85.61) .. (333.5,85.61) .. controls (336.54,85.61) and (339,88.07) .. (339,91.11) .. controls (339,94.15) and (336.54,96.61) .. (333.5,96.61) .. controls (330.46,96.61) and (328,94.15) .. (328,91.11) -- cycle ;
\draw  [sa switch,fill=SAfigureSwitch  ,fill opacity=1 ] (397.5,89.72) .. controls (397.5,86.68) and (399.96,84.22) .. (403,84.22) .. controls (406.04,84.22) and (408.5,86.68) .. (408.5,89.72) .. controls (408.5,92.76) and (406.04,95.22) .. (403,95.22) .. controls (399.96,95.22) and (397.5,92.76) .. (397.5,89.72) -- cycle ;
\draw  [sa switch,fill=SAfigureSwitch  ,fill opacity=1 ] (322.5,220.11) .. controls (322.5,217.07) and (324.96,214.61) .. (328,214.61) .. controls (331.04,214.61) and (333.5,217.07) .. (333.5,220.11) .. controls (333.5,223.15) and (331.04,225.61) .. (328,225.61) .. controls (324.96,225.61) and (322.5,223.15) .. (322.5,220.11) -- cycle ;
\draw  [sa switch,fill=SAfigureSwitch  ,fill opacity=1 ] (395.5,216.72) .. controls (395.5,213.68) and (397.96,211.22) .. (401,211.22) .. controls (404.04,211.22) and (406.5,213.68) .. (406.5,216.72) .. controls (406.5,219.76) and (404.04,222.22) .. (401,222.22) .. controls (397.96,222.22) and (395.5,219.76) .. (395.5,216.72) -- cycle ;
\draw    (181.83,222.72) .. controls (203,261.11) and (337,311.11) .. (483,216.11) ;
\fill [shift={(483,216.11)}, rotate = 146.95] [color={rgb, 255:red, 0; green, 0; blue, 0 }  ] (0,0) -- (8.5,-3.5) -- (8.5,3.5) -- cycle ;
\draw    (181.83,93.05) .. controls (182.99,167.73) and (424.57,160.19) .. (482.15,215.27) ;
\fill [shift={(483,216.11)}, rotate = 225] [color={rgb, 255:red, 0; green, 0; blue, 0 }  ] (0,0) -- (8.5,-3.5) -- (8.5,3.5) -- cycle ;
\draw [sa player one]   (15,31.22) -- (65,31.22) -- (65,81.22) -- (15,81.22) -- cycle ;
\draw  [sa switch,fill=SAfigureSwitch  ,fill opacity=1 ] (520.5,91.72) .. controls (520.5,88.68) and (522.96,86.22) .. (526,86.22) .. controls (529.04,86.22) and (531.5,88.68) .. (531.5,91.72) .. controls (531.5,94.76) and (529.04,97.22) .. (526,97.22) .. controls (522.96,97.22) and (520.5,94.76) .. (520.5,91.72) -- cycle ;

\draw (40.00,56.22) node [sa node label large] {$as_i$};
\draw (40.00,163.22) node [sa node label] {$\displaystyle ag_{i}$};
\draw (41.00,289.22) node [sa node label] {$\displaystyle ver_{i}$};
\draw (156.83,93.05) node [sa node label] {$\displaystyle x_{i}^{T}$};
\draw (156.83,222.72) node [sa node label] {$\displaystyle x_{i}^{F}$};
\draw (351,79.72) node [anchor=north west][inner sep=0.75pt]   [align=left] {$\displaystyle \mathit{Neg}_{i}$};
\draw (352,205.72) node [anchor=north west][inner sep=0.75pt]   [align=left] {$\displaystyle \mathit{Pos}_{i}$};
\draw (245,71.05) node [sa edge label] {$2,\dotsc,D+1$};
\draw (245,200.72) node [sa edge label] {$2,\dotsc,D+1$};
\draw (501.5,66.22) node [anchor=north west][inner sep=0.75pt]   [align=left] {$\displaystyle ret_{i}$};
\draw (202,49.05) node [sa edge label] {$1$};
\draw (207,180.72) node [sa edge label] {$1$};
\draw (486.5,206.22) node [anchor=north west][inner sep=0.75pt]   [align=left] {$\displaystyle next_{i}$};
\draw (238,119.05) node [sa edge label] {$D+2$};
\draw (238,251.72) node [sa edge label] {$D+2$};

\end{tikzpicture}}
    \caption{Existentially Quantified}
    \label{sa:fig:pspace-rs1-exists-quant}
\end{subfigure}
\caption[Proof of \Cref{sa:res:rs1-quant-is-pspace-hard}: Quantification constructions.]{Gadgets for quantified variables.}
\label{sa:fig:pspace-rs1-quantifiers}
\end{figure}
\begin{figure}[H]
\begin{subfigure}[htbp]{0.45\textwidth }
    \centering
\resizebox{\linewidth}{!}{\begin{tikzpicture}[x=0.75pt,y=0.75pt,yscale=-1,xscale=1]

\draw [sa switch]   (7,78.85) .. controls (7,65.04) and (18.19,53.85) .. (32,53.85) .. controls (45.81,53.85) and (57,65.04) .. (57,78.85) .. controls (57,92.65) and (45.81,103.85) .. (32,103.85) .. controls (18.19,103.85) and (7,92.65) .. (7,78.85) -- cycle ;
\draw    (57,78.85) -- (115.35,38.63) ;
\fill [shift={(117,37.5)}, rotate = 145.43] [color={rgb, 255:red, 0; green, 0; blue, 0 }  ] (0,0) -- (8.5,-3.5) -- (8.5,3.5) -- cycle ;
\draw    (57,78.85) -- (114.23,108.58) ;
\fill [shift={(116,109.5)}, rotate = 207.45] [color={rgb, 255:red, 0; green, 0; blue, 0 }  ] (0,0) -- (8.5,-3.5) -- (8.5,3.5) -- cycle ;
\draw  [dash pattern={on 4.5pt off 4.5pt}] (15,13.5) -- (232,13.5) -- (232,192.5) -- (15,192.5) -- cycle ;
\draw  [sa switch,fill=SAfigureSwitch  ,fill opacity=1 ] (230.75,99.8) .. controls (230.75,96.77) and (233.21,94.3) .. (236.25,94.3) .. controls (239.29,94.3) and (241.75,96.77) .. (241.75,99.8) .. controls (241.75,102.84) and (239.29,105.3) .. (236.25,105.3) .. controls (233.21,105.3) and (230.75,102.84) .. (230.75,99.8) -- cycle ;
\draw  [sa switch,fill=SAfigureSwitch  ,fill opacity=1 ] (111.5,37.5) .. controls (111.5,34.46) and (113.96,32) .. (117,32) .. controls (120.04,32) and (122.5,34.46) .. (122.5,37.5) .. controls (122.5,40.54) and (120.04,43) .. (117,43) .. controls (113.96,43) and (111.5,40.54) .. (111.5,37.5) -- cycle ;
\draw  [sa switch,fill=SAfigureSwitch  ,fill opacity=1 ] (110.5,109.5) .. controls (110.5,106.46) and (112.96,104) .. (116,104) .. controls (119.04,104) and (121.5,106.46) .. (121.5,109.5) .. controls (121.5,112.54) and (119.04,115) .. (116,115) .. controls (112.96,115) and (110.5,112.54) .. (110.5,109.5) -- cycle ;
\draw  [sa switch,fill=SAfigureSwitch  ,fill opacity=1 ] (158,37.11) .. controls (158,34.07) and (160.46,31.61) .. (163.5,31.61) .. controls (166.54,31.61) and (169,34.07) .. (169,37.11) .. controls (169,40.15) and (166.54,42.61) .. (163.5,42.61) .. controls (160.46,42.61) and (158,40.15) .. (158,37.11) -- cycle ;
\draw  [sa switch,fill=SAfigureSwitch  ,fill opacity=1 ] (160.5,108.5) .. controls (160.5,105.46) and (162.96,103) .. (166,103) .. controls (169.04,103) and (171.5,105.46) .. (171.5,108.5) .. controls (171.5,111.54) and (169.04,114) .. (166,114) .. controls (162.96,114) and (160.5,111.54) .. (160.5,108.5) -- cycle ;
\draw    (57,78.85) .. controls (30.27,225.02) and (187.55,131.71) .. (229.51,100.72) ;
\fill [shift={(230.75,99.8)}, rotate = 143.13] [color={rgb, 255:red, 0; green, 0; blue, 0 }  ] (0,0) -- (8.5,-3.5) -- (8.5,3.5) -- cycle ;
\draw    (163.5,37.11) -- (229.29,98.44) ;
\fill [shift={(230.75,99.8)}, rotate = 222.99] [color={rgb, 255:red, 0; green, 0; blue, 0 }  ] (0,0) -- (8.5,-3.5) -- (8.5,3.5) -- cycle ;
\draw    (166,108.5) -- (228.77,100.07) ;
\fill [shift={(230.75,99.8)}, rotate = 172.35] [color={rgb, 255:red, 0; green, 0; blue, 0 }  ] (0,0) -- (8.5,-3.5) -- (8.5,3.5) -- cycle ;

\draw  [dash pattern={on 4.5pt off 4.5pt}]  (117.1,17.5) -- (162.1,17.5) -- (162.1,56.5) -- (117.1,56.5) -- cycle  ;
\draw (139.6,37) node [sa gadget label] {$C_{k_{1}}$};
\draw  [dash pattern={on 4.5pt off 4.5pt}]  (120.1,89.5) -- (165.1,89.5) -- (165.1,128.5) -- (120.1,128.5) -- cycle  ;
\draw (142.6,109) node [sa gadget label] {$C_{k_{a_{i}}}$};
\draw (32.00,78.85) node [sa node label] {$\mathit{Pos}_{i}$};
\draw (79.0,37.0) node [anchor=north west][inner sep=0.75pt]   [align=left] {$\displaystyle 1$};
\draw (54.1,164.8) node [anchor=north west][inner sep=0.75pt]   [align=left] {$\displaystyle a_{i} +1,\dotsc ,D$};
\draw (62.5,88) node [anchor=north west][inner sep=0.75pt]   [align=left] {$\displaystyle a_{i}$};
\draw (106.1,52.7) node [anchor=north west][inner sep=0.75pt]  [font=\LARGE] [align=left] {...};

\end{tikzpicture}}
    \caption{$Pos_i$ consequence gadget}
    \label{sa:fig:pspace-rs1-pos-cons}
\end{subfigure}
\begin{subfigure}[htbp]{0.45\textwidth }
    \centering
\resizebox{\linewidth}{!}{\begin{tikzpicture}[x=0.75pt,y=0.75pt,yscale=-1,xscale=1]

\draw [sa switch]   (7,78.85) .. controls (7,65.04) and (18.19,53.85) .. (32,53.85) .. controls (45.81,53.85) and (57,65.04) .. (57,78.85) .. controls (57,92.65) and (45.81,103.85) .. (32,103.85) .. controls (18.19,103.85) and (7,92.65) .. (7,78.85) -- cycle ;
\draw    (57,78.85) -- (115.35,38.63) ;
\fill [shift={(117,37.5)}, rotate = 145.43] [color={rgb, 255:red, 0; green, 0; blue, 0 }  ] (0,0) -- (8.5,-3.5) -- (8.5,3.5) -- cycle ;
\draw    (57,78.85) -- (114.23,108.58) ;
\fill [shift={(116,109.5)}, rotate = 207.45] [color={rgb, 255:red, 0; green, 0; blue, 0 }  ] (0,0) -- (8.5,-3.5) -- (8.5,3.5) -- cycle ;
\draw  [dash pattern={on 4.5pt off 4.5pt}] (15,13.5) -- (232,13.5) -- (232,192.5) -- (15,192.5) -- cycle ;
\draw  [sa switch,fill=SAfigureSwitch  ,fill opacity=1 ] (230.75,99.8) .. controls (230.75,96.77) and (233.21,94.3) .. (236.25,94.3) .. controls (239.29,94.3) and (241.75,96.77) .. (241.75,99.8) .. controls (241.75,102.84) and (239.29,105.3) .. (236.25,105.3) .. controls (233.21,105.3) and (230.75,102.84) .. (230.75,99.8) -- cycle ;
\draw  [sa switch,fill=SAfigureSwitch  ,fill opacity=1 ] (111.5,37.5) .. controls (111.5,34.46) and (113.96,32) .. (117,32) .. controls (120.04,32) and (122.5,34.46) .. (122.5,37.5) .. controls (122.5,40.54) and (120.04,43) .. (117,43) .. controls (113.96,43) and (111.5,40.54) .. (111.5,37.5) -- cycle ;
\draw  [sa switch,fill=SAfigureSwitch  ,fill opacity=1 ] (110.5,109.5) .. controls (110.5,106.46) and (112.96,104) .. (116,104) .. controls (119.04,104) and (121.5,106.46) .. (121.5,109.5) .. controls (121.5,112.54) and (119.04,115) .. (116,115) .. controls (112.96,115) and (110.5,112.54) .. (110.5,109.5) -- cycle ;
\draw  [sa switch,fill=SAfigureSwitch  ,fill opacity=1 ] (158,37.11) .. controls (158,34.07) and (160.46,31.61) .. (163.5,31.61) .. controls (166.54,31.61) and (169,34.07) .. (169,37.11) .. controls (169,40.15) and (166.54,42.61) .. (163.5,42.61) .. controls (160.46,42.61) and (158,40.15) .. (158,37.11) -- cycle ;
\draw  [sa switch,fill=SAfigureSwitch  ,fill opacity=1 ] (160.5,108.5) .. controls (160.5,105.46) and (162.96,103) .. (166,103) .. controls (169.04,103) and (171.5,105.46) .. (171.5,108.5) .. controls (171.5,111.54) and (169.04,114) .. (166,114) .. controls (162.96,114) and (160.5,111.54) .. (160.5,108.5) -- cycle ;
\draw    (57,78.85) .. controls (30.27,225.02) and (187.55,131.71) .. (229.51,100.72) ;
\fill [shift={(230.75,99.8)}, rotate = 143.13] [color={rgb, 255:red, 0; green, 0; blue, 0 }  ] (0,0) -- (8.5,-3.5) -- (8.5,3.5) -- cycle ;
\draw    (163.5,37.11) -- (229.29,98.44) ;
\fill [shift={(230.75,99.8)}, rotate = 222.99] [color={rgb, 255:red, 0; green, 0; blue, 0 }  ] (0,0) -- (8.5,-3.5) -- (8.5,3.5) -- cycle ;
\draw    (166,108.5) -- (228.77,100.07) ;
\fill [shift={(230.75,99.8)}, rotate = 172.35] [color={rgb, 255:red, 0; green, 0; blue, 0 }  ] (0,0) -- (8.5,-3.5) -- (8.5,3.5) -- cycle ;

\draw  [dash pattern={on 4.5pt off 4.5pt}]  (117.1,17.5) -- (162.1,17.5) -- (162.1,56.5) -- (117.1,56.5) -- cycle  ;
\draw (139.6,37) node [sa gadget label] {$C_{j_{1}}$};
\draw  [dash pattern={on 4.5pt off 4.5pt}]  (120.1,89.5) -- (165.1,89.5) -- (165.1,128.5) -- (120.1,128.5) -- cycle  ;
\draw (142.6,109) node [sa gadget label] {$C_{j_{b_{i}}}$};
\draw (32.00,78.85) node [sa node label] {$\mathit{Neg}_{i}$};
\draw (79.0,37.0) node [anchor=north west][inner sep=0.75pt]   [align=left] {$\displaystyle 1$};
\draw (54.1,164.8) node [anchor=north west][inner sep=0.75pt]   [align=left] {$\displaystyle b_{i} +1,\dotsc ,D$};
\draw (62.5,88) node [anchor=north west][inner sep=0.75pt]   [align=left] {$\displaystyle b_{i}$};
\draw (106.1,52.7) node [anchor=north west][inner sep=0.75pt]  [font=\LARGE] [align=left] {...};

\end{tikzpicture}}
    \caption{$Neg_i$ consequence gadget}
    \label{sa:fig:pspace-rs1-neg-cons}
\end{subfigure}
\caption[Proof of \Cref{sa:res:rs1-quant-is-pspace-hard}: Consequence constructions.]{Gadgets for the positive and negative consequences of variable $x_i$.}
\label{sa:fig:pspace-rs1-consequences}
\end{figure}
\begin{figure}[H]
    \centering
\resizebox{0.28\textwidth}{!}{\begin{tikzpicture}[x=0.75pt,y=0.75pt,yscale=-1,xscale=1]

\draw [sa switch]   (12.5,56.5) .. controls (12.5,42.69) and (23.69,31.5) .. (37.5,31.5) .. controls (51.31,31.5) and (62.5,42.69) .. (62.5,56.5) .. controls (62.5,70.31) and (51.31,81.5) .. (37.5,81.5) .. controls (23.69,81.5) and (12.5,70.31) .. (12.5,56.5) -- cycle ;
\draw    (62.5,56.5) -- (214.02,37.75) ;
\fill [shift={(216,37.5)}, rotate = 172.94] [color={rgb, 255:red, 0; green, 0; blue, 0 }  ] (0,0) -- (8.5,-3.5) -- (8.5,3.5) -- cycle ;
\draw    (62.5,56.5) -- (111.09,105.09) ;
\fill [shift={(112.5,106.5)}, rotate = 225] [color={rgb, 255:red, 0; green, 0; blue, 0 }  ] (0,0) -- (8.5,-3.5) -- (8.5,3.5) -- cycle ;
\draw  [dash pattern={on 4.5pt off 4.5pt}] (21,11.5) -- (219,11.5) -- (219,144.5) -- (21,144.5) -- cycle ;
\draw [sa switch]   (112.5,106.5) .. controls (112.5,92.69) and (123.69,81.5) .. (137.5,81.5) .. controls (151.31,81.5) and (162.5,92.69) .. (162.5,106.5) .. controls (162.5,120.31) and (151.31,131.5) .. (137.5,131.5) .. controls (123.69,131.5) and (112.5,120.31) .. (112.5,106.5) -- cycle ;
\draw    (162.5,106.5) .. controls (210.52,90.66) and (178.16,50.32) .. (156.65,88.32) ;
\fill [shift={(156,89.5)}, rotate = 298.26] [color={rgb, 255:red, 0; green, 0; blue, 0 }  ] (0,0) -- (8.5,-3.5) -- (8.5,3.5) -- cycle ;
\draw  [sa switch,fill=SAfigureSwitch  ,fill opacity=1 ] (210.5,37.5) .. controls (210.5,34.46) and (212.96,32) .. (216,32) .. controls (219.04,32) and (221.5,34.46) .. (221.5,37.5) .. controls (221.5,40.54) and (219.04,43) .. (216,43) .. controls (212.96,43) and (210.5,40.54) .. (210.5,37.5) -- cycle ;

\draw (75,33.5) node [anchor=north west][inner sep=0.75pt]   [align=left] {$\displaystyle 1,2$};
\draw (67.5,84) node [anchor=north west][inner sep=0.75pt]   [align=left] {$\displaystyle 3$};
\draw (137.50,106.50) node [sa node label large] {$\overline d_l$};
\draw (37.50,56.50) node [sa node label] {$\displaystyle c_{l}$};

\end{tikzpicture}}
\caption[Proof of \Cref{sa:res:rs1-quant-is-pspace-hard}: Clause construction.]{Gadget for clauses $C_l$.}
    \label{sa:fig:pspace-rs1-clause}
\end{figure}

We now explain this construction in more detail.  Given $\varphi= C_1 \wedge C_2 \wedge \ldots \wedge C_m$,
to begin with, in polynomial time we enumerate our $n$ variables as $x_1,\ldots,x_n$ and for each we compute constants $a_i = | \{ l \in \{1,\ldots,m\} \mid  x_i \in C_l \} |$
and $b_i = | \{ l \in \{1, \ldots, m\} \mid \neg x_i \in C_l \} |$. 
Here $a_i$ is the number of clauses in which the literal $x_i$ appears, and $b_i$ is the number of clauses in which the literal $\neg x_i$ appears. 
We let $D=\max  \bigcup_i \{a_i,b_i\}$ be the maximum number of occurrences of any
literal.
The control vertices $\mathit{as}$, $\mathit{ag}$, and $\mathit{ver}$ abbreviate \emph{assignment}, \emph{agreement}, and \emph{verification}, respectively. We divide the game into three phases corresponding to these entries in $\Ord(o)$: the ``assignment'' phase, consisting of the time strictly before the $n+1$'th visit to the vertex $o$ where the switching node takes us to the node $\mathit{as}$, the ``agreement'' phase, consisting of the time after the assignment phase and strictly before the $((D+1)n+1)$-st visit to $o$, and the ``verification'' phase consisting of the time afterwards where the switching takes us to either $\mathit{ver}$ or $\overline d$. Each phase has the following objective:
\begin{itemize}
    \item \textbf{Assignment Phase} - In this phase the player and nature alternate in choosing values of $x_1,\ldots,x_n$ in sequence.
    \item \textbf{Agreement Phase} - In this phase, the player must continue to agree with the choices in the ``assignment'' phase. Each time we visit we go through a list of clauses which our choice of assignment to that variable does not satisfy.
    \item \textbf{Verification Phase} - In this phase we verify that the player acted honestly and did agree with the choices in the ``assignment'' phase by moving through each variable gadget.
\end{itemize}

These phases correspond to the three distinct entries to each of our quantified variable gadgets and we only use the entrance matching the phase we are in. We use ``pass'' to refer to a path from an entry to the exit, the ``initial pass'' is the one made in the ``assignment'' phase. Our gadgets function like:
\begin{itemize}
    \item \textbf{The Control Structure.} In this structure shown in \Cref{sa:fig:pspace-rs1-control} we enforce the phases using the switching behaviour at $o$. The nodes $\mathit{as}$ and $\mathit{ag}$ cycle through the $n$ quantified variable gadgets, visiting each once in the ``assignment'' phase and $D$ times in the ``agreement'' phase. The node $\mathit{ver}$ finally starts the verification process by moving to $\mathit{ver}_1$. We note any more visits to $o$ send us to $\overline d$. We note our quantified gadgets are connected with edges between $\mathit{as}$ and all $\mathit{as}_i$ and between $\mathit{ag}$ and all $\mathit{ag}_i$, return edges from $ret_i$ to $o$ and a chain of edges going from $\mathit{ver}$ to $\mathit{ver}_1$, $\mathit{next}_1$ to $\mathit{ver}_2$,..., and finally $\mathit{next}_n$ to $d$.
    \item \textbf{Quantified Variable Gadget.} We have two variations of this gadget shown in \Cref{sa:fig:pspace-rs1-exists-quant,sa:fig:pspace-rs1-rand-quant} which depend on whether $x_i$ is existentially or randomly quantified in $\varphi$, differing only in the node type of $\mathit{as}_i$. On the initial pass, the assignment is chosen by the player or uniformly at random respectively. The three entries correspond to the different phases of the game and we have two exits, $ret_i$ returns back to the $o$ and $\mathit{next}_i$ moves us on to the next variable's verification entry $\mathit{ver}_{i+1}$, or to $d$
    if $i=n$. The nodes $x_i^T$ and $x_i^F$ represent choosing an assignment of the variable $x_i$ on this pass, and the ``initial assignment'' is the one from the initial pass. The switching behaviour of $x_i^T$ and $x_i^F$ prevents $\mathit{next}_i$ being reached without $D+2$ visits to one of the two nodes, which forces $D$ visits to the respective Consequence gadget $Neg_i$ or $Pos_i$.
    \item \textbf{Consequences Gadget.} We have two consequences gadgets for each variable, $Neg_i$ and $Pos_i$, shown in \Cref{sa:fig:pspace-rs1-pos-cons,sa:fig:pspace-rs1-neg-cons}. $Neg_i$ (resp. $Pos_i$) enumerates the gadgets for clauses, $C_{j_1},\ldots,C_{j_{b_i}}$ (resp. $C_{k_1},\ldots,C_{k_{a_i}}$), where the literal $\neg x_i$ (resp. $x_i$) appears. When we choose an assignment of true (resp. false) these clauses are not immediately satisfied by our assignment. As any literal appears in at most $D$ clauses by visiting this gadget $D$ times we are guaranteed to go through each of the contained clause gadgets. If we have $b_i<D$ (resp. $a_i<D$) then any further edges proceed straight to the exit to ensure if we make exactly $D$ passes we visit each clause gadget exactly once. 
    \item \textbf{The Clause Gadget.} This is shown in \Cref{sa:fig:pspace-rs1-clause}. Here we check if it is possible to still satisfy a clause. Note we pass through the clause gadget for $C_l$ only in the following situations:
\begin{itemize}
    \item From a $Neg_i$ gadget where we have assigned $x_i$ true on this pass and $\neg x_i$ appears in $C_l$,
    \item From a $Pos_i$ gadget where we have assigned $x_i$ false on this pass and $x_i$ appears in $C_l$,
\end{itemize}
Thus as a consequence of our truth assignment to $x_i$ it does not witness the truth of $C_l$. Our clause $C_l$ has width $3$ and if our assignment is satisfying then we must have at least one of the $3$ literals as a witness to the truth of $C_l$. Thus our gadget acts as a simple counter of the number of literals in the clause which evaluates to false, after $3$ passes the clause switch sends the play to $\overline d_l$, because the assignment we have chosen does not
satisfy $C_l$. On the first and second passes, the counter is just incremented and we use this gadget to ensure the clause is satisfied. 
\end{itemize}

In the figures, $\mathit{ret}_i$ returns from variable gadget $i$ to $o$, while $\mathit{next}_i$ advances the verification path after successfully checking that gadget. We can prove that this instance has value $\val(G^\varphi,o,d)$ satisfying \Cref{sa:eq:val-gf-ssat}.

The construction has polynomial size: the control structure (\Cref{sa:fig:pspace-rs1-control}) contains only instances of the randomly and existentially quantified variable gadgets, the quantified-variable gadgets (\Cref{sa:fig:pspace-rs1-quantifiers}) contain only the Consequence gadgets $Pos_i$ and $Neg_i$ and the consequence gadgets (\Cref{sa:fig:pspace-rs1-neg-cons,sa:fig:pspace-rs1-pos-cons}) contain only Clause Gadgets (\Cref{sa:fig:pspace-rs1-clause}). Further the $ret_i$ exits and all exits of the consequence and clause gadgets may be treated as the node $o$, independent of the index $i$ or $l$ of the gadget, as each has an onward path containing only nodes of out-degree one leading to $o$.

\begin{theorem}\label[theorem]{sa:res:rs1-quant-is-pspace-hard}
{\tt $\{R,S,1\}$-Arrival-Quant} is \PSPACE{}-hard.
\end{theorem}
\begin{proof}
\noindent\emph{Idea.} The assignment phase realises the alternating existential/random choices of SSAT. The agreement phase forces every later traversal to use the assignment chosen on the first pass. The verification phase then reaches the target exactly for satisfying assignments. We prove these three properties and finally identify the resulting value with the SSAT value.

\paragraph{1. Size and termination.}
The quantities $a_i$ and $b_i$ count the positive and negative occurrences of
variable $x_i$, respectively; hence $D=\max_i\{a_i,b_i\}\le m$.  The control
structure and quantified-variable gadgets have $O(n)$ vertices.  There are
$2n$ consequence switching orders, each of length at most $D$ with padding,
and $m$ constant-size clause gadgets.  Including all cyclic-order entries,
vertices and edges, the encoding size is $O(nD+n+m+\sum_i(a_i+b_i))$,
which is polynomial in the size of the 3CNF formula.

Every play eventually enters an absorbing outcome.  Outside $d$ and the losing
sinks, the construction has no directed cycle avoiding the control vertex
$o$.  Each return to $o$ advances its finite switching order, which sends
any play that has not already terminated to the losing sink after the final
verification entry.  In particular, the construction cannot produce an
infinite play that avoids both outcomes.

\paragraph{2. Forced consistency.}
The switching orders at $o$, $\mathit{as}$, and $\mathit{ag}$ schedule one
assignment traversal of each variable gadget, then $D$ agreement traversals
of each, and finally one verification traversal of each along the
$\mathit{next}_i$ chain.  The final visit to $o$ sends the play to the losing
sink, not back to assignment.  A traversal of variable gadget $i$ uses
exactly one of the two branch switches $x_i^T,x_i^F$.

A branch switch first returns through $\mathit{ret}_i$, then uses the
corresponding consequence gadget on its next $D$ visits, and offers
$\mathit{next}_i$ only on its $(D+2)$-nd visit.  Therefore a path reaching
the final target must traverse each variable gadget exactly $D+2$ times,
and reach $\mathit{next}_i$ for each $i$.  Since those $D+2$ visits are
needed by \emph{one} branch switch before it can reach $\mathit{next}_i$,
all of them must use the same branch, namely the one chosen during
assignment.  In particular, an inconsistent choice in a later phase makes
reaching $d$ impossible.

This is a statement about paths that can reach $d$, not about every action
of every optimal strategy.  On a history from which reaching $d$ has become
impossible, a strategy's subsequent choices do not affect the value.
Player~1 can implement the consistent continuation using the switch positions
of the earlier $x_i^T,x_i^F$ nodes as a record of the assignment.  During
assignment, choices of later existential variables can depend on earlier
random outcomes, as required by the SSAT quantifier prefix.

\paragraph{3. Clause check.}
For a consistent assignment $V$, the $D$ agreement traversals of variable
$i$ visit $\mathit{Neg}_i$ exactly $D$ times when $V(x_i)=T$, and visit
$\mathit{Pos}_i$ exactly $D$ times when $V(x_i)=F$.  These consequence
orders enumerate, without repetition, all clauses containing the respective
\emph{false} literals $\neg x_i$ and $x_i$; padding entries return to the
control structure.  Thus each clause gadget $C_l$ is visited exactly once
for each of its false literals and no other times.  Since $C_l$ has three
distinct literals, its switch sends the play to its losing sink on its
third such visit precisely when all three literals are false.

If $V$ satisfies every clause, consistent agreement traversals never
trigger a clause losing sink.  The verification chain then reaches
$\mathit{next}_1,\ldots,\mathit{next}_n$ and finally $d$.  Conversely,
if $V$ falsifies a clause, its third false-literal visit reaches a losing
sink before verification can succeed.  Together with forced consistency,
this proves that a target-reaching play exists precisely for a satisfying
assignment and only when the later choices repeat its branches.

\paragraph{4. Value correspondence.}
The preceding claims show that the outcome depends only on the choices made at the nodes $\mathit{as}_i$ during the assignment phase. The target is reached exactly when the resulting valuation $V_\pi$ satisfies $\varphi$. Hence $\val(G,o,d)$ is the maximum probability of producing a satisfying valuation. Consider the tree of partial valuations $V:[n]\rightharpoonup\{T,F\}$ that assign an initial segment of the variables. At odd levels Player~1 chooses a child; at even levels the next child is chosen uniformly at random. A leaf is winning exactly when its total valuation satisfies $\varphi$. Therefore:

\begin{equation*}
    \val(G^\varphi,o,d)=\max_{x_1}\mathbb{E}_{x_2}\max_{x_3}\cdots
    \mathbb{E}_{x_n}\bigl[\chi[\varphi(x_1,\ldots,x_n)=\top]\bigr].
\end{equation*}

Since {\tt SSAT} is \PSPACE{}-complete \cite[Theorem~2]{Pap85} and the construction is a polynomial-time reduction to {\tt $\{R,S,1\}$-Arrival-Quant}, the latter problem is \PSPACE{}-hard.
\end{proof}
\FloatBarrier

 As an immediate consequence, we can also give an analogous \PSPACE{}-hardness for the \\{\tt $\{R,S,2\}$-Arrival-Quant} problem.

\begin{theorem}\label[theorem]{sa:res:rs2-quant-is-pspace-hard}
{\tt $\{R,S,2\}$-Arrival-Quant} is \PSPACE{}-hard.
\end{theorem}
\begin{proof}
We can modify the construction of \Cref{sa:res:rs1-quant-is-pspace-hard} by making the following changes to also derive a hardness result for {\tt $\{R,S,2\}$-Arrival-Quant}, we replace Player~1 with Player~2 and exchange the nodes $d$ and $\overline d$, including in the clause gadgets. These changes are shown in \Cref{sa:fig:pspace-rs2-control,sa:fig:pspace-rs2-exists-quant,sa:fig:pspace-rs2-clause}. By the same argument above we will construct an instance $G^{\prime\varphi}$ where:
Every play of $G^\varphi$ terminates (each cycle passes through $o$, which sends the
$((D+1)n+2)$-nd visit to $\overline d$), at $d$ or at
$\overline d/\overline d_l$. In $G'^{\varphi}$ Player~1 is replaced by Player~2 and
$d$ and the $\overline d$ nodes (all merged into one) swap roles.
For every strategy $\tau$ of the replaced player,
\[
\P^{G'}_\tau(\text{reach new target})=1-\P^{G}_\tau(\text{reach target}),
\]
so
\[
\val(G'^{\varphi})=1-\val(G^\varphi)=\min_{x_1}\mathbb E_{x_2}\min_{x_3}\cdots
\mathbb E_{x_n}\bigl[\chi[\varphi(x_1,\dots,x_n)=\bot]\bigr].
\]
Thus $\val(G'^{\varphi})<\frac12$ iff $\varphi\in\textsc{SSAT}$. Hence $\textsc{SSAT}$ reduces to the complement of the non-strict quantitative problem for node set $\{R,S,2\}$. Since $\mathsf{PSPACE}=\mathsf{coPSPACE}$ \cite{Pap94}, \Cref{sa:res:rec-quant-equal-prob} then implies that {\tt $\{R,S,2\}$-Arrival-Quant} is $\mathsf{PSPACE}$-hard.
\begin{figure}
    \centering
    \resizebox{0.95\textwidth}{!}{\input{figures_tikz_work/fig12.tex}}
    \caption[Proof of \Cref{sa:res:rs2-quant-is-pspace-hard}: modified overall control structure.]{Modifications to the control structure for $\{R,S,2\}$-Arrival-Quant. Compared with \Cref{sa:fig:pspace-rs1-control}, the nodes $d$ and $\overline d$ have changed places.}
    \label{sa:fig:pspace-rs2-control}
\end{figure}
\end{proof}

\begin{figure}[htbp]
\centering
\begin{subfigure}[t]{0.66\textwidth}
    \centering
    \resizebox{\linewidth}{!}{\begin{tikzpicture}[x=0.75pt,y=0.75pt,yscale=-1,xscale=1]

\draw  [dash pattern={on 4.5pt off 4.5pt}] (22,16) -- (527,16) -- (527,335.39) -- (22,335.39) -- cycle ;
\draw [sa player one]   (15,139.5) -- (65,139.5) -- (65,189.5) -- (15,189.5) -- cycle ;
\draw [sa player one]   (16,265.5) -- (66,265.5) -- (66,315.5) -- (16,315.5) -- cycle ;
\draw [sa switch]   (131.83,94.33) .. controls (131.83,80.53) and (143.03,69.33) .. (156.83,69.33) .. controls (170.64,69.33) and (181.83,80.53) .. (181.83,94.33) .. controls (181.83,108.14) and (170.64,119.33) .. (156.83,119.33) .. controls (143.03,119.33) and (131.83,108.14) .. (131.83,94.33) -- cycle ;
\draw [sa switch]   (131.83,224) .. controls (131.83,210.19) and (143.03,199) .. (156.83,199) .. controls (170.64,199) and (181.83,210.19) .. (181.83,224) .. controls (181.83,237.81) and (170.64,249) .. (156.83,249) .. controls (143.03,249) and (131.83,237.81) .. (131.83,224) -- cycle ;
\draw  [dash pattern={on 4.5pt off 4.5pt}] (330,71) -- (400,71) -- (400,111) -- (330,111) -- cycle ;
\draw  [dash pattern={on 4.5pt off 4.5pt}] (331,197) -- (401,197) -- (401,237) -- (331,237) -- cycle ;
\draw    (65.08,81.83) -- (129.87,93.97) ;
\fill [shift={(131.83,94.33)}, rotate = 190.61] [color={rgb, 255:red, 0; green, 0; blue, 0 }  ] (0,0) -- (8.5,-3.5) -- (8.5,3.5) -- cycle ;
\draw    (65.08,81.83) -- (130.98,222.19) ;
\fill [shift={(131.83,224)}, rotate = 244.85] [color={rgb, 255:red, 0; green, 0; blue, 0 }  ] (0,0) -- (8.5,-3.5) -- (8.5,3.5) -- cycle ;
\draw    (66,164) -- (130.46,95.79) ;
\fill [shift={(131.83,94.33)}, rotate = 133.38] [color={rgb, 255:red, 0; green, 0; blue, 0 }  ] (0,0) -- (8.5,-3.5) -- (8.5,3.5) -- cycle ;
\draw    (66,164) -- (130.36,222.65) ;
\fill [shift={(131.83,224)}, rotate = 222.35] [color={rgb, 255:red, 0; green, 0; blue, 0 }  ] (0,0) -- (8.5,-3.5) -- (8.5,3.5) -- cycle ;
\draw    (65,289) -- (131.18,96.22) ;
\fill [shift={(131.83,94.33)}, rotate = 108.95] [color={rgb, 255:red, 0; green, 0; blue, 0 }  ] (0,0) -- (8.5,-3.5) -- (8.5,3.5) -- cycle ;
\draw    (65,289) -- (130.4,225.39) ;
\fill [shift={(131.83,224)}, rotate = 135.8] [color={rgb, 255:red, 0; green, 0; blue, 0 }  ] (0,0) -- (8.5,-3.5) -- (8.5,3.5) -- cycle ;
\draw    (181.83,94.33) -- (326,92.42) ;
\fill [shift={(328,92.39)}, rotate = 179.24] [color={rgb, 255:red, 0; green, 0; blue, 0 }  ] (0,0) -- (8.5,-3.5) -- (8.5,3.5) -- cycle ;
\draw    (181.83,224) -- (326,221.43) ;
\fill [shift={(328,221.39)}, rotate = 178.98] [color={rgb, 255:red, 0; green, 0; blue, 0 }  ] (0,0) -- (8.5,-3.5) -- (8.5,3.5) -- cycle ;
\draw    (181.83,94.33) .. controls (202,-11.61) and (397.5,42) .. (520.5,94) ;
\fill [shift={(520.5,94)}, rotate = 202.92] [color={rgb, 255:red, 0; green, 0; blue, 0 }  ] (0,0) -- (8.5,-3.5) -- (8.5,3.5) -- cycle ;
\draw    (181.83,224) .. controls (182.99,159.71) and (418.13,160) .. (518.99,94.99) ;
\fill [shift={(520.5,94)}, rotate = 146.58] [color={rgb, 255:red, 0; green, 0; blue, 0 }  ] (0,0) -- (8.5,-3.5) -- (8.5,3.5) -- cycle ;
\draw    (403,91) -- (518.5,93.95) ;
\fill [shift={(520.5,94)}, rotate = 181.46] [color={rgb, 255:red, 0; green, 0; blue, 0 }  ] (0,0) -- (8.5,-3.5) -- (8.5,3.5) -- cycle ;
\draw    (401,218) -- (519.11,95.44) ;
\fill [shift={(520.5,94)}, rotate = 133.94] [color={rgb, 255:red, 0; green, 0; blue, 0 }  ] (0,0) -- (8.5,-3.5) -- (8.5,3.5) -- cycle ;
\draw [sa switch]   (483,217.39) .. controls (483,203.58) and (494.19,192.39) .. (508,192.39) .. controls (521.81,192.39) and (533,203.58) .. (533,217.39) .. controls (533,231.2) and (521.81,242.39) .. (508,242.39) .. controls (494.19,242.39) and (483,231.2) .. (483,217.39) -- cycle ;
\draw  [sa switch,fill=SAfigureSwitch  ,fill opacity=1 ] (328,92.39) .. controls (328,89.35) and (330.46,86.89) .. (333.5,86.89) .. controls (336.54,86.89) and (339,89.35) .. (339,92.39) .. controls (339,95.43) and (336.54,97.89) .. (333.5,97.89) .. controls (330.46,97.89) and (328,95.43) .. (328,92.39) -- cycle ;
\draw  [sa switch,fill=SAfigureSwitch  ,fill opacity=1 ] (397.5,91) .. controls (397.5,87.96) and (399.96,85.5) .. (403,85.5) .. controls (406.04,85.5) and (408.5,87.96) .. (408.5,91) .. controls (408.5,94.04) and (406.04,96.5) .. (403,96.5) .. controls (399.96,96.5) and (397.5,94.04) .. (397.5,91) -- cycle ;
\draw  [sa switch,fill=SAfigureSwitch  ,fill opacity=1 ] (322.5,221.39) .. controls (322.5,218.35) and (324.96,215.89) .. (328,215.89) .. controls (331.04,215.89) and (333.5,218.35) .. (333.5,221.39) .. controls (333.5,224.43) and (331.04,226.89) .. (328,226.89) .. controls (324.96,226.89) and (322.5,224.43) .. (322.5,221.39) -- cycle ;
\draw  [sa switch,fill=SAfigureSwitch  ,fill opacity=1 ] (395.5,218) .. controls (395.5,214.96) and (397.96,212.5) .. (401,212.5) .. controls (404.04,212.5) and (406.5,214.96) .. (406.5,218) .. controls (406.5,221.04) and (404.04,223.5) .. (401,223.5) .. controls (397.96,223.5) and (395.5,221.04) .. (395.5,218) -- cycle ;
\draw    (181.83,224) .. controls (203,262.39) and (337,312.39) .. (483,217.39) ;
\fill [shift={(483,217.39)}, rotate = 146.95] [color={rgb, 255:red, 0; green, 0; blue, 0 }  ] (0,0) -- (8.5,-3.5) -- (8.5,3.5) -- cycle ;
\draw    (181.83,94.33) .. controls (182.99,169.02) and (424.57,161.47) .. (482.15,216.56) ;
\fill [shift={(483,217.39)}, rotate = 225] [color={rgb, 255:red, 0; green, 0; blue, 0 }  ] (0,0) -- (8.5,-3.5) -- (8.5,3.5) -- cycle ;
\draw  [sa switch,fill=SAfigureSwitch  ,fill opacity=1 ] (520.5,94) .. controls (520.5,90.96) and (522.96,88.5) .. (526,88.5) .. controls (529.04,88.5) and (531.5,90.96) .. (531.5,94) .. controls (531.5,97.04) and (529.04,99.5) .. (526,99.5) .. controls (522.96,99.5) and (520.5,97.04) .. (520.5,94) -- cycle ;
\draw [sa player two]   (40.08,31.83) -- (65.08,81.83) -- (15.08,81.83) -- cycle ;

\draw (40.08,65.16) node [sa triangle label] {$as_{i}$};
\draw (40.00,164.50) node [sa node label] {$\displaystyle ag_{i}$};
\draw (41.00,290.50) node [sa node label] {$\displaystyle ver_{i}$};
\draw (156.83,94.33) node [sa node label] {$\displaystyle x_{i}^{T}$};
\draw (156.83,224.00) node [sa node label] {$\displaystyle x_{i}^{F}$};
\draw (351,81) node [anchor=north west][inner sep=0.75pt]   [align=left] {$\displaystyle \mathit{Neg}_{i}$};
\draw (352,207) node [anchor=north west][inner sep=0.75pt]   [align=left] {$\displaystyle \mathit{Pos}_{i}$};
\draw (230.0,72.33) node [anchor=north west][inner sep=0.75pt]   [align=left] {$\displaystyle 2,\dotsc ,D+1$};
\draw (230.0,200.33) node [anchor=north west][inner sep=0.75pt]   [align=left] {$\displaystyle 2,\dotsc ,D+1$};
\draw (501.5,64.5) node [anchor=north west][inner sep=0.75pt]   [align=left] {$\displaystyle ret_{i}$};
\draw (178,47) node [anchor=north west][inner sep=0.75pt]   [align=left] {$\displaystyle 1$};
\draw (183,183) node [anchor=north west][inner sep=0.75pt]   [align=left] {$\displaystyle 1$};
\draw (508,217.39) node [sa node label] {$\displaystyle next_{i}$};
\draw (200,104) node [anchor=north west][inner sep=0.75pt]   [align=left] {$\displaystyle D+2$};
\draw (217,237) node [anchor=north west][inner sep=0.75pt]   [align=left] {$\displaystyle D+2$};

\end{tikzpicture}}
    \caption{Existential choice with Player~2 in place of Player~1.}
    \label{sa:fig:pspace-rs2-exists-quant}
\end{subfigure}\hfill
\begin{subfigure}[t]{0.31\textwidth}
    \centering
    \resizebox{\linewidth}{!}{\begin{tikzpicture}[x=0.75pt,y=0.75pt,yscale=-1,xscale=1]

\draw [sa switch]   (12.5,56.5) .. controls (12.5,42.69) and (23.69,31.5) .. (37.5,31.5) .. controls (51.31,31.5) and (62.5,42.69) .. (62.5,56.5) .. controls (62.5,70.31) and (51.31,81.5) .. (37.5,81.5) .. controls (23.69,81.5) and (12.5,70.31) .. (12.5,56.5) -- cycle ;
\draw    (62.5,56.5) -- (214.02,37.75) ;
\fill [shift={(216,37.5)}, rotate = 172.94] [color={rgb, 255:red, 0; green, 0; blue, 0 }  ] (0,0) -- (8.5,-3.5) -- (8.5,3.5) -- cycle ;
\draw    (62.5,56.5) -- (236.6,112.88) ;
\fill [shift={(238.5,113.5)}, rotate = 197.95] [color={rgb, 255:red, 0; green, 0; blue, 0 }  ] (0,0) -- (8.5,-3.5) -- (8.5,3.5) -- cycle ;
\draw  [dash pattern={on 4.5pt off 4.5pt}] (21,11.5) -- (219,11.5) -- (219,144.5) -- (21,144.5) -- cycle ;
\draw [sa switch]   (238.5,113.5) .. controls (238.5,99.69) and (249.69,88.5) .. (263.5,88.5) .. controls (277.31,88.5) and (288.5,99.69) .. (288.5,113.5) .. controls (288.5,127.31) and (277.31,138.5) .. (263.5,138.5) .. controls (249.69,138.5) and (238.5,127.31) .. (238.5,113.5) -- cycle ;
\draw  [sa switch,fill=SAfigureSwitch  ,fill opacity=1 ] (210.5,37.5) .. controls (210.5,34.46) and (212.96,32) .. (216,32) .. controls (219.04,32) and (221.5,34.46) .. (221.5,37.5) .. controls (221.5,40.54) and (219.04,43) .. (216,43) .. controls (212.96,43) and (210.5,40.54) .. (210.5,37.5) -- cycle ;

\draw (75,33.5) node [anchor=north west][inner sep=0.75pt]   [align=left] {$\displaystyle 1,2$};
\draw (66.5,66) node [anchor=north west][inner sep=0.75pt]   [align=left] {$\displaystyle 3$};
\draw (263.50,113.50) node [sa node label] {$d$};
\draw (37.50,56.50) node [sa node label] {$\displaystyle c_{l}$};

\end{tikzpicture}}
    \caption{The clause exit to $d$ replaces the previous losing exit.}
    \label{sa:fig:pspace-rs2-clause}
\end{subfigure}
\caption{Further changes for $\{R,S,2\}$-Arrival-Quant relative to the $\{R,S,1\}$ construction.}
\end{figure}

\section{The {\tt $\{R,S\}$-Arrival} Qualitative Problems}\label{sa:sec:RSarrivalQual}

We first establish two general facts about qualitative reachability in finite stochastic games. These let us
separate the positive-probability and almost-sure questions from the switching representation: after expanding
the switching memory, the resulting game is finite, although exponentially larger than the input.

\begin{lemma}[Positive reachability]\label[lemma]{sa:lem:positive-reachability}
Let $\Gamma$ be a finite stochastic game with absorbing target $T$. Let $A$ be the attractor of $T$ for the
coalition of Player~1 and random states: $T\subseteq A$; a Player~1 or random state belongs to $A$ if it has a successor
in $A$; and a Player~2 state belongs to $A$ if all its successors belong to $A$. Then $\val_\Gamma(s)>0$ iff $s\in A$.
\end{lemma}
\begin{proof}
Let $A_0:=T$ and let $A_{i+1}$ be obtained from $A_i$ by one application of the predecessor rules defining $A$. For $s\in A$, define its \emph{attractor rank} to be the least $i$ with $s\in A_i$. Player~1 moves to a lower-rank state. From every
random state there is a positive-probability transition to lower rank, and Player~2 cannot prevent the rank from
decreasing. Hence the target is reached with positive probability. Conversely, outside $A$, Player~2 always has
a successor outside $A$, while Player~1 and random states have no successor in $A$. Player~2 can therefore keep the play
outside $A$, so the target is reached with probability zero.
\end{proof}

\begin{lemma}[Almost-sure reachability]\label[lemma]{sa:lem:almost-sure-reachability}
For a finite stochastic game with absorbing target $T$, the set $\mathrm{Val}_1$ of states of value $1$
is computable in time polynomial in $|S|$ (plus the size of the probabilities).
\end{lemma}
\begin{proof}
For $Y\subseteq S$ let $F(Y)$ be the least set $X$ such that $X\supseteq T\cap Y$ and, for
$s\in Y\setminus T$: $s\in X$ if
(Player~1) some successor of $s$ lies in $X$; (Player~2) all successors lie in $X$;
(random) all successors lie in $Y$ and some successor lies in $X$.
$F$ is monotone. Put $Y_0=S$, $Y_{i+1}=F(Y_i)$; the sequence decreases and stabilises at
$Y^*=F(Y^*)$ after at most $|S|$ steps, each polynomial. We claim $Y^*=\mathrm{Val}_1$.

\emph{$Y^*\subseteq\mathrm{Val}_1$.}
Since $Y^*=F(Y^*)$, each $s\in Y^*\setminus T$ has a \emph{fixpoint rank}, namely the stage at which it enters the least fixpoint.
Player~1 moves to a successor of lower rank (it lies in $Y^*$). Player~2 states in $Y^*$ have all successors
in $X\subseteq Y^*$, and random states in $Y^*$ have all successors in $Y^*$. Thus the play never
leaves $Y^*$, and from any state of rank $r\le |S|$ the rank decreases at every step with probability
at least $p_{\min}$, whatever Player~2 does. Hence $T$ is reached within $|S|$ steps with probability at
least $\delta=p_{\min}^{|S|}$ from every state of $Y^*$; repeating, $T$ is reached almost surely.

\emph{$\mathrm{Val}_1\subseteq Y^*$.}
We use the optimality equations for the value function:
$\val=1$ on $T$; $\val(s)=\max_{s'}\val(s')$ at Player~1 states;
$\val(s)=\min_{s'}\val(s')$ at Player~2 states; and
$\val(s)=\sum_{s'}P(s,s')\val(s')$ at random states. Hence a Player~2 or random state of value $1$ has all
successors of value $1$ (random probabilities are positive), and a Player~1 state of value $1$ has a successor
of value $1$. Let $\tau^*$ be the positional Player~2 strategy that picks, at each Player~2 state, a successor of
minimum value. Under $\tau^*$ and any Player~1 strategy, $\val(Z_t)$ is a supermartingale: at Player~1 states
$\val(s)\ge\val(s')$ for every successor $s'$; at Player~2 states equality holds for the chosen successor;
at random states equality holds in expectation.

We prove $\mathrm{Val}_1\subseteq Y_i$ for all $i$ by induction. The base case $i=0$ holds because $Y_0=S$. Assume
$\mathrm{Val}_1\subseteq Y_i$, put $X=F(Y_i)$ and suppose
$C:=\mathrm{Val}_1\setminus X\neq\emptyset$. As $T\subseteq X$ we have $C\cap T=\emptyset$, so every
$s\in C$ lies in $Y_i\setminus T$, and the closure rules defining $F(Y_i)$ apply:
\begin{enumerate}
\item[(a)] a Player~1 state $s\in C$ has no successor in $X$ (otherwise $s\in X$); so each successor lies in $C$ or outside $\mathrm{Val}_1$;
\item[(b)] a Player~2 state $s\in C$ has a successor $s'\notin X$; since $s$ has value $1$, $s'\in\mathrm{Val}_1$, hence $s'\in C$;
\item[(c)] a random state $s\in C$ has all successors in $\mathrm{Val}_1\subseteq Y_i$; none is in $X$
(otherwise $s\in X$); so all lie in $C$.
\end{enumerate}
Let $\tau$ be the Player~2 strategy that, at Player~2 states of $C$, moves to a successor in $C$ (possible by (b))
and plays $\tau^*$ elsewhere. Let
$\eta=\min_{u\notin\mathrm{Val}_1}(1-\val(u))\in(0,1]$ (take $\eta=1$ if
$\mathrm{Val}_1=S$) and define $f(s)=1-\eta$ for $s\in C$ and $f(s)=\val(s)$ for $s\notin C$.
Then $0\le f\le\val$ and $f=1$ on $T$. We check that $f(Z_t)$ is a supermartingale under $\tau$
and every Player~1 strategy:
\begin{itemize}
\item $s\in C$ Player~1: by (a) each successor is in $C$ ($f=1-\eta$) or outside $\mathrm{Val}_1$
(where $f=\val\le1-\eta$), so the bound is $1-\eta=f(s)$.
\item $s\in C$ Player~2: $\tau$ moves into $C$, so $f=1-\eta=f(s)$. 
$s\in C$ random: all successors lie in $C$ by (c), so $f=1-\eta=f(s)$.
\item $s\notin C\cup T$: here $\tau$ plays $\tau^*$, so
$\mathbb E[f(\text{next})]\le\mathbb E[\val(\text{next})]\le\val(s)=f(s)$ by the supermartingale property of $\val$.
\item $s\in T$: absorbing, $f=1$.
\end{itemize}
Let $\theta$ be the hitting time of $T$. For $s\in C$ and any Player~1 strategy $\sigma$, since $f\ge0$ and $f=1$ on $T$,
$\mathbb P_s^{\sigma,\tau}(\theta\le t)\le\mathbb E[f(Z_{t\wedge\theta})]\le f(s)=1-\eta$.
Letting $t\to\infty$, $\mathbb P_s^{\sigma,\tau}(\Reach\,T)\le1-\eta<1$ for all $\sigma$,
so $\val(s)<1$, contradicting $s\in C\subseteq\mathrm{Val}_1$. Hence
$\mathrm{Val}_1\subseteq F(Y_i)=Y_{i+1}$, and in the limit $\mathrm{Val}_1\subseteq Y^*$.
\end{proof}

We now apply these facts to the qualitative problems for stochastic arrival. The positive-probability question
is closely related to the corresponding deterministic switching game, whereas the almost-sure question can be
strictly harder once player control is present.

\begin{theorem}\label[theorem]{sa:res:rs-qual-1-is-pspace-hard}
The {\tt $\{R,S,1\}$-Arrival-Qual-$1$} problem is $\PSPACE{}$-hard. Consequently,
{\tt $\{R,S,1,2\}$-Arrival-Qual-$1$} is $\PSPACE{}$-hard.
\end{theorem}
\begin{proof}
Reduce from the PSPACE-complete restriction of QBF whose quantifier-free part is a 3CNF formula \cite{Pap94}. By inserting dummy quantified variables we may assume without loss of generality that the prefix is strictly alternating and that $n$ is even. Thus let
$\Psi=\exists x_1\forall x_2\cdots\exists x_{n-1}\forall x_n\,\varphi$, with $\varphi$ in 3CNF (see, e.g., \cite{Pap94}), and
use the construction from Theorem~\ref{sa:res:rs1-quant-is-pspace-hard}, interpreting each universal
quantifier as a random choice. The construction gives
\[
\val(G^\varphi)=\max_{x_1}\mathbb E_{x_2}\max_{x_3}\cdots\mathbb E_{x_n}
\chi[\varphi(x_1,\ldots,x_n)=\top].
\]
Evaluating this expression on the binary tree of partial assignments gives value $1$ exactly when the
corresponding QBF is true: Player~1 nodes take the maximum of their two children, while a random node has value
$1$ exactly when both children have value $1$. Hence QBF reduces to $\{R,S,1\}$-Arrival-Qual-$1$.
\end{proof}

\begin{proposition}\label[proposition]{sa:res:all-qual-in-exptime}
For every $\mathcal B\subseteq\{R,S,1,2\}$, both $\mathcal B$-Arrival-Qual-$0$ and
$\mathcal B$-Arrival-Qual-$1$ are in $\EXPTIME{}$.
\end{proposition}
\begin{proof}
Expand $G$ to $\mathrm{Exp}(G)$. In \emph{normal form} this has at most
$N\leq |V|\prod_{v\in V_S}|\Ord(v)|=2^{O(|G|)}$ states. By the positive-reachability lemma, Qual-$0$
is decidable by an attractor computation on $\mathrm{Exp}(G)$. By the almost-sure reachability lemma, Qual-$1$ is
decidable by the corresponding fixed-point computation. Both are polynomial in $N$, hence exponential in the
input size.
\end{proof}

We first classify the qualitative problems for $\{R,S\}$-Arrival. We then bound the expected number of traversals of each edge, which yields an exponential expected-termination bound. 

The following pair of reductions respectively gives Player~1 control of random nodes and randomises over Player~1 choices. The same positive-reachability principle is standard for finite MDPs and simple stochastic games; see, for example, \cite{Der70,Con92}. 

\begin{lemma}\label[lemma]{sa:res:rand-to-player}
Suppose $R\in\mathcal B$ and let $\mathcal B'=(\mathcal B\setminus\{R\})\cup\{1\}$. Then
$\mathcal B$-Arrival-Qual-$0$ is polynomial-time reducible to $\mathcal B'$-Arrival.
\end{lemma}
\begin{proof}
Give Player~1 control of all random nodes, preserving exactly those edges having positive probability.
By the positive-reachability attractor characterisation, the original value is positive exactly when the
start state belongs to the attractor for Player~1 together with random states. This is precisely the winning region
for Player~1 after the replacement, so the reduction is polynomial.
\end{proof}

\begin{lemma}\label[lemma]{sa:res:player-to-rand}
Suppose $1\in\mathcal B$ and $R\notin\mathcal B$, and let $\mathcal B'=(\mathcal B\setminus\{1\})\cup\{R\}$. Then
$\mathcal B$-Arrival is polynomial-time reducible to $\mathcal B'$-Arrival-Qual-$0$.
\end{lemma}
\begin{proof}
Replace each player-1 node by a uniformly random node over its outgoing edges, leaving all other node types unchanged. Consider the expanded finite games. In the randomised game, by \Cref{sa:lem:positive-reachability}, positive value is equivalent to membership of the target attractor in which random states use the existential predecessor rule and player-2 states use the universal predecessor rule. These random states are exactly the player-1 states of the original deterministic switching game, and every outgoing edge has positive probability. Hence this attractor is exactly the winning region of player~1 in the original expanded reachability game, including when player~2 is present. Therefore player~1 wins the original instance if and only if the randomised instance has positive value. The construction is polynomial.
\end{proof}

These reductions yield polynomial-time equivalences between several qualitative problems and the corresponding reachability switching games. The first of these gives us \NP{}-completeness for two qualitative problems as follows.

\begin{theorem}\label[theorem]{sa:res:rs(1)-Qual-0-NPc}
The three problems {\tt $\{R,S\}$-Arrival-Qual-$0$}, {\tt $\{S,1\}$-Arrival}, and {\tt $\{R,S,1\}$-Arrival-Qual-$0$} are polynomial-time equivalent and \NP{}-complete. 
\end{theorem}
\begin{proof}
By the results in \cite{FGMS21} we know that the {\tt $\{S,1\}$-Arrival} problem is \NP{}-complete, we will then show the other two problems are equivalent to this.
Firstly, {\tt $\{R,S\}$-Arrival-Qual-$0$} is polynomial-time reducible to {\tt $\{S,1\}$-Arrival} by \Cref{sa:res:rand-to-player} and, for the reverse, {\tt $\{S,1\}$-Arrival} is polynomial-time reducible to {\tt $\{R,S\}$-Arrival-Qual-$0$} by \Cref{sa:res:player-to-rand}. 
Similarly, {\tt $\{R,S,1\}$-Arrival-Qual-$0$} is polynomial-time reducible to {\tt $\{S,1\}$-Arrival} by \Cref{sa:res:rand-to-player} and the reverse reduction follows by containment. 
Thus all three problems are \NP{}-complete, since they are polynomial-time equivalent to the \NP{}-complete {\tt $\{S,1\}$-Arrival} problem.
\end{proof}

In the second case, we are not able to deduce a completeness result however our polynomial-time equivalence places both qualitative problems as being both \PSPACE{}-hard and in \EXPTIME{} using the results of \cite{FGMS21} on {\tt $\{S,1,2\}$-Arrival}. The result also opens up potentially new approaches to determine the exact complexity of {\tt $\{S,1,2\}$-Arrival} using these equivalent formulations.

\begin{theorem}\label[theorem]{sa:res:rs(1)2-qual-0-equiv-to-s12}
The three problems {\tt $\{R,S,1,2\}$-Arrival-Qual-$0$}, {\tt $\{S,1,2\}$-Arrival}, and {\tt $\{R,S,2\}$-Arrival-Qual-$0$} are polynomial-time equivalent.
\end{theorem}
\begin{proof}
By \Cref{sa:res:rand-to-player}, replacing random nodes by Player~1 nodes reduces each qualitative problem to {\tt $\{S,1,2\}$-Arrival}. Conversely, \Cref{sa:res:player-to-rand} reduces {\tt $\{S,1,2\}$-Arrival} to {\tt $\{R,S,2\}$-Arrival-Qual-$0$}, and containment gives the remaining reverse reduction. Thus all three problems are polynomial-time equivalent.
\end{proof}

The corresponding exchange between random and Player~2 vertices is subtler at threshold~$1$. A random cycle with a positive-probability exit is left almost surely, whereas Player~2 could choose to remain on it forever. We first compress the components responsible for this difference. We use $\mathrm{Succ}(v)$ for the successors as defined in Section~2.

\begin{definition}\label{sa:def:player2-end-components}
A nonempty set $C\subseteq V\setminus\{d,\overline d\}$ is a \emph{liberal end component} if $G[C]$ is strongly connected, every switching vertex in $C$ has all successors in $C$, and every random or Player~2 vertex in $C$ has at least one successor in $C$. It is a \emph{genuine end component} if $G[C]$ is strongly connected, every switching or random vertex in $C$ has all successors in $C$, and every Player~2 vertex in $C$ has at least one successor in $C$.
\end{definition}

Maximal components of either kind are pairwise disjoint: two intersecting components of the same kind have a union satisfying the same conditions. They can be found in polynomial time by repeatedly computing strongly connected components and deleting vertices violating the relevant successor conditions, until no further deletion is possible. We use this \emph{component-decomposition property} below.

\begin{lemma}[Exits from an escapable component]\label[lemma]{sa:lem:player2-component-compression}
Let $C$ be a liberal end component containing no genuine end component. Then, from any expanded state over $C$: (i) under every deterministic state-based Player~2 strategy the play leaves $C$ almost surely; and (ii) every exit edge of $C$ can be the first exit with positive probability under a suitable Player~2 strategy.
\end{lemma}
\begin{proof}
First we show that, from any expanded state over $C$, every vertex $u\in C$ can be reached by a finite partial play staying inside $C$, with suitable Player~2 choices and positive-probability random outcomes. Otherwise consider a bottom strongly connected component $H$ of internally reachable expanded states. Its projection $D$ onto $V$ omits $u$. Bottomness closes $D$ under all internal random and Player~2 successors. If $(v,q)\in H$ is a switching state, a closed walk returning to $(v,q)$ visits $v$ a positive multiple of its switching period and therefore traverses every switching successor. Since $C$ is liberal, all such successors lie in $C$, hence in $D$. Thus no edge of $G[C]$ leaves $D$, contrary to strong connectivity of $G[C]$.

For (i), fix a deterministic state-based Player~2 strategy. If there were positive probability of staying in $C$ forever, the resulting finite expanded Markov chain would have a reachable recurrent class contained in $C$. Its projection is strongly connected; recurrence closes it under random successors, and the same switching-period argument closes it under switching successors. The chosen Player~2 moves provide an internal successor at each Player~2 vertex. This would give a genuine end component, a contradiction. For (ii), use the internal-support observation to reach the source of the desired exit. Its source cannot be switching, since every switching successor belongs to $C$. At a random vertex the exit has positive probability; at a Player~2 vertex choose that edge. The entire finite witness has positive probability.
\end{proof}

\begin{lemma}[Random vertices to Player~2 after compression]\label[lemma]{sa:lem:qual1-random-to-player2}
There is a polynomial-time many-one reduction
\[
\{R,S,2\}\text{-Arrival-Qual-1}\ \leq_m^p\ \{S,2\}\text{-Arrival}.
\]
\end{lemma}
\begin{proof}
First contract every maximal genuine end component disjoint from $d$ to the absorbing losing sink $\overline d$. These are actual traps: Player~2 can keep the play away from $d$ on entering one, regardless of the current switching positions. In the remaining graph $G_0$, contract every maximal liberal end component $C$ to a Player~2 vertex $c_C$, with one option for each exit of $C$. Finally replace all remaining random vertices by Player~2 vertices, retaining precisely the positive-probability edges. Call the resulting $\{S,2\}$-instance $\mathsf{Adv}(G)$.

\emph{The quotient introduces no new trap.}
We claim that $\mathsf{Adv}(G)$ has no target-free end component other than $\overline d$. Suppose $E$ is such a component, let $\pi:V(G_0)\to V(\mathsf{Adv}(G))$ be the contraction map, and set $U=\pi^{-1}(E)$. Quotient paths through a macro-vertex lift to paths inside its component: internal strong connectivity supplies paths between any entry and exit source. Thus $G_0[U]$ is strongly connected. Every switching vertex in $U$ has all successors in $U$, while random and original Player~2 vertices have an internal successor, either by liberal closure inside a contracted component or by the quotient end-component condition. Hence $U$ is a liberal end component of $G_0$.

If $E$ contains an uncontracted vertex, that vertex belongs to a maximal liberal component and should already have been contracted. If $E$ contains at least two contracted vertices, $U$ strictly contains one of the maximal liberal components, also impossible. A singleton $E=\{c_C\}$ cannot be an end component because every outgoing edge of $c_C$ represents an exit from $C$, not a self-loop. Therefore no such $E$ exists, and every play in the finite expanded game $\mathsf{Adv}(G)$ reaches $d$ or $\overline d$.

\emph{Correctness.}
The original value is below one exactly when Player~2 can arrange a positive probability of avoiding $d$. By positional determinacy for finite reachability games, we may fix a deterministic state-based strategy on the expanded state space. With positive probability an avoiding run then enters a target-free recurrent class. Its projection is a genuine end component, reached by a finite positive-probability prefix, so a corresponding quotient play reaches $\overline d$.

Conversely, suppose Player~2 can reach $\overline d$ in $\mathsf{Adv}(G)$. Lift the finite quotient play through each contracted liberal component using \Cref{sa:lem:player2-component-compression}. Every lifted finite path has positive probability. After concatenation, delete loops between repeated \emph{expanded} states. This preserves a valid path to a genuine component (or the original losing sink) and ensures that no expanded Player~2 state is assigned conflicting choices. The choices extend to one deterministic state-based strategy; the resulting positive-probability event avoids $d$. Hence the original value is below one. We have shown $\val(G)=1$ precisely when Player~2 cannot reach $\overline d$ in $\mathsf{Adv}(G)$.
\end{proof}

\begin{lemma}[Player~2 to randomness after trap removal]\label[lemma]{sa:lem:qual1-player2-to-random}
There is a polynomial-time many-one reduction
\[
\{S,2\}\text{-Arrival}\ \leq_m^p\ \{R,S\}\text{-Arrival-Qual-1}.
\]
\end{lemma}
\begin{proof}
Given a $\{S,2\}$ instance, contract every maximal target-free end component to an absorbing losing sink $\overline d$. Here the component conditions amount to strong connectivity, closure under switching successors, and an internal successor at each Player~2 vertex. Player~2 can avoid $d$ indefinitely in the original game exactly when a finite play can reach one of these components: any infinite avoiding play in the finite expanded game has a recurrent cycle whose projection is a target-free end component; conversely Player~2 can stay inside such a component.

The quotient has no further target-free end component: the preimage of one would be a target-free end component of the original graph by the same path-lifting and successor-closure argument as above. It would contain an uncontracted vertex that should have been contracted, strictly enlarge an already maximal component, or be a singleton macro-vertex without a self-loop. Each is impossible. Consequently every play in the quotient's expanded game reaches $d$ or $\overline d$ in finite time.

Replace each remaining Player~2 vertex by a random vertex assigning positive probability to every outgoing edge, leaving switching orders unchanged. Each finite Player~2 choice sequence now occurs with positive probability. Thus the randomised instance reaches $\overline d$ with positive probability exactly when Player~2 could avoid $d$ in the original game. Because the randomised process reaches one of the two sinks almost surely, its value is one exactly when the original $\{S,2\}$ instance is winning. This gives the reduction.
\end{proof}

\begin{theorem}[Player~2 compression equivalence]\label[theorem]{sa:res:player2-qual1-equivalence}
The following three problems are polynomial-time equivalent and $\coNP{}$-complete:
\begin{center}
{\tt $\{R,S\}$-Arrival-Qual-1},\qquad {\tt $\{S,2\}$-Arrival},\\
{\tt $\{R,S,2\}$-Arrival-Qual-1}.
\end{center}
\end{theorem}
\begin{proof}
\Cref{sa:lem:qual1-random-to-player2} reduces $\{R,S,2\}$-Arrival-Qual-1 to $\{S,2\}$-Arrival and hence also reduces its $\{R,S\}$ subfamily. Conversely, \Cref{sa:lem:qual1-player2-to-random} reduces $\{S,2\}$-Arrival to $\{R,S\}$-Arrival-Qual-1, while inclusion of node types reduces that subfamily to $\{R,S,2\}$-Arrival-Qual-1. Thus all three problems are polynomial-time equivalent, and $\coNP{}$-completeness follows from \Cref{sa:res:S2-is-coNP-com}.
\end{proof}

The compression is specific to almost-sure reachability: it does not justify exchanging random and Player~2 vertices in the positive-reachability reductions.

We next establish an absorption-time bound for polynomially processed $\{R,S\}$ instances.

\begin{definition}[{cf. \cite[Definition 3]{DGKMW16}}]\label{sa:def:desperation}
Let $(V, E, o, d, \{V_S,V_R\},\Prob,\Ord)$ be an instance of generalised $\{R,S\}$-Arrival. If $(v,w)\in E$ is $d$-hopeful (\Cref{sa:def:hopeful}) we call its {\em $d$-desperation} the length of the shortest directed path from $w$ to $d$. This shortest-path distance is distinct from the attractor ranks appearing in the preceding qualitative proofs.
\end{definition}

 We proceed to generalise a lemma of \cite{DGKMW16} to the randomised setting. For the remainder of this termination argument we first put the instance in \emph{normal form}, and $n$ and $m$ denote the numbers of vertices and edges of this transformed instance. In particular, every non-absorbing random node with two successors chooses each with probability $1/2$, and every non-absorbing switching node alternates between its two successors. We then compute the set $H$ of $d$-hopeful vertices. If the start is outside $H$, the value is already $0$ and there is nothing to prove. Otherwise we restrict to $H$ together with a fresh absorbing losing sink $\overline d$, redirecting every edge (and every switching-order entry) from $H$ to $V\setminus H$ into $\overline d$. This preserves the probability of reaching $d$ and makes $\overline d$ the only reachable $d$-dead vertex. Since the normalisation to \emph{normal form} is polynomial, an exponential bound in this transformed $n$ is still exponential in the original input size. 

\begin{definition}\label{sa:def:rv-num-traversals}
Let $(G,o,d)$ be a processed instance in \emph{normal form} as above and $e\in E$ an edge. Define the random variable $T_e$ to be the number of traversals of $e$ before the first visit to either absorbing sink $d$ or $\overline d$.
\end{definition}

\begin{lemma}\label[lemma]{sa:lem:rs-hopeful-edge-bound}
Let $e$ be a hopeful edge of desperation $k$ in a processed $d$-hopeful instance in \emph{normal form} $\{R,S\}$-instance, where $k\le n-1$. From every initial expanded state $(v,q)$,
$\mathbb E[T_e]\le2^{k+1}-1$.
\end{lemma}
\begin{proof}
Induct on $k$. For $k=0$, $e$ enters $d$, so $T_e\le1$. Let $e=(v,w)$ have desperation $k>0$. Then $w\ne d$ is non-absorbing. Each traversal of $e$ is followed by a traversal of an out-edge of $w$. Let $f=(w,s)$ be a hopeful out-edge of desperation $k-1$, and let $f'$ be the other out-edge if there is one. Then $T_e\le T_f+T_{f'}$ (or $T_e\le T_f$ if $w$ has only one out-edge), and by induction $\mathbb E[T_f]\le2^k-1$.

If $w\in V_R$, let $Z_t$ be the expanded state at time $t$ and let $\theta$ be the absorption time. The fair choice made on leaving $w$ is independent of the event $\{Z_t=(w,q),t<\theta\}$ for each possible switch position $q$. Summing over $q$ and $t$, and using monotone convergence, gives
\[
 \mathbb E[T_f]=\frac12\,\mathbb E[\#\hbox{ visits to }w\hbox{ before }\theta]
 =\mathbb E[T_{f'}],
\]
with the identity valid also in $[0,\infty]$, so no finiteness assumption is being used. If $w\in V_S$, the two out-edges alternate from either initial switch position, hence $|T_f-T_{f'}|\le1$ on every run. Therefore in both cases
$\mathbb E[T_{f'}]\le\mathbb E[T_f]+1\le2^k$, and
\[
 \mathbb E[T_e]\le(2^k-1)+2^k=2^{k+1}-1.
\]
\end{proof}

Consequently $\mathbb E[T_e]\le2^n-1$ for every hopeful edge.

\begin{proposition}\label[proposition]{sa:res:rs-arrival-hopeful-terminates}
Let $(G,o,d)$ be a processed $d$-hopeful instance in \emph{normal form} $\{R,S\}$-instance with $n$ vertices and $m$ edges, and let $L$ be the
number of steps until absorption in $\{d,\overline d\}$. From every initial expanded state,
$\mathbb E[L]\le m2^n$.
\end{proposition}
\begin{proof}
Each step traverses an edge. Edges entering $\overline d$ are traversed at most once, so
$L\le1+\sum_eT_e$ over hopeful edges. By the lemma, each term has expectation at most $2^n-1$,
and hence $\mathbb E[L]\le1+m(2^n-1)\le m2^n$.
\end{proof}

\section{{\tt $\{R,S\}$-Arrival-Quant} is \PP{}-hard}
After considering several qualitative problems we now turn our attention to a particular quantitative one, {\tt $\{R,S\}$-Arrival-Quant}. Our qualitative classifications in \Cref{sa:res:rs(1)-Qual-0-NPc,sa:res:player2-qual1-equivalence} already imply the following hardness result.

\begin{corollary}\label[corollary]{sa:res:rs-quant-is-NP-coNP-hard}
The {\tt $\{R,S\}$-Arrival-Quant} problem is \NP{}-hard \& \coNP{}-hard, under many-one (Karp) reductions.
\end{corollary}
\begin{proof}
We begin by showing \NP{}-hardness, through reduction from the {\tt $\{R,S\}$-Arrival-Qual-$0$}, which is \NP{}-complete by \Cref{sa:res:rs(1)-Qual-0-NPc}. Considering a generalised instance $(G,o,d)$ of $\{R,S\}$-Arrival, where $G:=(V, E, \{V_R,V_S\},\Prob,\Ord)$, we construct a new instance where we add a new start vertex $o^\prime$ to $G$ as follows; let $G^\prime:=(V\cup\{o^\prime\}, E\cup\{(o^\prime,d),(o^\prime,o)\}, \{V_R\cup\{o^\prime\},V_S\},\Prob^\prime,\Ord)$ where the new start transitions to either the original start $o$ or the target $d$ uniformly at random. This is shown in \Cref{sa:fig:rs-np-hard}. The new random start gives $\val(G^\prime,o^\prime,d)=\frac{1}{2}(1+\val(G,o,d))$, thus is strictly greater than a half if and only if we had $\val(G,o,d)>0$. Thus we have a many-one reduction from a \NP{}-complete problem.

\begin{figure}
\centering
\begin{subfigure}[t]{0.49\textwidth}
  \centering
  \resizebox{\linewidth}{!}{\begin{tikzpicture}[x=0.75pt,y=0.75pt,yscale=-1,xscale=1]

\draw [sa switch]   (252.67,34.4) .. controls (252.67,26.12) and (259.38,19.4) .. (267.67,19.4) .. controls (275.95,19.4) and (282.67,26.12) .. (282.67,34.4) .. controls (282.67,42.68) and (275.95,49.4) .. (267.67,49.4) .. controls (259.38,49.4) and (252.67,42.68) .. (252.67,34.4) -- cycle ;
\draw [sa switch]   (85.67,72.4) .. controls (85.67,64.12) and (92.38,57.4) .. (100.67,57.4) .. controls (108.95,57.4) and (115.67,64.12) .. (115.67,72.4) .. controls (115.67,80.68) and (108.95,87.4) .. (100.67,87.4) .. controls (92.38,87.4) and (85.67,80.68) .. (85.67,72.4) -- cycle ;
\draw    (36,70.17) .. controls (48.81,79.03) and (56.76,94.69) .. (84.39,73.4) ;
\fill [shift={(85.67,72.4)}, rotate = 141.54] [color={rgb, 255:red, 0; green, 0; blue, 0 }  ] (0,0) -- (8.5,-3.5) -- (8.5,3.5) -- cycle ;
\draw  [dash pattern={on 4.5pt off 4.5pt}] (69,7.5) -- (306.4,7.5) -- (306.4,97) -- (69,97) -- cycle ;
\draw [sa random]   (26.35,36.3) -- (46.6,56.68) -- (26.35,77.05) -- (6.1,56.68) -- cycle ;
\draw    (38,47.17) .. controls (42.98,9.36) and (166.75,8.51) .. (251.39,34.01) ;
\fill [shift={(252.67,34.4)}, rotate = 197.01] [color={rgb, 255:red, 0; green, 0; blue, 0 }  ] (0,0) -- (8.5,-3.5) -- (8.5,3.5) -- cycle ;

\draw (267.67,34.40) node [sa node label] {$\displaystyle d$};
\draw (100.67,72.40) node [sa node label] {$\displaystyle o$};
\draw (73.4,26.2) node [anchor=north west][inner sep=0.75pt]  [font=\LARGE] [align=left] {G};
\draw (26.35,56.67) node [sa node label] {$\displaystyle o^{\prime }$};

\end{tikzpicture}}
  \caption{Reduction from {\tt $\{R,S\}$-Arrival-Qual-$0$}.}
  \label{sa:fig:rs-np-hard}
\end{subfigure}\hfill
\begin{subfigure}[t]{0.49\textwidth}
  \centering
  \resizebox{\linewidth}{!}{\begin{tikzpicture}[x=0.75pt,y=0.75pt,yscale=-1,xscale=1]

\draw [sa switch]   (206.67,75.4) .. controls (206.67,67.12) and (213.38,60.4) .. (221.67,60.4) .. controls (229.95,60.4) and (236.67,67.12) .. (236.67,75.4) .. controls (236.67,83.68) and (229.95,90.4) .. (221.67,90.4) .. controls (213.38,90.4) and (206.67,83.68) .. (206.67,75.4) -- cycle ;
\draw [sa switch]   (85.67,72.4) .. controls (85.67,64.12) and (92.38,57.4) .. (100.67,57.4) .. controls (108.95,57.4) and (115.67,64.12) .. (115.67,72.4) .. controls (115.67,80.68) and (108.95,87.4) .. (100.67,87.4) .. controls (92.38,87.4) and (85.67,80.68) .. (85.67,72.4) -- cycle ;
\draw    (36,70.17) .. controls (48.81,79.03) and (56.76,94.69) .. (84.39,73.4) ;
\fill [shift={(85.67,72.4)}, rotate = 141.54] [color={rgb, 255:red, 0; green, 0; blue, 0 }  ] (0,0) -- (8.5,-3.5) -- (8.5,3.5) -- cycle ;
\draw  [dash pattern={on 4.5pt off 4.5pt}] (69,7.5) -- (306.4,7.5) -- (306.4,97) -- (69,97) -- cycle ;
\draw [sa random]   (26.35,36.3) -- (46.6,56.68) -- (26.35,77.05) -- (6.1,56.68) -- cycle ;
\draw    (38,47.17) .. controls (42.95,9.55) and (217.46,10.15) .. (221.59,58.91) ;
\fill [shift={(221.67,60.4)}, rotate = 269.24] [color={rgb, 255:red, 0; green, 0; blue, 0 }  ] (0,0) -- (8.5,-3.5) -- (8.5,3.5) -- cycle ;

\draw (221.67,75.40) node [sa node label] {$\displaystyle \overline{d}$};
\draw (100.67,72.40) node [sa node label] {$\displaystyle o$};
\draw (73.4,26.2) node [anchor=north west][inner sep=0.75pt]  [font=\LARGE] [align=left] {G};
\draw (26.35,56.67) node [sa node label] {$\displaystyle o^{\prime }$};

\end{tikzpicture}}
  \caption{Reduction from {\tt $\{R,S\}$-Arrival-Qual-$1$}.}
  \label{sa:fig:rs-conp-hard}
\end{subfigure}
\caption{Many-one reductions to {\tt $\{R,S\}$-Arrival-Quant}.}
\label{sa:fig:rs-qual-to-quant}
\end{figure}

For \coNP{}-hardness we know by \Cref{sa:res:player2-qual1-equivalence} that {\tt $\{R,S\}$-Arrival-Qual-$1$} is \coNP{}-complete. Starting from $G=(V,E,\{V_R,V_S\},\Prob,\Ord)$, add a new random start $o^\prime$ and a fresh absorbing losing sink $\overline d$. Formally, take $V^\prime=V\cup\{o^\prime,\overline d\}$, add the edges $(o^\prime,o)$, $(o^\prime,\overline d)$ and $(\overline d,\overline d)$ to $E$, and take $V_R^\prime=V_R\cup\{o^\prime\}$ and $V_S^\prime=V_S\cup\{\overline d\}$, with switching order $\Ord(\overline d)=(\overline d)$. The new random start chooses $o$ or $\overline d$ with probability $\frac12$ each. This is shown in \Cref{sa:fig:rs-conp-hard}. The new random start gives $\val(G^\prime,o^\prime,d)=\frac{1}{2}\val(G,o,d)$, thus is greater than or equal to a half if and only if we had $\val(G,o,d)=1$. Hence we have a many-one reduction from a \coNP{}-complete problem to the {\tt $\{R,S\}$-Arrival-Quant-Eq} problem, which is polynomial time equivalent to {\tt $\{R,S\}$-Arrival-Quant} by \Cref{sa:res:rec-quant-equal-prob}.

\end{proof}

However, we can show a stronger $\PP{}$-hardness result using a construction similar to \Cref{sa:res:rs1-quant-is-pspace-hard} to construct a hard instance. \PP{} is a probabilistic complexity class introduced by Gill \cite{Gil74}; it contains both \NP{} and \coNP{} and is contained in \PSPACE{}. A language is in \PP{} if a polynomial-time probabilistic Turing machine accepts members with probability strictly greater than $\frac{1}{2}$ and accepts non-members with probability at most $\frac{1}{2}$. This acceptance probability can depend on the size of the member - meaning to amplify the probability using a majority vote over repetitions would, in general, require an exponential number of repeats to achieve a constant (say $\frac{2}{3}$s) correctness.

\problemStatement{Majority-SAT (MajSAT) \cite{Gil74,Pap94}}{
Instance={A CNF formula $\varphi$ over Boolean variables $x_1,\ldots,x_n$, with no fixed restriction on clause width.},
Problem={Decide whether more than half of the $2^n$ Boolean assignments satisfy $\varphi$.}
}

\begin{theorem}\label[theorem]{sa:res:rs-quant-is-PP-hard}
{\tt $\{R,S\}$-Arrival-Quant} is $\PP{}$-hard.
\end{theorem}
\begin{proof}
We reduce from {\tt MajSAT}, which is $\PP{}$-complete \cite{Gil74,Pap94}. Let
$\varphi=C_1\wedge\cdots\wedge C_m$ be an unrestricted-width CNF formula over variables $x_1,\ldots,x_n$.\footnote{The clause width is not fixed: for each fixed width, the majority-threshold SAT problem is in \P{} \cite{AkmalWilliams2021}.} We first perform value-preserving syntactic preprocessing: repeated copies of the same literal within a clause are removed, and any clause containing both $x_i$ and $\neg x_i$ is tautological and is deleted. If an empty clause occurs, then $\varphi$ is unsatisfiable and we output a fixed no-instance. Consequently every remaining clause contains $w_l$ distinct literals, and each false literal in a clause corresponds to one distinct incoming consequence edge in the construction below. Write $w_l$ for the width of $C_l$, let
\[
 a_i=|\{l:x_i\in C_l\}|,\qquad b_i=|\{l:\neg x_i\in C_l\}|,
 \qquad D=\max_i\{a_i,b_i\},
\]
and let $p_\varphi$ be the probability that a uniformly random valuation satisfies $\varphi$. We construct in polynomial time an $\{R,S\}$-arrival instance whose value is
\begin{equation}\label{sa:eqn:val-majsat}
    v=\frac12+\left(p_\varphi-\frac12\right)2^{-nD}.
\end{equation}
Consequently $v>1/2$ iff $p_\varphi>1/2$.

The construction is shown in Figures~\ref{sa:fig:ppRSControl}--\ref{sa:fig:ppRSclause}. It has an assignment phase followed, unless an inconsistency is detected, by a verification phase. The control switch sends the play through each variable gadget $D+1$ times. On every pass through gadget $i$, a fair random choice selects $x_i^T$ or $x_i^F$. The first visit to either branch goes through the consistency switch $\mathit{cons}_i$ and returns to the control structure. A later visit to the same branch enters its consequence gadget, whereas the first visit to the opposite branch reaches $\mathit{cons}_i$ for a second time and is sent to $\mathit{bad}_i$, and then to the global node $\mathit{bad}$. The vertex $\mathit{bad}$ records an inconsistent random assignment. It then sends the play to $d$ or $\bar d$ with probability $1/2$ each, so invalid assignments contribute an unbiased baseline: only consistently encoded assignments contribute to the excess above $1/2$.

The consequence gadgets use the corrected indexing convention from Section~\ref{sa:sec:pspacehardness}: $\mathit{Neg}_i$, entered when $x_i$ is assigned true, enumerates the $b_i$ clauses containing $\neg x_i$, while $\mathit{Pos}_i$, entered when $x_i$ is assigned false, enumerates the $a_i$ clauses containing $x_i$. Padding transitions return directly to the control structure, so exactly $D$ visits to one consequence gadget traverse each of its listed clause edges exactly once.

For a clause $C_l$ of width $w_l$, its clause gadget has an assignment entrance and a verification entrance. During the assignment phase, a traversal through the assignment entrance counts one literal of $C_l$ that is false under the currently selected valuation. After $w_l$ such traversals the switch is positioned so that the verification entrance leads to the losing state $\bar d_l$; if fewer than $w_l$ assignment traversals occurred, the verification entrance returns to the control structure. Thus, conditioned on consistent choices for every variable, the subsequent verification phase reaches $d$ exactly when the induced valuation satisfies every clause, and otherwise reaches some $\bar d_l$.

It remains to compute the probability of the three possible behaviours. For a fixed variable $i$, the $D+1$ independent fair choices are consistent exactly when all are true or all are false, which has probability
\[
 2\cdot2^{-(D+1)}=2^{-D}.
\]
The variables use independent random choices, so
\[
 \Pr(\hbox{verification is reached})=2^{-nD},
 \qquad
 \Pr(\hbox{an inconsistency reaches }\mathit{bad})=1-2^{-nD}.
\]
Conditioned on reaching verification, each of the $2^n$ valuations is equally likely: a particular valuation is realised by exactly one consistent outcome pattern and hence has unconditional probability $2^{-n(D+1)}$, so conditioning on the event of probability $2^{-nD}$ gives probability $2^{-n}$.

The inconsistent case contributes $\frac12(1-2^{-nD})$ to the probability of reaching $d$. In the consistent case, verification reaches $d$ with conditional probability $p_\varphi$. Hence
\[
 v=\frac12\bigl(1-2^{-nD}\bigr)+2^{-nD}p_\varphi
   =\frac12+\left(p_\varphi-\frac12\right)2^{-nD},
\]
as claimed. All switching orders and gadgets have size polynomial in the encoding of $\varphi$.
\end{proof}

\begin{figure}[htbp]
    \centering
\resizebox{\textwidth}{!}{\input{figures_tikz_work/fig17.tex}}
    \caption[Proof of \Cref{sa:res:rs-quant-is-PP-hard}: overall control structure.]{Overall ``Control'' structure of the instance.}
    \label{sa:fig:ppRSControl}
\end{figure}
\begin{figure}[htbp]
        \centering
\resizebox{0.8\textwidth}{!}{\begin{tikzpicture}[x=0.75pt,y=0.75pt,yscale=-1,xscale=1]

\draw  [dash pattern={on 4.5pt off 4.5pt}] (11.25,12.9) -- (516.25,12.9) -- (516.25,289.9) -- (11.25,289.9) -- cycle ;
\draw [sa switch]   (121.08,57.08) .. controls (121.08,43.28) and (132.28,32.08) .. (146.08,32.08) .. controls (159.89,32.08) and (171.08,43.28) .. (171.08,57.08) .. controls (171.08,70.89) and (159.89,82.08) .. (146.08,82.08) .. controls (132.28,82.08) and (121.08,70.89) .. (121.08,57.08) -- cycle ;
\draw [sa switch]   (121.08,186.75) .. controls (121.08,172.94) and (132.28,161.75) .. (146.08,161.75) .. controls (159.89,161.75) and (171.08,172.94) .. (171.08,186.75) .. controls (171.08,200.56) and (159.89,211.75) .. (146.08,211.75) .. controls (132.28,211.75) and (121.08,200.56) .. (121.08,186.75) -- cycle ;
\draw  [dash pattern={on 4.5pt off 4.5pt}] (319.25,33.75) -- (389.25,33.75) -- (389.25,85.95) -- (319.25,85.95) -- cycle ;
\draw  [dash pattern={on 4.5pt off 4.5pt}] (320.25,151.95) -- (390.25,151.95) -- (390.25,207.95) -- (320.25,207.95) -- cycle ;
\draw    (55.25,126.75) -- (119.71,58.54) ;
\fill [shift={(121.08,57.08)}, rotate = 133.38] [color={rgb, 255:red, 0; green, 0; blue, 0 }  ] (0,0) -- (8.5,-3.5) -- (8.5,3.5) -- cycle ;
\draw    (55.25,126.75) -- (119.61,185.4) ;
\fill [shift={(121.08,186.75)}, rotate = 222.35] [color={rgb, 255:red, 0; green, 0; blue, 0 }  ] (0,0) -- (8.5,-3.5) -- (8.5,3.5) -- cycle ;
\draw    (171.08,57.08) -- (315.25,55.17) ;
\fill [shift={(317.25,55.14)}, rotate = 179.24] [color={rgb, 255:red, 0; green, 0; blue, 0 }  ] (0,0) -- (8.5,-3.5) -- (8.5,3.5) -- cycle ;
\draw    (171.08,186.75) -- (315.25,184.18) ;
\fill [shift={(317.25,184.14)}, rotate = 178.98] [color={rgb, 255:red, 0; green, 0; blue, 0 }  ] (0,0) -- (8.5,-3.5) -- (8.5,3.5) -- cycle ;
\draw    (447.25,230.14) .. controls (432.17,101.9) and (440.68,71.84) .. (471.81,58.73) ;
\fill [shift={(473.25,58.14)}, rotate = 158.38] [color={rgb, 255:red, 0; green, 0; blue, 0 }  ] (0,0) -- (8.5,-3.5) -- (8.5,3.5) -- cycle ;
\draw [sa switch]   (473.25,58.14) .. controls (473.25,44.33) and (484.44,33.14) .. (498.25,33.14) .. controls (512.06,33.14) and (523.25,44.33) .. (523.25,58.14) .. controls (523.25,71.95) and (512.06,83.14) .. (498.25,83.14) .. controls (484.44,83.14) and (473.25,71.95) .. (473.25,58.14) -- cycle ;
\draw    (392.25,53.75) -- (471.18,178.45) ;
\fill [shift={(472.25,180.14)}, rotate = 237.67] [color={rgb, 255:red, 0; green, 0; blue, 0 }  ] (0,0) -- (8.5,-3.5) -- (8.5,3.5) -- cycle ;
\draw    (390.25,180.75) -- (470.25,180.16) ;
\fill [shift={(472.25,180.14)}, rotate = 179.57] [color={rgb, 255:red, 0; green, 0; blue, 0 }  ] (0,0) -- (8.5,-3.5) -- (8.5,3.5) -- cycle ;
\draw [sa switch]   (472.25,180.14) .. controls (472.25,166.33) and (483.44,155.14) .. (497.25,155.14) .. controls (511.06,155.14) and (522.25,166.33) .. (522.25,180.14) .. controls (522.25,193.95) and (511.06,205.14) .. (497.25,205.14) .. controls (483.44,205.14) and (472.25,193.95) .. (472.25,180.14) -- cycle ;
\draw  [sa switch,fill=SAfigureSwitch,fill opacity=1 ] (317.25,55.14) .. controls (317.25,52.1) and (319.71,49.64) .. (322.75,49.64) .. controls (325.79,49.64) and (328.25,52.1) .. (328.25,55.14) .. controls (328.25,58.18) and (325.79,60.64) .. (322.75,60.64) .. controls (319.71,60.64) and (317.25,58.18) .. (317.25,55.14) -- cycle ;
\draw  [sa switch,fill=SAfigureSwitch,fill opacity=1 ] (386.75,53.75) .. controls (386.75,50.71) and (389.21,48.25) .. (392.25,48.25) .. controls (395.29,48.25) and (397.75,50.71) .. (397.75,53.75) .. controls (397.75,56.79) and (395.29,59.25) .. (392.25,59.25) .. controls (389.21,59.25) and (386.75,56.79) .. (386.75,53.75) -- cycle ;
\draw  [sa switch,fill=SAfigureSwitch,fill opacity=1 ] (311.75,184.14) .. controls (311.75,181.1) and (314.21,178.64) .. (317.25,178.64) .. controls (320.29,178.64) and (322.75,181.1) .. (322.75,184.14) .. controls (322.75,187.18) and (320.29,189.64) .. (317.25,189.64) .. controls (314.21,189.64) and (311.75,187.18) .. (311.75,184.14) -- cycle ;
\draw  [sa switch,fill=SAfigureSwitch,fill opacity=1 ] (384.75,180.75) .. controls (384.75,177.71) and (387.21,175.25) .. (390.25,175.25) .. controls (393.29,175.25) and (395.75,177.71) .. (395.75,180.75) .. controls (395.75,183.79) and (393.29,186.25) .. (390.25,186.25) .. controls (387.21,186.25) and (384.75,183.79) .. (384.75,180.75) -- cycle ;
\draw    (171.08,186.75) .. controls (192.25,225.14) and (233.95,219.95) .. (408.66,250.14) ;
\fill [shift={(408.66,250.14)}, rotate = 189.81] [color={rgb, 255:red, 0; green, 0; blue, 0 }  ] (0,0) -- (8.5,-3.5) -- (8.5,3.5) -- cycle ;
\draw    (171.08,57.08) .. controls (184.45,111.95) and (441.95,82.95) .. (423.95,225.95) ;
\fill [shift={(423.95,225.95)}, rotate = 277.17] [color={rgb, 255:red, 0; green, 0; blue, 0 }  ] (0,0) -- (8.5,-3.5) -- (8.5,3.5) -- cycle ;
\draw [sa random]   (30.5,101.75) -- (55.25,126.75) -- (30.5,151.75) -- (5.75,126.75) -- cycle ;
\draw [sa switch]   (408.66,250.14) .. controls (408.66,236.33) and (419.86,225.14) .. (433.66,225.14) .. controls (447.47,225.14) and (458.66,236.33) .. (458.66,250.14) .. controls (458.66,263.95) and (447.47,275.14) .. (433.66,275.14) .. controls (419.86,275.14) and (408.66,263.95) .. (408.66,250.14) -- cycle ;
\draw    (447.25,230.14) -- (479.52,198.35) ;
\fill [shift={(480.95,196.95)}, rotate = 135.43] [color={rgb, 255:red, 0; green, 0; blue, 0 }  ] (0,0) -- (8.5,-3.5) -- (8.5,3.5) -- cycle ;

\draw (30.50,126.75) node [sa node label] {$\displaystyle as_{i}$};
\draw (146.08,57.08) node [sa node label] {$\displaystyle x_{i}^{T}$};
\draw (146.08,186.75) node [sa node label] {$\displaystyle x_{i}^{F}$};
\draw (336.1,49.1) node [anchor=north west][inner sep=0.75pt]   [align=left] {$\displaystyle Neg_{i}$};
\draw (336.6,174.1) node [anchor=north west][inner sep=0.75pt]   [align=left] {$\displaystyle Pos_{i}$};
\draw (207.08,35.08) node [anchor=north west][inner sep=0.75pt]   [align=left] {$\displaystyle 2,\dotsc ,D+1$};
\draw (201.08,163.08) node [anchor=north west][inner sep=0.75pt]   [align=left] {$\displaystyle 2,\dotsc ,D+1$};
\draw (497.25,180.14) node [sa node label] {$\displaystyle o$};
\draw (203.25,192.75) node [anchor=north west][inner sep=0.75pt]   [align=left] {$\displaystyle 1$};
\draw (208.25,67.25) node [anchor=north west][inner sep=0.75pt]   [align=left] {$\displaystyle 1$};
\draw (498.25,58.14) node [sa node label] {$\displaystyle bad_{i}$};
\draw (466.1,216.54) node [anchor=north west][inner sep=0.75pt]   [align=left] {1};
\draw (434.9,208.4) node [anchor=north west][inner sep=0.75pt]   [align=left] {2};
\draw (433.66,250.14) node [sa node label] {$\displaystyle cons_{i}$};

\end{tikzpicture}}
\caption[Proof of \Cref{sa:res:rs-quant-is-PP-hard}: random assignment construction.]{Gadget for random assignment of variables.}
\label{sa:fig:RSPPvar}
\end{figure}
\begin{figure}[htbp]
\begin{subfigure}[h]{0.49\textwidth }
    \centering
\resizebox{\linewidth}{!}{\begin{tikzpicture}[x=0.75pt,y=0.75pt,yscale=-1,xscale=1]

\draw [sa switch]   (6,75.85) .. controls (6,62.04) and (17.19,50.85) .. (31,50.85) .. controls (44.81,50.85) and (56,62.04) .. (56,75.85) .. controls (56,89.65) and (44.81,100.85) .. (31,100.85) .. controls (17.19,100.85) and (6,89.65) .. (6,75.85) -- cycle ;
\draw    (56,75.85) -- (162.54,52.92) ;
\fill [shift={(164.5,52.5)}, rotate = 167.86] [color={rgb, 255:red, 0; green, 0; blue, 0 }  ] (0,0) -- (8.5,-3.5) -- (8.5,3.5) -- cycle ;
\draw    (56,75.85) -- (158.95,160.23) ;
\fill [shift={(160.5,161.5)}, rotate = 219.34] [color={rgb, 255:red, 0; green, 0; blue, 0 }  ] (0,0) -- (8.5,-3.5) -- (8.5,3.5) -- cycle ;
\draw  [dash pattern={on 4.5pt off 4.5pt}] (14,10.5) -- (392,10.5) -- (392,289.84) -- (14,289.84) -- cycle ;
\draw  [sa switch,fill=SAfigureSwitch  ,fill opacity=1 ] (160.5,161.5) .. controls (160.5,158.46) and (162.96,156) .. (166,156) .. controls (169.04,156) and (171.5,158.46) .. (171.5,161.5) .. controls (171.5,164.54) and (169.04,167) .. (166,167) .. controls (162.96,167) and (160.5,164.54) .. (160.5,161.5) -- cycle ;
\draw  [sa switch,fill=SAfigureSwitch  ,fill opacity=1 ] (269.5,162.5) .. controls (269.5,159.46) and (271.96,157) .. (275,157) .. controls (278.04,157) and (280.5,159.46) .. (280.5,162.5) .. controls (280.5,165.54) and (278.04,168) .. (275,168) .. controls (271.96,168) and (269.5,165.54) .. (269.5,162.5) -- cycle ;
\draw    (56,75.85) .. controls (-27.58,377.33) and (387.81,263.99) .. (387.84,149.72) ;
\fill [shift={(387.8,148)}, rotate = 87.91] [color={rgb, 255:red, 0; green, 0; blue, 0 }  ] (0,0) -- (8.5,-3.5) -- (8.5,3.5) -- cycle ;
\draw    (280.5,162.5) -- (385.82,148.27) ;
\fill [shift={(387.8,148)}, rotate = 172.3] [color={rgb, 255:red, 0; green, 0; blue, 0 }  ] (0,0) -- (8.5,-3.5) -- (8.5,3.5) -- cycle ;
\draw [sa switch]   (412.8,123) .. controls (426.61,123) and (437.8,134.19) .. (437.8,148) .. controls (437.8,161.81) and (426.61,173) .. (412.8,173) .. controls (399,173) and (387.8,161.81) .. (387.8,148) .. controls (387.8,134.19) and (399,123) .. (412.8,123) -- cycle ;
\draw  [sa switch,fill=SAfigureSwitch  ,fill opacity=1 ] (162.5,191.5) .. controls (162.5,188.46) and (164.96,186) .. (168,186) .. controls (171.04,186) and (173.5,188.46) .. (173.5,191.5) .. controls (173.5,194.54) and (171.04,197) .. (168,197) .. controls (164.96,197) and (162.5,194.54) .. (162.5,191.5) -- cycle ;
\draw  [sa switch,fill=SAfigureSwitch  ,fill opacity=1 ] (164.5,52.5) .. controls (164.5,49.46) and (166.96,47) .. (170,47) .. controls (173.04,47) and (175.5,49.46) .. (175.5,52.5) .. controls (175.5,55.54) and (173.04,58) .. (170,58) .. controls (166.96,58) and (164.5,55.54) .. (164.5,52.5) -- cycle ;
\draw  [sa switch,fill=SAfigureSwitch  ,fill opacity=1 ] (273.5,53.5) .. controls (273.5,50.46) and (275.96,48) .. (279,48) .. controls (282.04,48) and (284.5,50.46) .. (284.5,53.5) .. controls (284.5,56.54) and (282.04,59) .. (279,59) .. controls (275.96,59) and (273.5,56.54) .. (273.5,53.5) -- cycle ;
\draw  [sa switch,fill=SAfigureSwitch  ,fill opacity=1 ] (166.5,82.5) .. controls (166.5,79.46) and (168.96,77) .. (172,77) .. controls (175.04,77) and (177.5,79.46) .. (177.5,82.5) .. controls (177.5,85.54) and (175.04,88) .. (172,88) .. controls (168.96,88) and (166.5,85.54) .. (166.5,82.5) -- cycle ;
\draw    (284.5,53.5) -- (386.33,146.65) ;
\fill [shift={(387.8,148)}, rotate = 222.45] [color={rgb, 255:red, 0; green, 0; blue, 0 }  ] (0,0) -- (8.5,-3.5) -- (8.5,3.5) -- cycle ;

\draw (31.00,75.85) node [sa node label] {$\mathit{Pos}_{i}$};
\draw (70.5,49) node [anchor=north west][inner sep=0.75pt]   [align=left] {$\displaystyle 1$};
\draw (134.1,242.8) node [anchor=north west][inner sep=0.75pt]   [align=left] {$\displaystyle a_{i} +1,\dotsc ,D$};
\draw  [dash pattern={on 4.5pt off 4.5pt}]  (167,142.48) -- (275,142.48) -- (275,218.48) -- (167,218.48) -- cycle  ;
\draw (221,180.48) node   [align=left] {\begin{minipage}[lt]{70.72pt}\setlength\topsep{0pt}

\end{minipage}};
\draw (62.5,90) node [anchor=north west][inner sep=0.75pt]   [align=left] {$\displaystyle a_{i}$};
\draw (208.1,103.7) node [anchor=north west][inner sep=0.75pt]  [font=\LARGE] [align=left] {...};
\draw (412.80,148.00) node [sa node label] {$\displaystyle o$};
\draw (223,161) node [anchor=north west][inner sep=0.75pt]  [font=\Large] [align=left] {$\displaystyle C_{k_{a_{i}}}$};
\draw (176,152) node [anchor=north west][inner sep=0.75pt]   [align=left] {$\displaystyle c_{k_{a_{i}}}$};
\draw (175,180) node [anchor=north west][inner sep=0.75pt]   [align=left] {$\displaystyle ver_{k_{a_{i}}}$};
\draw  [dash pattern={on 4.5pt off 4.5pt}]  (171,33.48) -- (279,33.48) -- (279,109.48) -- (171,109.48) -- cycle  ;
\draw (225,71.48) node   [align=left] {\begin{minipage}[lt]{70.72pt}\setlength\topsep{0pt}

\end{minipage}};
\draw (227,52) node [anchor=north west][inner sep=0.75pt]  [font=\Large] [align=left] {$\displaystyle C_{k_{a_{i}}}$};
\draw (180,43) node [anchor=north west][inner sep=0.75pt]   [align=left] {$\displaystyle c_{k_{a_{i}}}$};
\draw (179,71) node [anchor=north west][inner sep=0.75pt]   [align=left] {$\displaystyle ver_{k_{a_{i}}}$};
\draw (96.0,82.0) node [anchor=north west][inner sep=0.75pt]  [font=\LARGE] [align=left] {...};

\end{tikzpicture}}
    \caption{$Pos_i$ consequence gadget for $x_i$}
    \label{sa:fig:ppRSPos}
\end{subfigure}
\begin{subfigure}[htbp]{0.49\textwidth }
    \centering
\resizebox{\linewidth}{!}{\begin{tikzpicture}[x=0.75pt,y=0.75pt,yscale=-1,xscale=1]

\draw [sa switch]   (6,75.85) .. controls (6,62.04) and (17.19,50.85) .. (31,50.85) .. controls (44.81,50.85) and (56,62.04) .. (56,75.85) .. controls (56,89.65) and (44.81,100.85) .. (31,100.85) .. controls (17.19,100.85) and (6,89.65) .. (6,75.85) -- cycle ;
\draw    (56,75.85) -- (162.54,52.92) ;
\fill [shift={(164.5,52.5)}, rotate = 167.86] [color={rgb, 255:red, 0; green, 0; blue, 0 }  ] (0,0) -- (8.5,-3.5) -- (8.5,3.5) -- cycle ;
\draw    (56,75.85) -- (158.95,160.23) ;
\fill [shift={(160.5,161.5)}, rotate = 219.34] [color={rgb, 255:red, 0; green, 0; blue, 0 }  ] (0,0) -- (8.5,-3.5) -- (8.5,3.5) -- cycle ;
\draw  [dash pattern={on 4.5pt off 4.5pt}] (14,10.5) -- (392,10.5) -- (392,289.84) -- (14,289.84) -- cycle ;
\draw  [sa switch,fill=SAfigureSwitch,fill opacity=1 ] (160.5,161.5) .. controls (160.5,158.46) and (162.96,156) .. (166,156) .. controls (169.04,156) and (171.5,158.46) .. (171.5,161.5) .. controls (171.5,164.54) and (169.04,167) .. (166,167) .. controls (162.96,167) and (160.5,164.54) .. (160.5,161.5) -- cycle ;
\draw  [sa switch,fill=SAfigureSwitch,fill opacity=1 ] (269.5,162.5) .. controls (269.5,159.46) and (271.96,157) .. (275,157) .. controls (278.04,157) and (280.5,159.46) .. (280.5,162.5) .. controls (280.5,165.54) and (278.04,168) .. (275,168) .. controls (271.96,168) and (269.5,165.54) .. (269.5,162.5) -- cycle ;
\draw    (56,75.85) .. controls (-27.58,377.33) and (387.81,263.99) .. (387.84,149.72) ;
\fill [shift={(387.8,148)}, rotate = 87.91] [color={rgb, 255:red, 0; green, 0; blue, 0 }  ] (0,0) -- (8.5,-3.5) -- (8.5,3.5) -- cycle ;
\draw    (280.5,162.5) -- (385.82,148.27) ;
\fill [shift={(387.8,148)}, rotate = 172.3] [color={rgb, 255:red, 0; green, 0; blue, 0 }  ] (0,0) -- (8.5,-3.5) -- (8.5,3.5) -- cycle ;
\draw [sa switch]   (412.8,123) .. controls (426.61,123) and (437.8,134.19) .. (437.8,148) .. controls (437.8,161.81) and (426.61,173) .. (412.8,173) .. controls (399,173) and (387.8,161.81) .. (387.8,148) .. controls (387.8,134.19) and (399,123) .. (412.8,123) -- cycle ;
\draw  [sa switch,fill=SAfigureSwitch,fill opacity=1 ] (162.5,191.5) .. controls (162.5,188.46) and (164.96,186) .. (168,186) .. controls (171.04,186) and (173.5,188.46) .. (173.5,191.5) .. controls (173.5,194.54) and (171.04,197) .. (168,197) .. controls (164.96,197) and (162.5,194.54) .. (162.5,191.5) -- cycle ;
\draw  [sa switch,fill=SAfigureSwitch,fill opacity=1 ] (164.5,52.5) .. controls (164.5,49.46) and (166.96,47) .. (170,47) .. controls (173.04,47) and (175.5,49.46) .. (175.5,52.5) .. controls (175.5,55.54) and (173.04,58) .. (170,58) .. controls (166.96,58) and (164.5,55.54) .. (164.5,52.5) -- cycle ;
\draw  [sa switch,fill=SAfigureSwitch,fill opacity=1 ] (273.5,53.5) .. controls (273.5,50.46) and (275.96,48) .. (279,48) .. controls (282.04,48) and (284.5,50.46) .. (284.5,53.5) .. controls (284.5,56.54) and (282.04,59) .. (279,59) .. controls (275.96,59) and (273.5,56.54) .. (273.5,53.5) -- cycle ;
\draw  [sa switch,fill=SAfigureSwitch,fill opacity=1 ] (166.5,82.5) .. controls (166.5,79.46) and (168.96,77) .. (172,77) .. controls (175.04,77) and (177.5,79.46) .. (177.5,82.5) .. controls (177.5,85.54) and (175.04,88) .. (172,88) .. controls (168.96,88) and (166.5,85.54) .. (166.5,82.5) -- cycle ;
\draw    (284.5,53.5) -- (386.33,146.65) ;
\fill [shift={(387.8,148)}, rotate = 222.45] [color={rgb, 255:red, 0; green, 0; blue, 0 }  ] (0,0) -- (8.5,-3.5) -- (8.5,3.5) -- cycle ;

\draw (31.00,75.85) node [sa node label] {pos$\displaystyle _{i}$};
\draw (70.5,49) node [anchor=north west][inner sep=0.75pt]   [align=left] {$\displaystyle 1$};
\draw (134.1,242.8) node [anchor=north west][inner sep=0.75pt]   [align=left] {$\displaystyle b_{i} +1,\dotsc ,D$};
\draw  [dash pattern={on 4.5pt off 4.5pt}]  (167,142.48) -- (275,142.48) -- (275,218.48) -- (167,218.48) -- cycle  ;
\draw (221,180.48) node   [align=left] {\begin{minipage}[lt]{70.72pt}\setlength\topsep{0pt}

\end{minipage}};
\draw (62.5,90) node [anchor=north west][inner sep=0.75pt]   [align=left] {$\displaystyle b_{i}$};
\draw (208.1,103.7) node [anchor=north west][inner sep=0.75pt]  [font=\LARGE] [align=left] {...};
\draw (412.80,148.00) node [sa node label] {$\displaystyle o$};
\draw (223,161) node [anchor=north west][inner sep=0.75pt]  [font=\Large] [align=left] {$\displaystyle C_{k_{b_{i}}}$};
\draw (176,152) node [anchor=north west][inner sep=0.75pt]   [align=left] {$\displaystyle c_{k_{b_{i}}}$};
\draw (175,180) node [anchor=north west][inner sep=0.75pt]   [align=left] {$\displaystyle ver_{k_{b_{i}}}$};
\draw  [dash pattern={on 4.5pt off 4.5pt}]  (171,33.48) -- (279,33.48) -- (279,109.48) -- (171,109.48) -- cycle  ;
\draw (225,71.48) node   [align=left] {\begin{minipage}[lt]{70.72pt}\setlength\topsep{0pt}

\end{minipage}};
\draw (227,52) node [anchor=north west][inner sep=0.75pt]  [font=\Large] [align=left] {$\displaystyle C_{k_{b_{i}}}$};
\draw (180,43) node [anchor=north west][inner sep=0.75pt]   [align=left] {$\displaystyle c_{k_{b_{i}}}$};
\draw (179,71) node [anchor=north west][inner sep=0.75pt]   [align=left] {$\displaystyle ver_{k_{b_{i}}}$};
\draw (96.0,82.0) node [anchor=north west][inner sep=0.75pt]  [font=\LARGE] [align=left] {...};

\end{tikzpicture}}
    \caption{$Neg_i$ consequence gadget for $x_i$}
    \label{sa:fig:ppRSNeg}
\end{subfigure}
\caption[Proof of \Cref{sa:res:rs-quant-is-PP-hard}: consequence gadgets.]{Gadgets for Consequences}
\label{sa:fig:RSPPChecks}
\end{figure}
\begin{figure}[htbp]
    \centering
\resizebox{0.62\textwidth}{!}{\begin{tikzpicture}[x=0.75pt,y=0.75pt,yscale=-1,xscale=1]

\draw [sa switch]   (12.5,56.5) .. controls (12.5,42.69) and (23.69,31.5) .. (37.5,31.5) .. controls (51.31,31.5) and (62.5,42.69) .. (62.5,56.5) .. controls (62.5,70.31) and (51.31,81.5) .. (37.5,81.5) .. controls (23.69,81.5) and (12.5,70.31) .. (12.5,56.5) -- cycle ;
\draw    (62.5,56.5) -- (267,49.57) ;
\fill [shift={(269,49.5)}, rotate = 178.06] [color={rgb, 255:red, 0; green, 0; blue, 0 }  ] (0,0) -- (8.5,-3.5) -- (8.5,3.5) -- cycle ;
\draw    (62.5,56.5) -- (111.09,105.09) ;
\fill [shift={(112.5,106.5)}, rotate = 225] [color={rgb, 255:red, 0; green, 0; blue, 0 }  ] (0,0) -- (8.5,-3.5) -- (8.5,3.5) -- cycle ;
\draw  [dash pattern={on 4.5pt off 4.5pt}] (21,11.5) -- (276,11.5) -- (276,215.5) -- (21,215.5) -- cycle ;
\draw [sa switch]   (112.5,106.5) .. controls (112.5,92.69) and (123.69,81.5) .. (137.5,81.5) .. controls (151.31,81.5) and (162.5,92.69) .. (162.5,106.5) .. controls (162.5,120.31) and (151.31,131.5) .. (137.5,131.5) .. controls (123.69,131.5) and (112.5,120.31) .. (112.5,106.5) -- cycle ;
\draw    (162.5,106.5) -- (271.18,57.32) ;
\fill [shift={(273,56.5)}, rotate = 155.65] [color={rgb, 255:red, 0; green, 0; blue, 0 }  ] (0,0) -- (8.5,-3.5) -- (8.5,3.5) -- cycle ;
\draw    (161,118) -- (181.9,149.83) ;
\fill [shift={(183,151.5)}, rotate = 236.71] [color={rgb, 255:red, 0; green, 0; blue, 0 }  ] (0,0) -- (8.5,-3.5) -- (8.5,3.5) -- cycle ;
\draw [sa switch]   (171,175.5) .. controls (171,161.69) and (182.19,150.5) .. (196,150.5) .. controls (209.81,150.5) and (221,161.69) .. (221,175.5) .. controls (221,189.31) and (209.81,200.5) .. (196,200.5) .. controls (182.19,200.5) and (171,189.31) .. (171,175.5) -- cycle ;
\draw    (221,175.5) .. controls (270.5,206.19) and (260.21,124.17) .. (220.22,166.19) ;
\fill [shift={(219,167.5)}, rotate = 312.34] [color={rgb, 255:red, 0; green, 0; blue, 0 }  ] (0,0) -- (8.5,-3.5) -- (8.5,3.5) -- cycle ;
\draw [sa switch]   (12.5,172.5) .. controls (12.5,158.69) and (23.69,147.5) .. (37.5,147.5) .. controls (51.31,147.5) and (62.5,158.69) .. (62.5,172.5) .. controls (62.5,186.31) and (51.31,197.5) .. (37.5,197.5) .. controls (23.69,197.5) and (12.5,186.31) .. (12.5,172.5) -- cycle ;
\draw    (62.5,172.5) -- (117.46,126.78) ;
\fill [shift={(119,125.5)}, rotate = 140.24] [color={rgb, 255:red, 0; green, 0; blue, 0 }  ] (0,0) -- (8.5,-3.5) -- (8.5,3.5) -- cycle ;
\draw [sa switch]   (296.8,25) .. controls (310.61,25) and (321.8,36.19) .. (321.8,50) .. controls (321.8,63.81) and (310.61,75) .. (296.8,75) .. controls (283,75) and (271.8,63.81) .. (271.8,50) .. controls (271.8,36.19) and (283,25) .. (296.8,25) -- cycle ;

\draw (150,39) node [sa edge label] {$1,2,\dotsc,w_l-1$};
\draw (87,82) node [sa edge label] {$w_l$};
\draw (37.50,56.50) node [sa node label large] {$\displaystyle c_{l}$};
\draw (177,83) node [sa edge label] {$1$};
\draw (174,125) node [sa edge label] {$2$};
\draw (37.50,172.50) node [sa node label large] {$\displaystyle ver_{l}$};
\draw (196.00,175.50) node [sa node label large] {$\displaystyle \bar d_{l}$};
\draw (296.80,50.00) node [sa node label large] {$\displaystyle o$};
\draw (137.50,106.50) node [sa node label large] {$\displaystyle sat_{l}$};

\end{tikzpicture}}
    \caption[Proof of \Cref{sa:res:rs-quant-is-PP-hard}: clause gadgets.]{Clause gadget for a clause $C_l$ with width $w_l$.}
    \label{sa:fig:ppRSclause}
\end{figure}

\FloatBarrier
\section{{\tt $\{R,S\}$-Arrival-Quant} is in \PSPACE{}}
\label{sec:rs-quant-pspace}

We now prove a \PSPACE{} upper bound for the quantitative problem with random
and switching vertices. The bit-level space argument uses classical links between parallel determinant computation and space-bounded probabilistic computation \cite{BorodinCookPippenger1983}.  We regard the input as specifying the exponentially larger Markov chain
$Exp(G)$ and work with its transition matrix indirectly.  Expanded states have
polynomial-length descriptions, so entries can be computed as needed without
constructing the expanded chain or storing its transition matrix.  We represent the required hitting probability
by an entry of a matrix inverse and reduce comparison with $\frac12$ to the sign
of the determinant of one succinctly represented integer matrix.

Throughout this section, let $(G,o,d)$ be an instance of generalised
$\{R,S\}$-Arrival, with
\[
       G=(V,E,\{V_R,V_S\},\Prob,\Ord).
\]
Let $Q$ be the set of switching positions and let $q^0\in Q$ be the initial
switching position.  An expanded state $(v,q)$ has a binary description of
length polynomial in the encoding length of $G$, although $|Q|$ may be
exponential.

Collapse all expanded target states $(d,q)$, $q\in Q$, to one symbol
$\dagger$ and define
\[
   \mathcal J:=((V\setminus\{d\})\cross Q)\cup\{\dagger\}.
\]
We enumerate the switching configurations as mixed-radix tuples and order
$(V\setminus\{d\})\cross Q$ lexicographically, followed by $\dagger$.  Both converting an index
to the corresponding element of $\mathcal J$ and converting an element of
$\mathcal J$ to its index use polynomial space.  Write $N:=|\mathcal J|$.
There is a polynomial $p$ such that
\[
       N\leq 2^{p(|G|)}.
\]

Call $\dagger$ \emph{live}.  For
$x\in\mathcal J\setminus\{\dagger\}$, call $x$ \emph{live} if the support graph
of the expanded Markov chain contains a directed path from $x$ to a state whose
vertex is $d$, equivalently to $\dagger$ after the above identification.  Thus
liveness is exactly positive-probability reachability of the target.
Let $\widehat M(x,y)$ denote the one-step transition probability in the expanded
chain, with all transitions into a target state accumulated into the column
$\dagger$.  Define
$M\in[0,1]^{\mathcal J\cross\mathcal J}$ by
\begin{equation}\label{sa:eqn:mod-Trans-M}
 M(x,y):=
 \begin{cases}
   \widehat M(x,y), & \text{if $x$ and $y$ are live and $x\neq\dagger$},\\
   0,               & \text{otherwise}.
 \end{cases}
\end{equation}
In particular, the row of $\dagger$ is zero.  Transitions into non-live states
are discarded; this cannot change the probability of eventually reaching $d$,
since every state on a run prefix that is followed by a visit to $d$ is itself
live.

Let $D$ be the product of the positive denominators, written in lowest terms,
of all rational probabilities $\Prob(v,w)$ occurring in the input.  If there
are no non-integral probabilities, take $D=1$.  Since these denominators are
part of the input, the binary representation of $D$ has length at most the sum
of their binary lengths and hence polynomial in $|G|$.  Every denominator
occurring in an entry of $M$ divides $D$.

\begin{lemma}\label{sa:res:compute-M-elems}
Given $(G,o,d)$ and row and column indices $i,j\in[N]$, the entry $M_{i,j}$ from
\eqref{sa:eqn:mod-Trans-M} can be computed in polynomial space.  Consequently,
for
\begin{equation}\label{eq:def-L-section6}
        L:=D(I-M),
\end{equation}
the entry $L_{i,j}$, and hence any requested bit of a fixed explicit binary
encoding of $L$, can be computed in polynomial space.  Every entry of $L$ is an
integer with a number of bits polynomial in $|G|$.
\end{lemma}

\begin{proof}
From $i$ and $j$ we first recover the corresponding elements
$x,y\in\mathcal J$ using the fixed enumeration above.  This requires only
$O(\log N)$ bits, hence polynomial space in $|G|$.

It remains to decide whether $x$ and $y$ are live and, when they are, to
compute the local transition probability $\widehat M(x,y)$.  Liveness is a
reachability question in an exponentially large directed graph whose vertices
have polynomial-length encodings and whose successors can be computed directly
from the definition of $Exp(G)$.  Nondeterministically guess a path of length at
most the number of expanded states, storing only the current expanded state and
a binary step counter.  This uses polynomial space.  By Savitch's theorem
\cite{Savitch1970}, nondeterministic polynomial space equals deterministic
polynomial space, so liveness can be decided in \PSPACE{}.

The one-step probability $\widehat M(x,y)$ is local.  At a probabilistic vertex
$(v,q)$, switching positions do not change and the probability of a successor
$(w,q)$ is $\Prob(v,w)$.  At a switching vertex there is a unique successor
expanded state, obtained by following the edge selected by the current
switching position and updating that position; its transition probability is
$1$.  If the successor vertex is $d$, the transition is instead placed in the
column $\dagger$.  Hence $M_{i,j}$ is computable in polynomial space.

The integer $D$ itself is computable with polynomially many bits by multiplying
the input denominators.  Since every denominator in $M$ divides $D$, the formula
$L=D(I-M)$ gives an integer $L_{i,j}$ using polynomial space.  Moreover
$|M_{i,j}|\leq1$, so $|L_{i,j}|\leq D$ off the diagonal and
$|L_{i,i}|\leq D$ on the diagonal as well.  Thus the entries have polynomial
bit length.  Fixing a sufficiently large polynomial-width signed binary field
for each matrix entry, a requested bit of the ordinary explicit row-major
encoding of $L$ can therefore be recovered in polynomial space.
\end{proof}

\begin{lemma}\label{sa:res:M-sub-stochastic}
The matrix $M$ is substochastic and there exists $K\geq1$ such that every row
of $M^K$ has sum strictly smaller than $1$.  Consequently,
\[
       \lim_{m\to\infty}M^m=0,
\]
the matrix $I-M$ is nonsingular, and the Neumann series converges with
\begin{equation}\label{eq:neumann-series-section6}
       (I-M)^{-1}=\sum_{m=0}^{\infty}M^m.
\end{equation}
\end{lemma}

\begin{proof}
Every row of $M$ is obtained from a probability distribution by possibly
removing transitions to non-live states, or it is the all-zero row.  Hence $M$
is substochastic.

Consider a live state $x\neq\dagger$.  Choose a shortest path in the support
graph
\[
       x=x_0,x_1,\ldots,x_{\ell_x}=\dagger.
\]
Since the path is shortest, $\dagger$ occurs only at its final position, and
every $x_i$ with $i<\ell_x$ is live because the suffix from $x_i$ reaches the
target.  Therefore every edge of this path is retained by $M$ and has strictly
positive transition probability.  Let $\delta_x>0$ be the product of these
probabilities.  Starting from $x$, at least $\delta_x$ probability mass is at
$\dagger$ after $\ell_x$ steps and is removed at the next multiplication,
because the row of $\dagger$ is zero.  Hence the row of $M^{\ell_x+1}$ indexed
by $x$ has sum at most $1-\delta_x<1$.  Rows indexed by non-live states, and
the row of $\dagger$, already have sum $0$ after one step.

Since $\mathcal J$ is finite, choose
\[
       K:=\max\bigl(\{1\}\cup
       \{\ell_x+1:x\text{ is live and }x\neq\dagger\}\bigr).
\]
For a substochastic matrix, row sums of successive powers are nonincreasing:
if $\mathbf 1$ is the all-one column vector, then
$M\mathbf 1\leq\mathbf 1$, and hence
$M^{m+1}\mathbf 1\leq M^m\mathbf 1$.  It follows that every row of $M^K$ has
sum strictly below $1$.  Let
\[
       c:=\lVert M^K\rVert_\infty<1,
\]
where the $\infty$-norm is the maximum row sum.  For $m=aK+b$ with
$0\leq b<K$,
\[
       \lVert M^m\rVert_\infty
       \leq \lVert M^K\rVert_\infty^a\lVert M^b\rVert_\infty
       \leq c^a,
\]
because every power of a substochastic matrix has maximum row sum at most $1$.
Thus $M^m\to0$.  Moreover,
\[
  \sum_{m=0}^{\infty}\lVert M^m\rVert_\infty
  \leq \sum_{a=0}^{\infty}\sum_{b=0}^{K-1}c^a
  =\frac{K}{1-c}<\infty,
\]
so the Neumann series converges absolutely in this matrix norm.  Taking limits
in the finite geometric identity
\[
       (I-M)\sum_{m=0}^{k}M^m=I-M^{k+1}
\]
gives \eqref{eq:neumann-series-section6}; in particular $I-M$ is nonsingular.
\end{proof}

\begin{lemma}\label{lem:succinct-determinant-pspace}
Let $I$ be a succinct input of length $s$ describing an $N\cross N$
integer matrix $A$.  Suppose there is a polynomial $p$ such that:
\begin{enumerate}
    \item $N\leq2^{p(s)}$;
    \item every entry of $A$ has at most $p(s)$ signed binary digits; and
    \item given $I$ and $i,j\in[N]$, the entry $A_{i,j}$ is computable
    in space polynomial in $s$.
\end{enumerate}
Then the three-valued sign
\[
       \operatorname{sgn}(\det A)\in\{-1,0,+1\}
\]
is computable in space polynomial in $s$ from $I$.
\end{lemma}

\begin{proof}
Using the polynomial $p$ in the lemma statement, fix a conventional explicit
row-major binary matrix encoding with every signed entry padded to $p(s)$
binary digits.  This padded representation and any standard self-delimiting binary
representation are mutually translatable in logarithmic space, so the choice
does not affect the space bound below.  Write this encoding as
$\operatorname{enc}(A)$ and let $m:=|\operatorname{enc}(A)|$.  Since $A$ has dimension at most
exponential in $s$ and polynomial-bit entries,
\[
       m=N^{O(1)}\cdot\operatorname{poly}(s)
       \leq2^{\operatorname{poly}(s)},
       \qquad
       \log m=\operatorname{poly}(s).
\]

For an explicitly represented binary integer matrix, the determinant function
belongs to uniform $\mathsf{NC}^2$ and therefore to
$\mathsf{FSPACE}(O(\log^2 m))$.  Lohrey and Schmidt-Schau{\ss} state this
bit-level space consequence explicitly in their preliminaries
\cite[Section~3]{LohreySchmidtSchauss2017}.  For the classical Boolean
$\mathsf{NC}^2$ implementation of determinant, see also Borodin, Cook and
Pippenger~\cite{BorodinCookPippenger1983}.

We simulate an explicit-matrix determinant transducer with
$\operatorname{enc}(A)$ supplied as a virtual read-only input.  The simulated
machine uses
$O(\log^2 m)=\operatorname{poly}(s)$ work space, and its current input-head
position uses $O(\log m)$ bits.  Whenever the simulation requests an input bit,
we decode that position into the corresponding row, column, and bit position
inside the fixed-width entry field, compute $A_{i,j}$ from $I$ using assumption~(3), and
return the requested bit.  Formatting and padding bits are immediate.  The
workspace used for this input-bit computation is reused when the simulated
machine resumes.  Thus the signed binary representation of $\det A$ can be
streamed in space polynomial in $s$, without constructing the explicit matrix or
storing the determinant.

To obtain the three-valued sign, use the usual signed binary output encoding
(or convert to sign-and-magnitude in logarithmic space).  While the output is
streamed, retain the sign and one Boolean flag recording whether a nonzero
magnitude bit has appeared.  At the end, the determinant is zero exactly when
that flag is false; otherwise the retained sign distinguishes the positive and
negative cases.  Hence $\operatorname{sgn}(\det A)$ is computable in
space polynomial in $s$.
\end{proof}

\begin{theorem}\label{sa:res:rs-quant-in-pspace}
The {\tt $\{R,S\}$-Arrival-Quant} problem is in \PSPACE{}.
\end{theorem}

\begin{proof}
By Proposition~\ref{sa:res:rec-quant-only-p-half}, it suffices to decide the
strict-threshold problem
\[
       \val(G,o,d)>\frac12.
\]
If $o=d$, the answer is immediate, so suppose $o\neq d$ and let
$\iota:=(o,q^0)\in\mathcal J$ be the initial expanded state.

We first identify the value with one entry of $(I-M)^{-1}$.  Since the row of
$\dagger$ in $M$ is zero, $(M^m)_{\iota,\dagger}$ is precisely the probability
of first reaching the target at step $m$.  No run that eventually reaches $d$
can previously enter a non-live state.  Hence discarding transitions to
non-live states in \eqref{sa:eqn:mod-Trans-M} does not alter the probability of
reaching $d$.  Using \eqref{eq:neumann-series-section6},
\begin{equation}\label{eq:value-as-inverse-entry}
       \val(G,o,d)
       =\sum_{m=0}^{\infty}(M^m)_{\iota,\dagger}
       =(I-M)^{-1}_{\iota,\dagger}.
\end{equation}

Let $L=D(I-M)$ as in \eqref{eq:def-L-section6} and put
$\Delta:=\det L$.  By Lemma~\ref{sa:res:M-sub-stochastic}, $M^m\to0$.
Hence every eigenvalue $\lambda$ of $M$ satisfies $|\lambda|<1$: if
$Mv=\lambda v$ for a nonzero vector $v$, then
$M^m v=\lambda^m v\to0$.  Therefore $I-tM$ is nonsingular for every
$t\in[0,1]$.  The real-valued function
$t\mapsto\det(I-tM)$ is continuous and never zero on this interval, and its
value at $t=0$ is $1$.  Consequently
\begin{equation}\label{eq:positive-determinant}
       \det(I-M)>0,
       \qquad
       \Delta=D^N\det(I-M)>0.
\end{equation}

Let $\kappa:=C_{\dagger,\iota}(L)$ denote the $(\dagger,\iota)$ cofactor of
$L$, including its usual sign.  Since
\[
       L^{-1}=\frac{\operatorname{adj}(L)}{\Delta}
       \qquad\text{and}\qquad
       (I-M)^{-1}=D L^{-1},
\]
Equation~\eqref{eq:value-as-inverse-entry} gives
\begin{equation}\label{eq:value-cofactor-section6}
       \val(G,o,d)
       =D(L^{-1})_{\iota,\dagger}
       =\frac{D\kappa}{\Delta}.
\end{equation}
Using \eqref{eq:positive-determinant},
\begin{equation}\label{eq:threshold-cofactor-section6}
       \val(G,o,d)>\frac12
       \quad\Longleftrightarrow\quad
       2D\kappa>\Delta.
\end{equation}

We do not compute or compare $\kappa$ and $\Delta$ separately.  Let $e_x$
denote the standard basis vector indexed by $x\in\mathcal J$ and define
\[
       L':=L-2D\,e_{\dagger}e_{\iota}^{\mathsf T}.
\]
Thus $L'$ differs from $L$ only in row $\dagger$ and column $\iota$.  The
determinant is affine in each individual matrix entry, with coefficient equal
to the corresponding cofactor.  The minor defining that cofactor does not
contain the modified entry, and therefore
\begin{equation}\label{eq:single-determinant-section6}
       \det L'=\Delta-2D\kappa.
\end{equation}
Combining \eqref{eq:threshold-cofactor-section6} and
\eqref{eq:single-determinant-section6},
\begin{equation}\label{eq:strict-threshold-single-det}
       \val(G,o,d)>\frac12
       \quad\Longleftrightarrow\quad
       \det L'<0.
\end{equation}
In particular, the equality case $\val(G,o,d)=\frac12$ corresponds to
$\det L'=0$ and is correctly rejected by the strict threshold test.

The entry oracle for $L'$ is explicit:
\[
       L'_{i,j}
       =L_{i,j}-2D\,\mathbf 1[(i,j)=(\dagger,\iota)].
\]
The entries of $L'$ still have polynomially many bits and are computable in
polynomial space by Lemma~\ref{sa:res:compute-M-elems}.  Therefore
Lemma~\ref{lem:succinct-determinant-pspace} computes the three-valued sign of
$\det L'$ in \PSPACE{}.  By \eqref{eq:strict-threshold-single-det}, this decides
{\tt $\{R,S\}$-Arrival-Quant}.
\end{proof}

\FloatBarrier
\section{Conclusion}
\label{sec:conclusion}

We have studied a family of stochastic generalisations of {\tt Arrival} obtained by allowing probabilistic and player-controlled vertices in addition to switching vertices. This gives a common framework containing the original {\tt Arrival} problem, Reachability Switching Games, Markov chains, Markov decision processes, and simple stochastic games as special cases.

For the combination of random and switching vertices, we have shown that the positive-probability and almost-sure reachability problems are \NP{}-complete and \coNP{}-complete, respectively. The quantitative threshold problem is \PP{}-hard and belongs to \PSPACE{}. End-component compression additionally shows that {\tt $\{R,S\}$-Arrival-Qual-1}, {\tt $\{S,2\}$-Arrival}, and {\tt $\{R,S,2\}$-Arrival-Qual-1} are polynomial-time equivalent and hence \coNP{}-complete. Introducing player-controlled vertices yields the further hardness results summarised in \Cref{tab:complexity}; in particular, the quantitative variants are \PSPACE{}-hard, and Qualitative-$1$ is \PSPACE{}-hard whenever Player~1 is available.

\section{Further Work}
\label{sec:future-work}

The most immediate open problem arising from our results is the complexity of quantitative {\tt $\{R,S\}$-Arrival}. We have shown that the problem is \PP{}-hard and belongs to \PSPACE{}, leaving a substantial gap between the lower and upper bounds. It is natural to ask whether the problem is in \PP{}; one plausible conjecture is that the \PP{} lower bound gives the correct complexity. Closing this gap appears to require a better understanding of how the long-term evolution of the switch configuration can be represented without explicitly expanding the exponentially large state space.

There are analogous gaps once player-controlled vertices are present. Quantitative {\tt $\{R,S,1\}$-Arrival} and {\tt $\{R,S,2\}$-Arrival} are \PSPACE{}-hard but are currently only known here to lie in \EXPTIME{}, while the two-player quantitative problem {\tt $\{R,S,1,2\}$-Arrival} has the general upper bound $\NEXPTIME\cap\coNEXPTIME$. It would be interesting to determine whether these upper bounds can be sharpened, and in particular whether the one-player quantitative variants belong to \PSPACE{}. More generally, one can ask whether strategies and values in these succinctly represented games admit representations that avoid explicit construction of the expanded stochastic game. For qualitative-$1$ with Player~1, the gap between \PSPACE{}-hardness and an \EXPTIME{} upper bound also remains open. In particular, it is unknown whether {\tt $\{R,S,1\}$-Arrival-Qual-1} is polynomial-time equivalent to {\tt $\{S,1,2\}$-Arrival}: directly replacing random choices with adversarial Player~2 choices is not sound at the almost-sure threshold, as Player~2 may enforce infinite paths of probability zero.

A second direction is suggested by recent algorithms for structurally restricted {\tt Arrival} instances. Haslebacher gives an algorithm parameterised by treewidth through the {\tt G-Arrival} framework \cite{Has25}, while G\"{a}rtner, Haslebacher and Hoang obtain polynomial-time algorithms for ladder instances \cite{GHH26}. Related progress has also been made for {\tt Multi-Arrival} on tree-like multigraphs \cite{GHM25}. It is natural to ask which of these structural tractability results survive after probabilistic vertices are introduced. For example, the complexity of quantitative {\tt $\{R,S\}$-Arrival} on graphs of bounded treewidth, or on graph classes on which ordinary {\tt Arrival} is known to be tractable, appears to be a natural intermediate question. The parallel results for {\tt Arrival} and simple stochastic games in \cite{Has25,GHH26} make this particularly suggestive for the mixed switching--stochastic setting studied here.

There is also a natural recursive extension of the present framework. {\tt Recursive Arrival} \cite{W23} was motivated by the analogy between {\tt Arrival} and state-transition models for which recursive versions were already well established: Recursive State Machines \cite{AEY01}, Recursive Markov Chains \cite{EY09}, and Recursive Markov Decision Processes and Recursive Stochastic Games \cite{EY15}. Combining that construction with the $\mathcal{B}$-Arrival framework of the present paper would give a natural notion of \emph{recursive $\mathcal{B}$-Arrival}. Under the local component semantics used for {\tt Recursive Arrival}, such a model could contain recursive calls together with switching, probabilistic, and player-controlled vertices. With only switching vertices it would recover {\tt Recursive Arrival}, while after removing switching vertices the corresponding choices of node types would lead back towards recursive Markov chains, recursive MDPs, and recursive stochastic games. Whether this combined model has an interesting complexity theory of its own, rather than simply inheriting the difficulty of its special cases, remains to be seen.

Beyond reachability, one can ask analogous questions for other objectives. For example, rather than asking whether a target is eventually reached, one could ask whether a vertex is visited infinitely often under optimal play. Parity games, mean-payoff games, and energy games provide other important graph-game models with related complexity-theoretic phenomena \cite{Z98,Jur98,EM79,ZP96,BFL+08,CAHS03}. The framework studied in this paper does not directly subsume these games: we vary the local transition mechanism while retaining a reachability objective, whereas these games obtain much of their structure from objectives defined over an infinite play. A possible direction is therefore to retain switching vertices but replace reachability by parity, mean-payoff, energy, or other $\omega$-regular objectives, and ask how the succinct state encoded by the switches changes the complexity of the resulting games.

Finally, one could consider variants with imperfect information. A simple example would be a model in which the initial switch configuration is not known to the players and the initial position of a switching vertex is revealed only when that vertex is first visited. This would separate the information available to the players from the hidden state maintained by the switching mechanism, and may give substantially different strategy and complexity questions.

\section*{Acknowledgements}
We thank the anonymous reviewers, including a referee of the conference submission who suggested a matrix-inversion route to the \PSPACE{} upper bound in \Cref{sa:res:rs-quant-in-pspace}; the present proof develops that suggestion with an explicit bit-level space argument.
%


\begin{thebibliography}{99}
\providecommand{\doi}[1]{\url{https://doi.org/#1}}
\bibitem{ACD22}
Auger D, Coucheney P, Duhaze L.
\newblock Polynomial Time Algorithm for ARRIVAL on Tree-like Multigraphs.
\newblock In: \emph{47th International Symposium on Mathematical Foundations of Computer Science (MFCS 2022)}.
\newblock Edited by Szeider, Stefan and Ganian, Robert and Silva, Alexandra.
\newblock Schloss Dagstuhl -- Leibniz-Zentrum f{\"u}r Informatik, 2022.
\newblock \doi{10.4230/LIPIcs.MFCS.2022.12}.
\bibitem{AEY01}
Alur R, Etessami K, Yannakakis M.
\newblock Analysis of Recursive State Machines.
\newblock In: \emph{Computer Aided Verification (CAV 2001)}.
\newblock Springer, pp. 207--220, 2001.
\newblock \doi{10.1007/3-540-44585-4_18}.
\bibitem{AkmalWilliams2021}
Akmal S, Williams R.
\newblock MAJORITY-3SAT (and Related Problems) in Polynomial Time.
\newblock In: \emph{62nd IEEE Annual Symposium on Foundations of Computer Science (FOCS 2021)}.
\newblock IEEE, pp. 1033--1043, 2021.
\newblock \doi{10.1109/FOCS52979.2021.00103}.

\bibitem{BaierKatoen2008}
Baier C, Katoen J-P.
\newblock \emph{Principles of Model Checking}.
\newblock MIT Press, 2008.
\newblock ISBN 9780262026499.
\bibitem{Ber84}
Berkowitz SJ.
\newblock On Computing the Determinant in Small Parallel Time Using a Small Number of Processors.
\newblock \emph{Information Processing Letters}, 1984.
\newblock \textbf{18}(3):147--150.
\newblock \doi{10.1016/0020-0190(84)90018-8}.
\bibitem{BFL+08}
Bouyer P, Fahrenberg U, Larsen KG, Markey N, Srba J.
\newblock Infinite Runs in Weighted Timed Automata with Energy Constraints.
\newblock In: \emph{Formal Modeling and Analysis of Timed Systems}.
\newblock Springer Berlin Heidelberg, pp. 33--47, 2008.
\newblock \doi{10.1007/978-3-540-85778-5_4}.
\bibitem{BorodinCookPippenger1983}
Borodin A, Cook SA, Pippenger N.
\newblock Parallel Computation for Well-Endowed Rings and Space-Bounded Probabilistic Machines.
\newblock \emph{Information and Control}, 1983.
\newblock \textbf{58}(1--3):113--136.
\newblock \doi{10.1016/S0019-9958(83)80060-6}.
\bibitem{CAHS03}
Chakrabarti A, de Alfaro L, Henzinger TA, Stoelinga M.
\newblock Resource Interfaces.
\newblock In: \emph{Embedded Software. EMSOFT}.
\newblock Springer, Berlin, Heidelberg, pp. 117--133, 2003.
\newblock \doi{10.1007/978-3-540-45212-6_9}.
\bibitem{Con92}
Condon A.
\newblock The Complexity of Stochastic Games.
\newblock \emph{Inf. Comput.}, 1992.
\newblock \textbf{96}(2):203--224.
\newblock \doi{10.1016/0890-5401(92)90048-K}.
\bibitem{Csa76}
Csanky L.
\newblock Fast Parallel Matrix Inversion Algorithms.
\newblock \emph{SIAM Journal on Computing}, 1976.
\newblock \textbf{5}(4):618--623.
\newblock \doi{10.1137/0205040}.
\bibitem{Der70}
Derman C.
\newblock Finite State Markovian Decision Processes.
\newblock Academic Press, Inc., USA, 1970.
\newblock \url{https://dl.acm.org/doi/10.5555/578852}.
\bibitem{DGKMW16}
Dohrau J, G{\"{a}}rtner B, Kohler M, Matou{\v{s}}ek J, Welzl E.
\newblock Arrival: A zero-player graph game in NP \ensuremath{\cap} coNP.
\newblock In: \emph{A Journey through Discrete Mathematics: A Tribute to Jiri Matousek}.
\newblock Springer, 2017.
\newblock \doi{10.1007/978-3-319-44479-6_14}.
\bibitem{EM79}
Ehrenfeucht A, Mycielski J.
\newblock Positional strategies for mean payoff games.
\newblock \emph{International Journal of Game Theory}, 1979.
\newblock \textbf{8}(2):109--113.
\newblock \doi{10.1007/BF01768705}.
\bibitem{EY09}
Etessami K, Yannakakis M.
\newblock Recursive Markov Chains, Stochastic Grammars, and Monotone Systems of Nonlinear Equations.
\newblock \emph{Journal of the ACM}, 2009.
\newblock \textbf{56}(1):1:1--1:66.
\newblock \doi{10.1145/1462153.1462154}.
\bibitem{EY15}
Etessami K, Yannakakis M.
\newblock Recursive Markov Decision Processes and Recursive Stochastic Games.
\newblock \emph{Journal of the ACM}, 2015.
\newblock \textbf{62}(2):11:1--11:69.
\newblock \doi{10.1145/2699431}.
\bibitem{FGMS19}
Fearnley J, Gordon S, Mehta R, Savani R.
\newblock Unique End of Potential Line.
\newblock \emph{46th International Colloquium on Automata, Languages, and Programming (ICALP 2019)}, 2019.
\newblock \textbf{132}:56:1–56:15.
\newblock \url{https://arxiv.org/abs/1811.03841}.
\bibitem{FGMS21}
Fearnley J, Gairing M, Mnich M, Savani R.
\newblock Reachability Switching Games.
\newblock \emph{Log. Methods Comput. Sci.}, 2021.
\newblock \textbf{17}(2):10:1--10:29.
\newblock \doi{10.23638/LMCS-17(2:10)2021}.
\bibitem{GHH+18}
G{\"{a}}rtner B, Hansen TD, Hub{\'{a}}cek P, Kr{\'{a}}l K, Mosaad H, Sl{\'{i}}vov{\'{a}} V.
\newblock ARRIVAL: Next Stop in CLS.
\newblock \emph{45th International Colloquium on Automata, Languages, and Programming}, 2018.
\newblock \textbf{107}:60:1–60:13.
\newblock \url{https://arxiv.org/abs/1802.07702}.
\bibitem{GHH21}
G{\"{a}}rtner B, Haslebacher S, Hoang HP.
\newblock A Subexponential Algorithm for ARRIVAL.
\newblock \emph{48th International Colloquium on Automata, Languages, and Programming}, 2021.
\newblock \textbf{198}:69:1–69:14.
\newblock \doi{10.4230/LIPIcs.ICALP.2021.69}.
\bibitem{GHH26}
G{\"{a}}rtner B, Haslebacher S, Hoang HP.
\newblock Sinks and Ladders: ARRIVAL and SSG with Two Vertices per Level.
\newblock In: \emph{13th International Conference on Fun with Algorithms (FUN 2026)}.
\newblock LIPIcs, vol. 366, pp. 19:1--19:16, 2026.
\newblock \doi{10.4230/LIPIcs.FUN.2026.19}.
\bibitem{GHM25}
Ghorbani E, Leander Hoff J, Mnich M.
\newblock A Quasi-Polynomial Time Algorithm for Multi-Arrival on Tree-Like Multigraphs.
\newblock In: \emph{42nd International Symposium on Theoretical Aspects of Computer Science (STACS 2025)}.
\newblock LIPIcs, vol. 327, pp. 39:1--39:19, 2025.
\newblock \doi{10.4230/LIPIcs.STACS.2025.39}.
\bibitem{Gil74}
Gill JT.
\newblock Computational complexity of probabilistic Turing machines.
\newblock In: \emph{Proceedings of the sixth annual ACM symposium on Theory of computing  - STOC '74}.
\newblock ACM Press, pp. 91--95, 1974.
\newblock \doi{10.1145/800119.803889}.
\bibitem{Has25}
Haslebacher S.
\newblock ARRIVAL: Recursive Framework \& $\ell_1$-Contraction.
\newblock In: \emph{52nd International Colloquium on Automata, Languages, and Programming (ICALP 2025)}.
\newblock LIPIcs, vol. 334, pp. 95:1--95:17, 2025.
\newblock \doi{10.4230/LIPIcs.ICALP.2025.95}.
\bibitem{HJ85}
Horn RA, Johnson CR.
\newblock Matrix Analysis.
\newblock Cambridge University Press, 1985.
\newblock \doi{10.1017/CBO9780511810817}.
\bibitem{Jur98}
Jurdzinski M.
\newblock Deciding the Winner in Parity Games is in UP\ensuremath{\cap} coUP.
\newblock \emph{Inf. Process. Lett.}, 1998.
\newblock \textbf{68}(3):119--124.
\bibitem{Kar17}
Karthik CS.
\newblock Did the train reach its destination: The complexity of finding a witness.
\newblock \emph{Information Processing Letters}, 2017.
\newblock \textbf{121}:17--21.
\newblock \doi{10.1016/j.ipl.2017.01.004}.
\bibitem{LohreySchmidtSchauss2017}
Lohrey M, Schmidt-Schau{\ss} M.
\newblock Processing Succinct Matrices and Vectors.
\newblock \emph{Theory of Computing Systems}, 2017.
\newblock \textbf{61}(2):322--351.
\newblock \doi{10.1007/s00224-015-9666-9}.
\bibitem{Man21}
Manuell G.
\newblock A simple lower bound for ARRIVAL.
\newblock \emph{CoRR}, 2021.
\newblock \textbf{abs/2108.06273}.
\newblock \url{https://arxiv.org/abs/2108.06273}.
\bibitem{Pap85}
Papadimitriou CH.
\newblock Games Against Nature.
\newblock \emph{J. Comput. Syst. Sci.}, 1985.
\newblock \textbf{31}(2):288--301.
\bibitem{Pap94}
Papadimitriou CH.
\newblock Computational complexity.
\newblock Addison-Wesley, 1994.
\bibitem{Savitch1970}
Savitch WJ.
\newblock Relationships Between Nondeterministic and Deterministic Tape Complexities.
\newblock \emph{Journal of Computer and System Sciences}, 1970.
\newblock \textbf{4}(2):177--192.
\newblock \doi{10.1016/S0022-0000(70)80006-X}.

\bibitem{W22}
Webster T.
\newblock The Stochastic Arrival Problem.
\newblock In: \emph{Reachability Problems}.
\newblock Springer, vol. 13608, pp. 93--107, 2022.
\newblock \doi{10.1007/978-3-031-19135-0_7}.
\bibitem{W23}
Webster T.
\newblock The Recursive Arrival Problem.
\newblock \emph{Electronic Proceedings in Theoretical Computer Science}, 2023.
\newblock \textbf{390}:168--184.
\newblock \doi{10.4204/EPTCS.390.11}.
\bibitem{Z98}
Zielonka W.
\newblock Infinite games on finitely coloured graphs with applications to automata on infinite trees.
\newblock \emph{Theoretical Computer Science}, 1998.
\newblock \textbf{200}(1-2):135--183.
\newblock \doi{10.1016/S0304-3975(98)00009-7}.
\bibitem{ZP96}
Zwick U, Paterson M.
\newblock The Complexity of Mean Payoff Games on Graphs.
\newblock \emph{Theor. Comput. Sci.}, 1996.
\newblock \textbf{158}(1\&2):343--359.

\end{thebibliography}
\end{document}